\documentclass[12pt]{article}
\usepackage{amssymb}
\usepackage{amsmath}
\usepackage{bm}
\usepackage{a4}
\usepackage{graphicx}
\usepackage{comment}
\usepackage{color}

\DeclareMathSymbol{\mrq}{\mathord}{operators}{`'}

\evensidemargin \oddsidemargin
\newcommand{\RR}{\mathbb{R}}
\newcommand{\ZZ}{\mathbb{Z}}
\newcommand{\NN}{\mathbb{N}}

\def\1{{\bf 1}}
\def\0{{\bf 0}}

\def \C {{\cal C}}
\def \D {{\cal D}}
\newcommand{\bD}{\overline{\cal D}}
\def \E {{\cal E}}

\def \G {{\cal G}}
\def \G {{\cal G}}
\def \I {{\cal I}}

\def \LL {{\cal L}}
\newcommand{\bL}{\overline{\cal L}}

\def \U {{\cal U}}

\newcommand{\UM}{\U_1^{{\scriptscriptstyle M}}}

\newcommand{\Um}{\U_1^{{\scriptscriptstyle m}}}

\def \xip {\xi_{{\scriptscriptstyle +}}}
\newcommand{\sm}{s_{{\scriptscriptstyle -}}}
\def \sp {s_{{\scriptscriptstyle +}}}
\def \spm {s_{{\scriptscriptstyle \pm}}}

\newcommand{\smm}{s_{{\scriptscriptstyle m}}}
\newcommand{\sM}{s_{{\scriptscriptstyle M}}}

\newcommand{\Be}{{\bm \epsilon}}
\newcommand{\Bep}{{\bm \epsilon}^{\scriptscriptstyle \perp}}

\newcommand{\bi}{\mathbf{i}}
\newcommand{\bj}{\mathbf{j}}
\newcommand{\bk}{\mathbf{k}}
\newcommand{\bom}{{\bar\omega}}

\newcommand{\q}{v}
\newcommand{\ty}{\breve{y}}
\newcommand{\cy}{\check{y}}
\newcommand{\MM}{\bm{L}}

\newcommand{\br}{\bm{r}}
\newcommand{\rr}{\bar{\bm{r}}}

\newcommand{\ti}{y}
\newcommand{\MN}{\bm{N}}

\newcommand{\cpsi}{{\check{\psi}}}
\newcommand{\tpsi}{\widetilde{\psi}}

\newcommand{\hD}{\hat\Delta}
\newcommand{\hs}{\hat s}
\newcommand{\hH}{\hat H}
\newcommand{\hJ}{\hat J}

\newcommand{\hbx}{\hat\bx}
\newcommand{\hxe}{\hat  x_e}
\newcommand{\hye}{\hat  y_e}
\newcommand{\hze}{\hat  z_e}
\newcommand{\hDO}{\Delta\!^{{\scriptscriptstyle (0)}}}

\newcommand{\hDOu}{\Delta_u}
\newcommand{\hDd}{\Delta_d}

\newcommand{\hsU}{s ^{{\scriptscriptstyle (1)}}}

\def\nn{\nonumber \\}
\newcommand{\bKp}{{\bm K}^{\scriptscriptstyle \perp}}

\newcommand{\bx}{{\bm x}}
\newcommand{\bxp}{{\bm x}^{\scriptscriptstyle \perp}}
\newcommand{\bX}{{\bm X}}
\newcommand{\bXp}{{\bm X}^{\scriptscriptstyle \perp}}
\newcommand{\bu}{{\bm u}}
\newcommand{\bup}{{\bm u}^{\scriptscriptstyle \perp}}

\newcommand{\bv}{{\bm v}}

\newcommand{\Bap}{{\bm \alpha}^{\scriptscriptstyle \perp}}
\newcommand{\bb}{{\bm \beta}}

\newcommand{\BD}{{\bm \Delta}}

\newcommand{\bchi}{{\bm \chi}}
\newcommand{\blambda}{{\bm \lambda}}

\newcommand{\bE}{{\bm E}}
\newcommand{\bEp}{{\bm E}^{\scriptscriptstyle \perp}}
\newcommand{\bB}{{\bm B}}
\newcommand{\bBp}{{\bm B}^{\scriptscriptstyle \perp}}
\newcommand{\bA}{{\bm A}}
\newcommand{\bAp}{{\bm A}\!^{\scriptscriptstyle \perp}}
\newcommand{\bjp}{{\bm j}^{\scriptscriptstyle \perp}}

\newcommand{\Bp}{{\bm p}}
\newcommand{\Bpp}{{\bm p}^{\scriptscriptstyle \perp}}

\newcommand{\DM}{\Delta_{{\scriptscriptstyle M}}}

\newcommand{\Dm}{\Delta_{{\scriptscriptstyle m}}}
\newcommand{\xiH}{\xi_{{\scriptscriptstyle H}}}
\newcommand{\bxiH}{\bar\xi_{{\scriptscriptstyle H}}}
\newcommand{\tH}{t_{{\scriptscriptstyle H}}}
\newcommand{\gammaM}{\gamma^{{\scriptscriptstyle M}}}

\newcommand{\be}{\begin{equation}}
\newcommand{\ee}{\end{equation}}
\newcommand{\bea}{\begin{eqnarray}}
\newcommand{\eea}{\end{eqnarray}}
\newcommand{\ba}{\begin{array}}
\newcommand{\ea}{\end{array}}
\newtheorem{prop}{Proposition}
\newtheorem{lemma}{Lemma}

\def\sq{\mbox{\rlap{$\sqcap$}$\sqcup$}}
\newenvironment{proof}[1]{\vspace{5pt}\noindent{\bf Proof #1}\hspace{6pt}}%
{\hfill\sq}
\newcommand{\bp}{\begin{proof}}
\newcommand{\ep}{\end{proof}\par\vspace{10pt}\noindent}
\begin{document}

\title{On the impact of short laser pulses on cold diluted plasmas
%A plane-wave model of the impact of  %intense and 
%short laser pulses on diluted %low-density 
%plasmas
}

\author{Gaetano Fiore$^{1,3,*}$\footnote{Corresponding author. Email: gaetano.fiore@na.infn.it}, \   Sergio De Nicola$^{2,3}$, \  Tahmina Akhter$^{2,3}$,  \
Renato Fedele$^{2,3}$, \\  Du\v{s}an Jovanovi\'c$^{4,5}$, 
   \\    %  \and
$^{1}$ Dip. di Matematica e Applicazioni, Universit\`a di Napoli ``Federico II'',\\
%Complesso Universitario  M. S. Angelo, Via Cintia, 80126 Napoli, Italy\\         %\and
$^{2}$  Dip. di Fisica, Universit\`a di Napoli ``Federico II'', \\
Complesso Universitario  M. S. Angelo, Via Cintia, 80126 Napoli, Italy\\         %\and
$^{3}$         INFN, Sez. di Napoli, Complesso  MSA,  Via Cintia, 80126 Napoli, Italy\\ 
$^{4}$  Inst. of Physics, University of Belgrade, %Pregrevica 118, 
11080 Belgrade, Serbia\\
$^{5}$  Texas A \& M University at Qatar, 23874 Doha, Qatar 
}

\date{}

\maketitle

\begin{abstract}

We analytically study the impact of a (possibly, very intense) short laser pulse onto an inhomogeneous cold diluted plasma at rest, in particular: the duration of the hydrodynamic regime; the formation and the features of plasma waves (PWs); their  wave-breakings (WBs); the motion of test electrons injected in the PWs.

If the pulse is a plane wave travelling in the 
$z$-direction, and the initial plasma density (IPD) depends only on $z$, then suitable matched bounds on the maximum and relative variations of the IPD, as well as the intensity and duration of the pulse, ensure a strictly hydrodynamic evolution of the electron fluid  during its whole interaction with the pulse,
while ions can be regarded as immobile. This evolution is ruled by a family (parametrized by $Z\ge 0$) of decoupled systems of non-autonomous Hamilton equations with 1 degree of freedom, which determine how electrons initially located in the layer $Z\le z<Z+dZ$ move; $\xi=ct-z$ replaces time $t$ as the independent variable. This family of ODEs is obtained by reduction from the Lorentz-Maxwell and continuity PDEs for the electrons' fluid within the spacetime region where the change of the pulse is negligible. After the laser-plasma interaction the Jacobian of the map from Lagrangian to Eulerian coordinates is {\it linear-quasi-periodic} in $\xi$. We determine spacetime locations and features of the first wave-breakings of the wakefield PWs, the motion of test electrons (self-)injected in the PWs. The energy of those trapped in a single PW trough grows linearly with the distance gone, where the IPD is constant. 

If the pulse  has cylindrical symmetry and a not too small radius, the same conclusions hold for the part of the plasma enclosed within the causal cone swept by it. 

This computationally light approach may help in a preliminary study of extreme acceleration mechanisms of electrons (LWFA, etc.), before 2D or 3D PIC simulations.

\end{abstract}


\section{Introduction}  

Interactions between ultra-intense  laser pulses and plasmas lead to  very interesting phenomena
\cite{Kruer19,SprEsaTin90PRA,SprEsaTin90PRL,Mac13}, 
in particular  laser wakefield acceleration  (LWFA)  \cite{Tajima-Dawson1979,Sprangle1988,EsaSchLee09,TajNakMou17} and other extremely compact acceleration mechanisms of charged particles.  Therefore huge investments are being devoted\footnote{We just mention the EU-funded project {\it Eupraxia} \cite{Eupraxia19AIP,Eupraxia19JPCS,Eupraxia20EPJ}.} to develop 
new, table-top accelerators on the base of such mechanisms, with a number of extremely important applications  in particle physics, materials science, biology, medicine, industry, environmental remediation, etc.
In general,  the equations ruling  these phenomena, i.e. Maxwell equations coupled to those of a  kinetic  theory
for the plasma  electrons and ions,
 can be  solved only numerically via  particle-in-cell (PIC) techniques, which involve huge and expensive computations
for each choice of the free parameters.
Sometimes,  treating the plasma as a multicomponent  fluid and solving
  the (simpler) associated hydrodynamic equations (analytically or via  multifluid codes, e.g. QFluid  \cite{TomEtAl17}; or
via hybrid kinetic/fluid  codes) provides essentially the same accuracy of predictions; 
but in general it is not known a priori in which conditions, or spacetime regions, this is possible. 

Most fluid models studied analytically assume a constant IPD; of course, this is not suitable to describe the pulse-plasma ``impact'', i.e. what occurs at the vacuum-plasma interface when  (and shortly after) the pulse reaches the edge of the plasma, as the IPD is not constant there. An exception is Ref. \cite{BraEtAl08}, which analyzes in detail what happens for a few piecewise linear IPDs. 
We believe that impacts of very short laser pulses onto a cold diluted plasma  at rest 
(or onto matter which is locally  ionized  into a plasma by the front of the pulse itself)
deserve  a deeper understanding because, among other things,  they may 
 generate: i) a plasma wave (PW) 
\cite{AkhPol56,Daw59,GorKir1987,BerMur90}
or even a {\it ion bubble}\footnote{Namely, a region containing only ions, because all electrons have been expelled out of it.} \cite{RosBreKat91,MorAnt96,PukMey2002,KosPukKis2004,LuEtAl2006,LuHuaZhoEtAl06},
producing
the LWFA, i.e.  accelerating  a small bunch of  electrons (that are injected in a PW trough
trailing the pulse; they are dubbed as {\it witness})  to very high energy,  in the forward direction; ii) the {\it slingshot effect}  \cite{FioFedDeA14,FioDeN16,FioDeN16b},
i.e. the backward acceleration and expulsion of energetic electrons from the vacuum-plasma interface, during or just after the impact and before vacuum heating \cite{Bru87-88,GibbBel92}.
Pursuing the research line of \cite{FioCat18,FioDeAFedGueJov22}, here we show that,  
with the help of the  improved  plane  Lagrangian model  of Ref. \cite{Fio14JPA,Fio18JPA} and very little computational power,
 we can obtain, for a broad class of inhomogeneous IPD, important information about  such impacts, in particular: a characterization of the hydrodynamic regime (HR); the formation and the features of PWs;  wave-breaking (WB); 
LWFA of {\it self-injected} test electrons
(which is very welcome if it can be controlled well).

We proceed in two steps: first a rather rigorous analysis  (sections \ref{plasmaelectrons-in-plane-model}-\ref{self-injected-electrons}) of the associated limit (1D) plane problem via the mentioned 1D Lagrangian 
(in the sense of non-Eulerian) model \cite{Fio14JPA,Fio18JPA}
(which we recap and slightly enrich  in section \ref{Setup}, to make the paper self-contained); then an
estimate of the deviations of the real (transversely dependent, i.e. 2D or 3D) problem from the plane one
(sections \ref{finiteR}, \ref{conclu}). 

\begin{figure}[htbp]
\includegraphics[height=5cm]{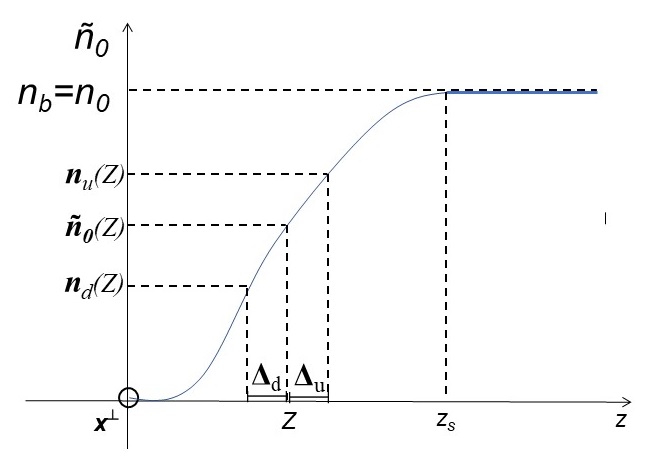}\hfill \includegraphics[height=5cm]{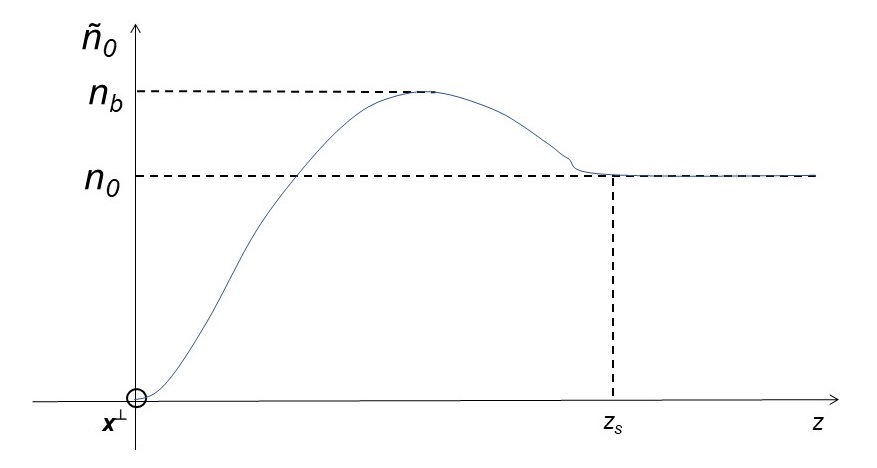}
\caption{Two examples of bounded initial density profiles with a final plateau:
 $\widetilde{n_0}(z)=n_0$ for $z\ge z_s$. In the left one we also illustrate the meaning of the functions $n_u(z),n_d(z)$ (see section \ref{during}). 
}
\label{fig1}
\end{figure}

\medskip
In the mentioned plane model one assumes that the plasma is initially neutral, unmagnetized and at rest with zero densities in the region  \ $z\!<\! 0$.
More precisely, the $t\!=\!0$ initial conditions for  the electron fluid  Eulerian density $n_e$ and velocity $\bv_e$ are of the type
\be 
\bv_e(0 ,\!\bx)\!=\!\0, \qquad n_e(0,\!\bx)\!=\!\widetilde{n_0}(z), 
 \label{asyc}
\ee
where the initial electron (as well as proton) density $\widetilde{n_0}(z)$ fulfills (see fig. \ref{fig1})
\be 
 \widetilde{n_0}(z)\!=\!0 \:\:\mbox{if }\: z\!\le\! 0, \qquad
0\!<\!\widetilde{n_0}(z)\!\le\! n_b  \quad \mbox{if }\: z\!>\!0
 \label{n_0bounds}
\ee
for some constant $n_b\!>\!0$. 
Here we assume that $\widetilde{n_0}(z)$ is continuous
and at least piecewise differentiable, with a suitably bounded derivative; 
moreover, we  make the results of sections \ref{test-electron}, \ref{self-injected-electrons} easier and more explicit by further assuming that $\widetilde{n_0}(z)$ becomes a constant $n_0$ for  $z\ge z_s>0$ (cf. fig. \ref{fig1}). As for the laser pulse,
one assumes that before the impact it is a sufficiently short [see (\ref{Lncond}) and (\ref{Lncond'}) below] free plane transverse wave  travelling  in the $z$-direction, i.e.
the electric and magnetic fields $\bE,\bB$  (the {\it pump}) are of the form
\be
\bE (t, \bx)=\bEp (t, \bx)=\Be^{{\scriptscriptstyle\perp}}\!(ct\!-\!z),\qquad  \bB=\bB^{{\scriptscriptstyle\perp}}=
\bk\!\times\!\bEp \qquad \mbox{if } t\le 0            
      \label{pump}
\ee
($\bk$ is the unit vector in the $z$-direction, and the superscript $\perp$ denotes vector components orthogonal to $\bk$); we choose the time origin so that $\xi\!=\!0$ is the left extreme of the support $S_{\Be^{{\scriptscriptstyle\perp}}}$ of $\Be^{{\scriptscriptstyle\perp}}\!(\xi)$, 
i.e. the pulse reaches the plasma at time $t\!=\!0$, and denote by $l$ the pulse length, i.e. the size of
the smallest interval $[0,l]$ containing $S_{\Be^{{\scriptscriptstyle\perp}}}$. 
One describes the plasma as a fully relativistic collisionless fluid of electrons and a static fluid of ions (section \ref{planemodel});
by continuity and the smallness of  the mass ratio $m/m_I$ (for all ions $I$), such a hydrodynamic regime (HR) is justified for a sufficiently short time lapse, because at the impact time $t\!=\!0$ the plasma is made of two static fluids, the ions' and the electrons'. 
One imposes that  $\bE,\bB$ and the plasma dynamical variables fulfill the Lorentz-Maxwell and continuity equations, but neglects  the depletion of the pump, i.e. describes the transverse EM field still by (\ref{pump});
again, this is  justified for a sufficiently short time lapse, by continuity and the plane symmetry of the problem. 
The specific time lapse for both is determined   {\it a posteriori},  by self-consistency.  
This allows one to  reduce the system of Lorentz-Maxwell and continuity 
partial differential equations (PDEs)  into ordinary differential equations (ODEs), more precisely into 
the family of {\it decoupled systems of Hamilton 
equations for systems with 1 degree of freedom}  (\ref{heq1}) parametrized by the
initial longitudinal coordinate $Z$ of the electrons \cite{Fio14JPA,Fio18JPA} 
(for shorter presentations see \cite{Fio14,Fio16b,Fio17PPLA,FioCat19,Fio21JPCS}).
In these ODEs one adopts: the light-like coordinate 
\ $\xi=ct\!-\!z$ \ as the independent variable in the Lagrangian description, in place of  time $t$;
the transverse components $\Bpp$ and the light-like component 
\ $p^0\!-\!cp^z\equiv mc^2 s$ \ (in place of the longitudinal one $p^z$) of the momentum 
as unknowns depending on $\xi$, both for a single electron (section \ref{1-particle}) and for the
generic electron fluid element  (section \ref{planemodel})\footnote{As the usual {\it quasistatic approximation} (see e.g.\cite{SprEsaTin90PRA,SprEsaTin90PRL}), our 
1D Lagrangian model: assumes that the plasma is initially cold, neutral, unmagnetized, at rest and that the plane
 pulse fulfills (\ref{Lncond});  neglects the motion of ions and 
the depletion of the pump.
On the other hand, to reduce the PDEs into (Hamiltonian) ODEs and thus simplify their resolution 
it adopts  $(\xi,Z)$  instead of $(\xi,t)$ as independent variables (and hence a Lagrangian instead of a Eulerian description of the electron fluid), as well as the longitudinal coordinate $z_e$ and 
the light-like component $s$ of its momentum  (rather than the longitudinal one) as unknowns; the $Z$-electron potential energy  (in the Lagrangian description), which
replaces the scalar potential in the Eulerian description, is explicitly expressed 
in terms of the unknown  $z_e$; finally, neither Fourier analysis of the pump nor a frequency-dependent refractive index is needed in its formulation. 
}. 
These Hamilton equations are non-autonomous only for $0\!<\!\xi\!<\!l$; 
for $\xi\ge l$ they can be solved also by quadrature, using the energy
integrals of motion $\hH(\xi,Z)=\hH(l,Z)=:h(Z)=$ const. Solving them 
yields the motions of the $Z$-electrons'  fluid elements, which are fully represented through their worldlines (WLs) in Minkowski space. In fig.  \ref{Worldlinescrossings} we have displayed 
the projections onto the $z,ct$ plane of these WLs for a specific set of input data; as we can see, the PW
emerges from them as a collective effect. 
Mathematically, the PW features can be derived passing to the Eulerian description of the electron fluid
(section \ref{plasmaelectrons-in-plane-model}). 

As it is well-known, when reached by the pulse, electrons start oscillating transversally (i.e. in the $x,y$-directions)
and drifting  in the positive $z$-direction, respectively pushed by  the electric and magnetic parts of the Lorentz force due to the pulse (see section \ref{during}); thermal effects may be neglected provided the pulse gives the electrons
a kinetic energy much greater than their initial thermal one, thus justifying the assumption $\bv_e(0 ,\!\bx)\!=\!\0$.
Thereafter, electrons  start oscillating also longitudinally (i.e. in $z$-direction), pushed by the restoring electric force due to charge separation. The reader can recognize such initial longitudinal motions e.g. in
fig. \ref{graphsb}b and from the electron WLs reported in fig. \ref{Worldlinescrossings}. 

It turns out that the electron fluid dynamics is simpler if the pulse is
{\it essentially short} \cite{FioDeAFedGueJov22}, what we shall assume henceforth. By this we mean that the pulse overcomes each plasma electron before the $z$-displacement $\Delta$ of the latter  reaches a negative minimum for the first time, as it occurs for the WL $\lambda_2$ in fig. \ref{Worldline}a.
An essentially short pulse will be said to be {\it strictly short} \cite{FioDeAFedGueJov22} if it overcomes each electron before  $\Delta$
becomes negative for the first time, as it occurs for the WL $\lambda_1$ in fig. \ref{Worldline}a. These conditions are  formalized in Definition (\ref{Lncond'}). 
By Proposition 1 of \cite{FioDeAFedGueJov22}, if  a pulse  is symmetric under inversion about its center 
(i.e., $\Be^{{\scriptscriptstyle\perp}}\!(\xi)=\Be^{{\scriptscriptstyle\perp}}\!(l\!-\!\xi)$) and not so intense to induce relativistic electrons   motions, then it
is strictly short, essentially short if its duration $l/c$  does not respectively exceed $1/2$, $1$  times the nonrelativistic (NR) plasma oscillation period $t_{{\scriptscriptstyle H}}^{{\scriptscriptstyle nr}}\!\equiv\!\sqrt{\pi m/n_b e^2}$ associated to the maximum $n_b$ of $\widetilde{n_0}$; namely,
\be
G:=\sqrt{\frac{n_b e^2}{\pi mc^2}}l \: \le \: \left\{\!\!\ba{ll} 1/2 &\quad\Rightarrow\quad  \mbox{NR strictly short pulse},\\[6pt] 1 &\quad\Rightarrow\quad  \mbox{NR essentially short pulse};\ea\right. 
   \label{Lncond}
\ee
here  $-e,m$ are the electron charge and mass, $c$ is the speed of light\footnote{If the pulse is  a slowly modulated monochromatic wave (\ref{modulate})
with wavelength $\lambda=2\pi/k$, then (\ref{Lncond}) implies a fortiori $\frac{4\pi e^2}{mc^2}n_b \lambda^2\ll1$,  so that the plasma is {\it underdense}.}. 
Since the relativistic oscillation period $\tH$ is not independent of the oscillation amplitude, but grows with the latter, which in turn grows with the pulse intensity,  conditions (\ref{Lncond'}) can be fulfilled also with a larger $G$. 
Condition (\ref{Lncond'}b) is  compatible with maximizing the  oscillation amplitude, and thus also the energy transfer from the pulse to the plasma wave  \cite{SprEsaTin90PRL,FioFedDeA14}; in the  NR regime this is achieved through a suitable $l$ such that 
$G\:\sim \: 1/2$.
As we make no extra assumptions on the Fourier spectrum or the polarization of $\Be^{{\scriptscriptstyle\perp}}$, our results can be applied  to all essentially short pulses, ranging from almost monochromatic to so called ``impulses"  (i.e. with one, or even a `fraction' of a, cycle) \cite{Aki96,CouEtAl06,Mor10,MouMirKhaSer14}.
For the same reason our equations do not involve 
dispersion relations, frequency-dependent refractive indices, etc.

Our main goal here is to explore by an apriori analysis the {\it general consequences} of our plane model for {\it all} input data  (i.e. all pair of functions $\widetilde{n_0},\Be^{{\scriptscriptstyle\perp}}$) fulfilling the above conditions,
choosing some specimen input data to illustrate them,
rather than just to solve the ODE's (\ref{heq1}) and describe the results for  such data. Our motivation is to provide
criteria that allow a preliminary selection of input data suitable for specific purposes, e.g. an efficient LWFA, before performing a detailed analysis of the dynamics (via numerical resolution of the family (\ref{heq1}) 
for $Z$ in a sufficiently fine lattice, or via multifluid simulations, or PIC ones, etc.).

The HR breaks where worldlines intersect, leading to WB of the PW. 
No WB occurs as long as the Jacobian  $\hJ$  of the transformation 
 from the Lagrangian  to the Eulerian coordinates remains positive (section \ref{Wave-breakings}).
If the initial density is uniform, $\widetilde{n_0}(Z)\!=\!n_0\!=\!$ const, 
not only the initial conditions (\ref{heq2}), but also equations  (\ref{heq1}) 
become $Z$-independent; consequently, also their solutions become $Z$-independent,
and $\hJ\equiv 1$ at all $\xi$. Otherwise,  for $\xi\ge l$ \ $\hat J(\xi,Z)$ 
 is {\it linear-quasiperiodic}
(LQP) in $\xi$ with  period $\xiH\equiv c\tH$ (section \ref{Jsigma-after}). In general, we say that a function $f$ 
of a variable $y$ is LQP if it can be written in the form
\be
f(y)=a(y)+y \, b(y)
 \label{lin-pseudoper}
\ee
where $a,b$ are periodic in $y$ with the same period, and $b$ has zero mean over  a period; 
$b(y)$  oscillates between positive and negative values, and so does the second term,
 which dominates as $y\to \infty$, with $y$ acting as a modulating amplitude.
Since $b$ for the decomposition (\ref{lin-pseudoper}) of $\hat J$ vanishes 
identically only if $\widetilde{n_0}=$ const, we recover the well-known result \cite{Daw59,BulEtAl98,BraEtAl08} that 
if $\widetilde{n_0}$ is inhomogeneous WB is unavoidable after a sufficiently long time.
 Moreover, we find (section \ref{Bounds0l})
an approximation in closed form for $\hat J$ that is good at least for small $\xi$, in particular $\xi\le l$,
and allows to determine apriori bounds for $\hJ$ based on the input data.
If $n_b$ or the relative variations of $\widetilde{n_0}(Z)$ are sufficiently small,  
then $\hat J>0$, and there is no wave-breaking
during the laser-plasma interaction (WBDLPI), i.e. for $0\le \xi\le l$: the HR holds at least for $0\le \xi\le l$.
After  the laser-plasma interaction we can more easily characterize and control the HR and the WB spacetime regions and features by means of (\ref{lin-pseudoper}).

 In section \ref{test-electron} we find the equations of motion, with $\xi$ as the `time' variable, of  test particles (TP) injected in the PW  and the qualitative behaviour of their solutions; differently from the standard treatments (see e.g. \cite{EsaSprKraTin96,BraEtAl08}), in our model TP cannot dephase, because their speed
is always smaller than the phase velocity $c$ of the PW.
 In particular, we find (section \ref{self-injected-electrons}) that the maximal energy
of the electrons  self-injected (via WB) in the PW and trapped by a single trough {\it grows approximately linearly} with the distance gone, cf. eq. (\ref{s_i^m<0|s_iz_i}).

In section \ref{finiteR} we 
use causality and  geometric arguments to qualitatively adapt the predictions
to the finiteness of the spot size $R$ in the ``real world". In section \ref{conclu} we 
discuss the results and draw the conclusions. In the appendix  (section \ref{App}) we have concentrated  some lengthy proofs or more technical results.
As an illustration and a test of our model, we apply it to the data  considered in the PIC simulations of
Ref. \cite{BraEtAl08}; we find consistent results.

\section{Setup and plane model}
\label{Setup}

We start by fixing some conventions: we shall say that a function $f(u)$ grows (resp. decreases) in a certain interval $I$
if $u'>u$ implies $f(u')\ge f(u)$ (resp. $f(u')\ge f(u)$); that $f(u)$ grows (resp. decreases) 
{\it strictly} in $I$ if $u'>u$ implies $f(u')> f(u)$ (resp. $f(u')<f(u)$).
We shall abbreviate $\dot f\equiv df/dt$, $\partial_z\equiv \partial/\partial z$, $\partial_Z\equiv \partial/\partial Z$, etc.

\subsection{Reformulation of the dynamics of a single charged particle}
\label{1-particle}

The  equations of motion of a charged particle in a given  external
electromagnetic  (EM) field is non-autonomous and highly nonlinear in the unknowns \  $\bx(t)$, \ $\Bp(t)=mc\,\bu(t)$:
\bea
\ba{l}
\displaystyle\dot\Bp(t)=q\bE[ct,\bx(t)] + \frac{\Bp(t) }{\sqrt{m^2c^2\!+\!\Bp^2(t)}}  \times q\bB[ct,\bx(t)] 
,\\[6pt]  
\displaystyle
\frac{\dot \bx(t)}c =\frac{\Bp(t) }{\sqrt{m^2c^2\!+\!\Bp^2(t)}} ,
\ea
\label{EOM}
\eea
Here \ $m,q,\bx,\Bp$ \ are the  rest mass, electric charge, position
and   relativistic momentum of the particle. We use  Gauss CGS units. 
In terms of EM potential 4-vector  $(A^\mu)=(A^0,-\bA)$  the electric and magnetic field read
\ $\bE=-\partial_t\bA/c-\nabla A^0$ and $\bB=\nabla\!\times\!\bA$. \
$\bx\!=\!x\bi\!+\!y\bj\!+\!z\bk\!=\bxp\!+\!z\bk$ will be the decomposition of $\bx$ in the cartesian coordinates of the laboratory frame.
As usual it is convenient to use dimensionless variables: \
$\bb\!\equiv\!\bv/c\!\equiv\!\dot \bx/c$, \  the Lorentz relativistic factor  
$\gamma\!\equiv\!dt/d\tau\!=\!1/\sqrt{1\!-\! \bb^2}$ ($\tau$ is the proper time of the particle), the 4-velocity \ $u\!=\!(u^0\!,\bu)
\!\equiv\!(\gamma,\!\gamma \bb)\!=\!\left(\!\frac {p^0}{mc^2},\!\frac {{\bf p}}{mc}\!\right)$, i.e. the dimensionless version of the 4-momentum; whence
$\gamma=\sqrt{1\!+\! \bu^2}$.

\begin{figure}
\centering
\includegraphics[width=8cm]{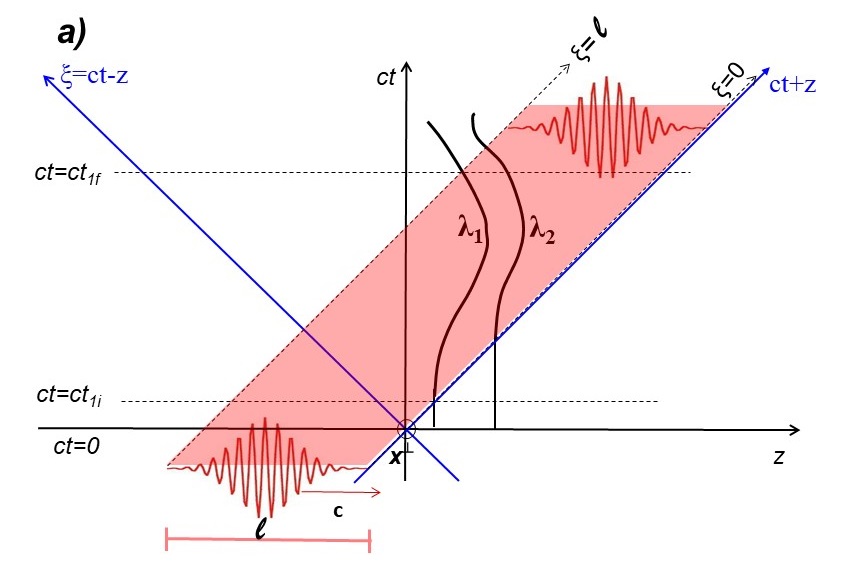}\hfill \includegraphics[width=8cm]{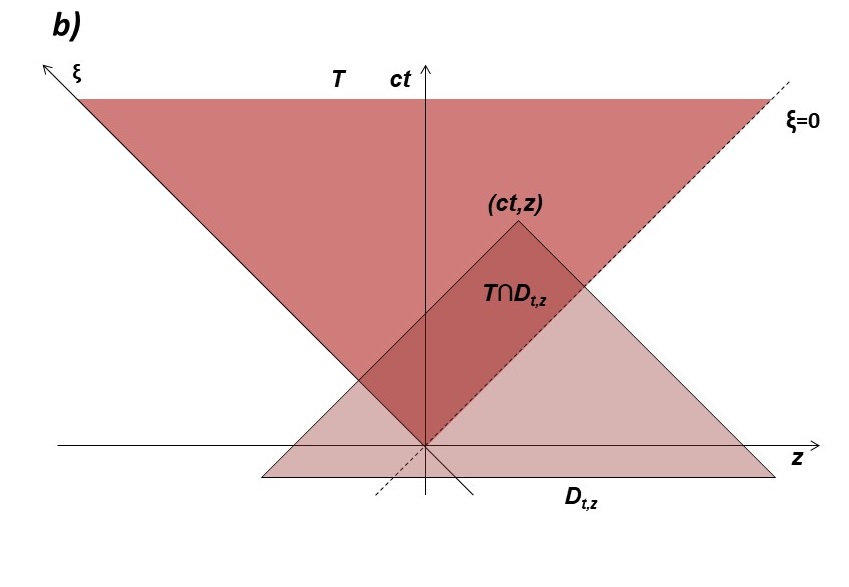}
\caption{a)  Two examples $\lambda_1,\lambda_2$ of particle worldlines (WLs) in Minkowski space (after projection onto the $z,ct$ plane), and their intersections with the support (pink) of a plane EM wave of total length $l$ moving in the positive $z$ direction.  Since each WL intersects once every hyperplane $\xi\!=$ const (beside every hyperplane $t\!=$ const),
we can use $\xi$ rather than $t$ as a parameter along it. While the instants $t_i,t_f$ of intersection with the front and the end of the EM wave depend on the particular WL
(in the picture we have pinpointed those for $\lambda_1$), the corresponding light-like coordinates are the same for all WLs: $\xi_i=0$, $\xi_f=l$. \ \ b)  The 2-dim future causal cone
$T\!=\!\{ (t,\!z)\: |\:  ct \!>\! |z| \}$ of the origin, the past causal cone
$D\!_{t,z}\!=\!\{ (t'\!,\!z')\: |\:  ct\!-\!ct' \!>\! |z\!\!-\!\!z'| \}$ of the point $(t,z)\in  T$, and their intersection.}
\label{Worldline}       
\end{figure}

Since no particle can reach the speed of light $c$ [$|\dot\bx|\!<\!c$ by (\ref{EOM}b)], \ $\tilde \xi(t)\!\equiv\!ct\!-\!z(t)$  is strictly growing,
and we can adopt $\xi\!=\!ct\!-\!z$ as a  parameter alternative to $t$ on the worldline $\lambda$ 
of the particle  in Minkowski spacetime (see fig. \ref{Worldline}a); in other words, this is possible because $\lambda$ intersects exactly once not only every hyperplane $ct\!=\!$ const, but also every hyperplane  $\xi\!=\!$ const. We denote as $\hat \bx(\xi)$ the position of the particle as a function of $\xi$; it is determined  by the
equation $\hat \bx(\xi)=\bx(t)$. More generally we shall put a caret
to distinguish the dependence of a dynamical variable on $\xi$ rather than on $t$, 
and denote $\hat f(\xi, \hat \bx)\equiv
f[(\xi\!+\!  \hat z)/c,  \hat \bx]$ for any given function $f(t,\bx)$; 
we shall also abbreviate $\dot f\!\equiv\! df/dt$, $\hat f'\!\equiv\! d\hat f/d\xi$ (total derivatives). Note that
$|\dot z|\!<\!c$ leads only to $\hat z'>-1/2$ (whereas there is no upper bound for $\hat z'$).

It is convenient to transform (\ref{EOM}) by both the  change of the independent variable  $t\mapsto \xi$
and the change of dependent (and unknown) variable $u^z\mapsto s$, where
$s$ is the  {\it light-like relativistic factor}, or shortly \  {\it $s$-factor} \ \cite{Fio18JPA},
i.e. the light-like component $u^-$ of $u$: 
\be
s\equiv\gamma\!- u^z=u^-=\gamma(1-\beta^z)>0;        \label{defs0}
\ee
clearly, $s$ is  positive-definite. The first reason is that $\gamma,\bu,\bb$ are {\it rational}  functions of $\bu^{{\scriptscriptstyle\perp}},s$:
\be
\gamma\!=\!\frac {1\!+\!{\bu^{{\scriptscriptstyle\perp}}}^2\!\!+\!s^2}{2s}, 
\qquad\quad u^z\!=\!\frac {1\!+\!{\bu^{{\scriptscriptstyle\perp}}}^2\!\!-\!s^2}{2s}, 
 \qquad\quad \bb\!=\! \frac{\bu}{\gamma}                                     \label{u_es_e}.
\ee
These relations hold also with the caret on all variables.
Note also that \ $s\!=\!d\xi/d(c\tau)$. \ Replacing \ $d/dt\mapsto(c s/ \gamma)d/d\xi$, \ using \  
$p^z\!=\!m dz/d\tau$, \ (\ref{u_es_e}), putting a caret on all dynamical variables, and abbreviating
$v\equiv \hat\bu^{\scriptscriptstyle\perp}{}^2$, we transform (\ref{EOM}) into the system \cite{Fio18JPA}
\bea
\ba{l}
 \displaystyle\hat\bu^{\scriptscriptstyle \perp}{}' =\frac q{mc^2}\!\left[\left(\frac 12+\frac {1\!+\!v}{2\hat s^2}\right)\!\hat\bE+\frac{\hat\bu}{\hat s}\!\times\!\hat\bB\right]^{\scriptscriptstyle \perp}\!, \\[12pt]
 \displaystyle \hat s'  = \frac {q}{mc^2}\!\left[\frac{\hat\bu^{\scriptscriptstyle \perp}}{\hat s}\!\cdot\!\hat\bEp\!-\!\hat E^z\!-\!\frac{(\hat\bu^{\scriptscriptstyle \perp}\!\times\!\hat\bBp)^z}{\hat s}\right], \\[12pt]
\hat \bxp{}'
=\displaystyle\frac {\hat\bu^{\scriptscriptstyle \perp}}{\hat s}, \qquad \qquad  \hat z'
=\displaystyle\frac {1\!+\!v}{2\hat s^2}\!-\!\frac 12, %\label{eqx}
\ea
\label{equps0}
\eea
which is {\it rational} in the  unknowns \ $\hat\bu^{\scriptscriptstyle \perp}(\xi),\hat s(\xi),\hat\bx(\xi)$.
Equations (\ref{EOM}), (\ref{equps0}) are resp. equivalent \cite{Fio18JPA} to the Hamilton equations of the particle with $t$ and $\xi$ as the independent
variable; the corresponding Hamiltonians $H=\gamma+A^0(t,\bx)q/mc^2$,
$\hat H=\hat\gamma+\hat A^0(\xi,\bx)q/mc^2$ (which we have made dimensionless dividing them by $mc^2$) are obtained by Legendre transform from 
Lagrangians $L, \LL$ that in turn arise from the decomposition of the same
action functional $S(\lambda)$ using resp. $t,\xi$ as a parameter on the worldline $\lambda$.
By (\ref{u_es_e}) $\hat H$ is rational in $\hat\bu^{\scriptscriptstyle \perp},\hat s$.
If the EM field is the sum of a plane travelling wave in the $z$-direction (the pump)  and a 
longitudinal electric field 
\be
\bE(t ,\bx)=\Bep(ct\!-\!z)+\bk E^z(z),\qquad \bB(t ,\bx)=\bk\times\Bep(ct\!-\!z),
\label{EBfields0}
\ee
(the EM field inside a plasma (\ref{n_0bounds}) hit by
a pulse (\ref{pump}) has this form as long as depletion is negligible, see  (\ref{explE'})), then (\ref{equps0}a) becomes \ $\hat\bu^{\scriptscriptstyle \perp}{}' (\xi)=q\Bep(\xi)/mc^2$, \  which is solved by 
\be
\hat\bu^{\scriptscriptstyle \perp} (\xi)=-\frac q{mc^2}\Bap(\xi) +\bKp,\qquad \quad\mbox{where }\:
\Bap(\xi)\equiv -\!\int^{\xi}_{ -\infty }\!\!\!d\zeta\:\Bep(\zeta)           \label{hatuperp}
\ee
($\bKp$ is an integration constant); replacing this the rest of  (\ref{equps0}) simplifies into
\be
\hat z' =\frac {1\!+\!\hat\bu^{\scriptscriptstyle \perp}{}^2}{2\hat s^2}\!-\!\frac 12, \qquad 
-\hat s'(\xi)  = \frac {q}{mc^2}\hat E^z(\hat z),            \label{reduced}
\ee
which is a Hamiltonian system in 1 dimension in the two (canonically conjugated)
unknowns $\hat z,-\hat s$ (if in particular $E^z=0$ then $\hat s=s_0=$ const and also
$\hat z$ is obtained by integration).  In fact, once (\ref{reduced}) is solved 
one determines from (\ref{equps0}c) 
also $\hat\bx^{\scriptscriptstyle \perp}(\xi)$, and thus the whole \ $\hat\bx(\xi)$:
\bea
&\hat\bx(\xi)=\bx_0+\hat \BD\!(\xi), \qquad &\mbox{where}\quad \hat \BD\!(\xi)\!\equiv\!\!\displaystyle\int^\xi_{0}\!\!\! d\eta \,\frac{\hat\bu(\eta)}{\hat s(\eta)}.          \label{hatsol}
\eea
The Cauchy problem (\ref{reduced}) with initial conditions $\big(\hat z(\xi_0),\hat s(\xi_0)\big)=(z_0,s_0)$
is equivalent to
\bea
\hat z(\xi)=z_0\!+\!\int^\xi_{\xi_0}\!\! d\zeta
\left[\frac {1\!+ \!v(\zeta)}{2\hat s^2(\zeta)}\!-\!\frac 12\right]\!, \qquad  
\hat s(\xi)=s_0\!- \!\int^\xi_{\xi_0}\!\!d\zeta\,\frac{qE_s^z[\hat z(\zeta)]}{mc^2}.
\label{heq1rint} 
\eea
 Clearly, the function defined by 
\be
\hat t(\xi)\equiv \hat t(\xi_0)+\!\!\int^\xi_{\xi_0}\! \frac{d\eta}c \,\frac{\hat\gamma(\eta)}{\hat s(\eta)}
=\hat t(\xi_0)+\!\!\int^\xi_{\xi_0}\! \frac{d\eta}{2c} 
\left[\frac{1\!+\!{\bu^{{\scriptscriptstyle\perp}}}^2(\eta)}{\hat s^2(\eta)}+1\right]
=\frac{\xi+\hat z(\xi)}c               \label{hatt}
\ee 
is strictly increasing.  
Inverting it gives $\tilde\xi(t)$; setting \ $\bx(t)=\hat\bx[\tilde\xi(t)]$
one finally obtains  the  original unknown. By (\ref{defs0}) $s$ grows with $u^z$, and $s\to 0,\infty$ resp. if $u^z\to\infty,-\infty$;  in the NR regime $|\bu|\ll 1$ and $s\simeq 1$.
Note that if in a solution of eq. (\ref{equps0}) 
$\hat s(\xi)$ vanishes at least as fast as \ $\sqrt{\xi_f\!-\!\xi}$ as $\xi\uparrow \xi_f<\infty$, 
then the {\it physical} solution expressed as a function  of  $t$ is
defined for {\it all} $t<\infty$, although  as a function of $\xi$ it is defined for $\xi<\xi_f$, because the time corresponding to $\xi_f$ is $t_f\equiv \hat t(\xi_f) =\infty$. 

Solving (\ref{EOM}), even only numerically, is much more difficult than 
solving (\ref{equps0}), because  the unknown $z(t)$ appears
in the argument of $\Bep$ (which for the problems considered here is a rapidly varying function). 
The same occurs in the more general situation where the EM field is the sum of a plane travelling wave in 
the $z$-direction and a generic static EM field  \cite{Fio18JPA}.
This illustrates one main advantage of our approach.
Another one is that we can more easily compute the  {\it energy gain}  due to the interaction with pump 
in an interval $[\xi_0,\xi_1]$:
\be
\E\equiv \hat H(\xi_1)\!-\!\hat H(\xi_0)= \int^{\xi_1}_{\xi_0}\!\!\! d\eta\,
\frac{v'(\eta)}{2\hat s(\eta) }=\int^{\xi_1}_{\xi_0}\!\!\! d\eta\,\frac q{mc^2}\Bep(\eta)\!\cdot\!\frac{\hat\bu^{\scriptscriptstyle \perp}(\eta)}{\hat s(\eta) }.
\label{EnergyGain}
\ee 
The final $\E$ is obtained   choosing $[\xi_0,\xi_1]$ 
as the support $[0,l]$ of $\Bep$. In the standard approach the computation is more complicated,
because the unknown $z(t)$ is present in the argument of $\Bep$, and  the time $t_1$
when the wave (laser pulse) overcomes the particle is unknown as well.

In applications the most common pump is a modulated monochromatic wave\footnote{
The elliptic polarization in (\ref{modulate}) is ruled by $\psi,\varphi_1,\varphi_2$; it  reduces to a linear one in the direction 
of ${\bm a}:=\bi\,\cos\psi+\bj\,\sin\psi$ if
$\varphi_1=\varphi_2$, to a circular one if $|\cos\psi|=|\sin\psi|=1/\sqrt{2}$ and $\varphi_1=\varphi_2\pm\pi/2$.},
\be
\Bep\!(\xi)\!=\!\underbrace{\epsilon(\xi)}_{\mbox{modulation}}
\underbrace{[\bi \cos\psi\,\sin (k\xi\!+\!\varphi_1)\!+\!\bj \sin\psi\sin (k\xi\!+\!\varphi_2)]}_{\mbox{carrier wave $\Be_o^{{\scriptscriptstyle \perp}}\!(\xi)$}},
 \label{modulate}
\ee
where $\bi=\nabla x$,  $\bj=\nabla y$, $k$ is the
 wave number, and the modulating amplitude $\epsilon(\xi)\ge 0$ has support $[0,l]$.
If $\Bep$ is slowly modulated, i.e. \ $|\epsilon'|\!\ll\! |k\epsilon|$ for $\xi\in[0,l]$, \ 
then\footnote{In fact, for a generic regular function $f$ vanishing at  $\xi\!=\!-\infty$  integration by parts gives
\be
\int^\xi_{-\infty}\!\!\!\!\!\! d\eta\: f(\eta)e^{ik\eta} = -\frac ik f(\xi)e^{ik\xi}+O\!\left(\frac 1{k^2}\right) ; \label{oscillestimate}
\ee
the remainder $O(\frac 1{k^2})$ is `small' if $|f'|\!\ll\! |k f|$, see \cite{Fio18JPA} appendix A.4  for details.
(\ref{oscillestimate}) applied to  (\ref{modulate}) gives  (\ref{alphaapprox}).
}
\be
\Bap(\xi)\!\simeq\!  \frac{\epsilon(\xi)}{k}\,\Bep_o\left(\xi\!+\!\frac{\pi}{2k}\right)=  \frac{\epsilon(\xi)}{k^2} \,\Bep_o{}'(\xi),
 \label{alphaapprox}
\ee
in particular $\Bap\!(\xi),\bup\!(\xi),v(\xi) \!\simeq\!   0$ if $\xi\!>\!l$. Moreover, by (\ref{heq1rint}) $\hat s$ is practically insensitive\footnote{In fact, 
the fast oscillations of $ v$  induce
by the  integration in (\ref{heq1rint}a) much smaller relative oscillations of $\hat z$, because $v/\hat s^2\!\ge\!0$ 
and its integral is a growing function of $\xi$; the  integration in  (\ref{heq1rint}b) averages the residual small oscillations of $E_s^z[\hat z(\xi)]$ to yield an essentially smooth $\hat s(\xi)$.  \   If e.g. \ $\Bep$ is linearly polarized, then
\bea
v(\xi) \simeq  (1\!-\!\cos2k\xi)\, v_a(\xi), \qquad 
v_a(\xi) \simeq \frac 12\left[\frac{q\,\epsilon(\xi)}{kmc^2}\right]^2 .       \label{v_a}
\eea
}  
 (see e.g. fig. \ref{graphsb}, \ref{graphs2},  \ref{graphs2'})  to the rapid oscillations of the pump $\Bep$, because it essentially depends on the average-over-a-cycle $v_a$ of $v$, hence on $\epsilon$;
this is the third important advantage of using $\hat s$ as an unknown.
The functions $\hat \gamma(\xi), \hat\bb(\xi), \hat\bu(\xi),...$, which 
are recovered via (\ref{u_es_e}), due to the form of the latter
 do not share the same remarkable property, nor do $\gamma(t), \bb(t), \bu(t),..$;
see the graphs of the plasma examples  below. By (\ref{alphaapprox}), $\hat s'\!\ll\! |k \hat s|$  eq.
(\ref{hatsol}) in turn yields
\bea
\hat \BD\!^{{\scriptscriptstyle \perp}}(\xi)  \: \simeq \:
\frac{-q}{m(kc)^2}\frac{\Bep(\xi)}{\hat s(\xi)}.
\eea
For the same reasons, replacing $v$ by $v_a$ in (\ref{heq1rint}),  (\ref{EnergyGain})  does not change the integrals significantly, but makes $\hat s,\E$ much easier to compute.

\subsection{Plane problem eqs: EM pulse hitting a cold plasma at rest}
\label{planemodel}

Next, we apply the previous changes of independent and dependent variables to a plasma as considered in the Introduction. 
We denote as \ $\bx_e(t,\bX)$  \ the position at time $t$
of the electrons' fluid element initially located at $\bX\!\equiv\!(X,Y,Z)$, as $\hat \bx_e(\xi,\bX)$ the position   of the same material element  as a function of $\xi$.
 For brevity   we shall refer to the electrons initially contained: in such a
fluid element as to the  ``$\bX$ electrons"; in fluid elements with arbitrary $X,Y$
and specified $Z$, or  with $\bX$ in a specified
region $\Omega$, respectively as the ``$Z$ electrons" or the ``$\Omega$ electrons''.
The function \ $\bx_e$ \ 
is required to have continuous second derivatives  (at least piecewise, while having continuous first derivatives) and for each $t$ 
the map $\bx_e(t, \cdot):\bX\mapsto \bx$ is required  to be one-to-one;
equivalently,  \ $\hat\bx_e$ \ 
is required to have continuous second derivatives  (at least piecewise, while having continuous first derivatives) and for each $\xi$ the 
map \ $\hat\bx_e(\xi, \cdot):\bX\mapsto \bx$ is required  to be one-to-one.
For each $t$ (resp. $\xi$) let \ $\bX_e(t, \cdot):\bx\mapsto \bX$ 
\ be the inverse of $ \bx_e(t,\cdot)$ [resp.  
$\hat \bX_e(\xi, \cdot):\bx\mapsto \bX$ \ be the inverse of $\hat \bx_e(\xi,\cdot)$].
Clearly, 
\bea
\ba{l}
\bX_e(t, \!\bx)=\hat \bX_e(ct\!-\!z, \!\bx). 
\label{clear}
\ea
\eea
As said, we assume that   (\ref{asyc}), (\ref{pump}) hold. 
 This implies   $\Bap(\xi)=\0$ if $\xi\!\le\!0$.
As we regard ions as immobile, the proton density will be $ n_p(t,\bx)=\widetilde{n_0}(z)$ for all $t$.  
Since the problem is independent of $x,y$, also the EM field, densities and
velocities obtained solving the Maxwell+plasma equations will depend only on $t,z$; 
similarly, the displacements  \ $\BD_e\equiv \bx_e(t,\!\bX\!) -\!\bX$ will 
actually depend only on $t,Z$ [and their ``hatted" counterparts 
$\hat\BD_e\equiv \hat\bx_e(\xi,\bX)\!-\!\bX$ only on $\xi,Z$].
Hence we can partially fix the gauge by choosing
the EM potential $A=(A^0,\bA)$ to depend only on $t,z$ as well, and its
transverse part as the physical observable
\be
\bA\!^{{\scriptscriptstyle\perp}}(t,z)=-\int^{t}_{-\infty }\!\!\!\!c\,dt' \,
\bE^{{\scriptscriptstyle\perp}}(t',z);                         \label{Apup}
\ee 
$\bAp$ alone  determines both $\bEp,\bB$ 
through \ $c\bEp=-\partial_t\bAp$, \ $\bB=\bBp=\bk\times\partial_z\bAp$. \ 
The Eulerian electrons' momentum $\Bp_e(t,z)$ satisfies  eq.
(\ref{EOM}), where one has to replace $\bx(t)\mapsto\bx_e(t,\bX)$,
 $\dot \Bp\mapsto d\Bp_e/dt\equiv$ {\it total}  derivative; as it is known, under the
present assumptions the transverse component of (\ref{EOM}a)  becomes
$\frac{d\Bp^{{\scriptscriptstyle\perp}}_e}{dt}=\frac ec \frac{d\bAp}{dt}$ and with the trivial initial conditions implies
\be
\Bp^{{\scriptscriptstyle\perp}}_e=\frac ec \bAp \qquad\quad 
\mbox{i.e. }\quad \bup_e=\frac e{mc^2} \bAp.              \label{bupExpl}
\ee
This equation allows to trade $\bup_e$  for $\bAp$ as an unknown function. 
Of course, (\ref{bupExpl}) reduces to  (\ref{hatuperp}) by
the replacement $\hat\bu^{\scriptscriptstyle \perp}_e(\xi,Z)\mapsto
\hat\bu^{\scriptscriptstyle \perp}$.
From (\ref{EBfields0})  it follows $\bAp(t,z)=\Bap(ct\!-\!z)$ if $ t\le 0$.

Since the problem is independent of $x,y,X,Y$,  the  local conservation \ $n_e\,dz=\widetilde{n_0}\,dZ$ \ of the number of electrons
(whence the continuity equation  $\frac{d n_e}{dt}\!+\!n_e\nabla_{\bx}\!\cdot \bv_e=0$),
takes the form
\bea
 n_e(t,z)=\widetilde{n_0}\!\left[Z_e(t,\!z)\right] \,\partial_z  Z_e(t,z).
\label{n_h}
\eea
Another important simplification is that  the Maxwell equations \ $\nabla\!\!\cdot\!\bE\!-\!4\pi j^0\!=\!\partial_z E^z\!-4\pi e(n_p-n_e)\!=\!0$, \
$ \partial_tE^z/c\!\!+\!4\pi j^z \!\!=\!( \nabla\!\!\times\!\bB)^z\!\!=\!0$ (the current density
being \ $\bj=-en_e\bb_e$) \ imply  \cite{Fio14JPA}
\be
  E^z(t,z)=4\pi e \left\{\widetilde{N}(z)-\widetilde{N}[ Z_e(t,z)]\right\}   , \quad   \qquad\mbox{where }\:
\widetilde{N}(Z)\equiv\int^{Z}_0\!\!\! d \zeta\,\widetilde{n_0}(\zeta).
\label{explE}
\ee
The Lagrangian counterpart of (\ref{explE}) for fixed fluid element resp. reads 
\be
\widetilde{E}^z(t,Z)=4\pi e \left\{\widetilde{N}[ z_e(t,Z)]-\widetilde{N}(Z)\right\},
\qquad
\hat E^z(\xi,Z)=4\pi e \left\{\widetilde{N}[ \hze(\xi,Z)]-\widetilde{N}(Z)\right\},
\label{explE'}
\ee
if we adopt $t$ or $\xi$ as an independent variable. These formulae
show that $E^z$ does not depend directly on $t$ (or $\xi$), but only through $z_e$,
i.e. leads to a conservative force as $E^z(z)$ in (\ref{EBfields0}).

Relations (\ref{n_h}-\ref{explE}) allow to compute $n_e,E^z$  explicitly in terms of 
the assigned initial density $\widetilde{n_0}$
and of the still unknown  $ Z_e(t,z)$,  i.e. of the longitudinal motion; 
thereby  they further reduce the number of unknowns. The remaining ones are $\bA\!^{{\scriptscriptstyle\perp}},\bx_e$ and $u_e^z$ - or equivalently $s$. 

$\bAp$ is coupled to the current  through \ $\Box\bAp=4\pi\bjp$ (in the Landau gauges).
By causality $\bj\!=\!\0$ if  $ct\!\le\! |z|$; more generally, \
$\bjp\!=-e n_e\bb_e=-e n_e\bup_e/ \gamma_e=- e^2 n_e\bAp/ mc^2\gamma_e$, \  by (\ref{bupExpl}).
Including the initial conditions for $\bAp$ and using the Green function of $\frac 1 {c^2}\partial_t^2\!-\!\partial_z^2$ one finds that this amounts to the integral equation (68) of \cite{Fio18JPA}, namely
\bea
\bA\!^{{\scriptscriptstyle\perp}}(t,z)-\Bap(ct\!-\!z)=
 -  \frac{2\pi e^2}{mc^2} \! \int_{T\cap  D\!_{t,z}}\!\!\!\!\!\!\!\!\!\!\!\!  cdt' \,  dz'\,
 \left(\!\frac{n_e\bAp}{\gamma}\!\right)\!\left(t',z'\right),           \label{inteq1}
\eea
where $T\!=\!\{ (t,\!z)\: |\:  ct \!>\! |z| \}$ is the 2-dim future causal cone at the origin, and $D\!_{t,z}\!\equiv\!\{ (t'\!,\!z')\: |\:  ct\!-\!ct' \!>\! |z\!\!-\!\!z'| \}$ is the past causal cone at the point $(t,z)$. If $t\!\le\! 0$ 
or $ct\!\le\! z$ then $T\cap  D\!_{t,z}$ is empty, and the right-hand side (rhs) of (\ref{inteq1}) is zero.
Otherwise $T\cap  D\!_{t,z}$ is a rectangle as in fig. \ref{Worldline}b.

Using  (\ref{bupExpl})   and abbreviating \
  $v\!\equiv\!\hat\bu^{\scriptscriptstyle \perp}_e{}^2\!=\![ e\hat\bAp/{mc^2}]^2$, \
$\hat\Delta(\xi,Z)\!\equiv\!\hat\Delta^z(\xi,Z)\!=\!\hze(\xi,Z)\!-\!Z$, \
the remaining equations   (\ref{reduced}) to solve  take the form  \cite{FioDeN16}
\bea
\hat\Delta'(\xi,Z)=\displaystyle\frac {1\!+\!v}{2\hat s^2}\!-\!\frac 12, \qquad 
\hat s'(\xi,Z)=K\left\{\!
\widetilde{N}\left[Z\!+\!\hat\Delta\right] \!-\! \widetilde{N}(Z)\!\right\} , \qquad  K:=\frac{4\pi e^2}{mc^2}; \label{heq1} 
\eea
they are equipped with the initial conditions
\ $\hat\Delta(-Z,\!Z)\!=\!0$, $\hat s(-Z,\!Z)\!=\! 1$.  Since for 
$ct\!\le\!z$ \ $\bAp\!=\!\0$, then $v,\hat\Delta,\hat s-1$ remain zero until $\xi=0$, 
and we can shift the initial conditions to
\bea
&&  \: \hat \Delta(0,\!Z)\!=\!0,  \qquad\qquad
 \hat s(0,\!Z)\!=\! 1. \qquad\qquad\qquad\qquad  \label{heq2}
\eea
Note that equations (\ref{heq1}) have a built-in back-reaction mechanism preventing $\hat s(\xi,\!Z)$ to vanish anywhere, consistently with the definition (\ref{defs0}): in fact, if {\it ad absurdum} $\hat s(\xi,\!Z)$ went to zero in some limit $\xi\to\xi_b$, then 
the rhs of the first equation would blow up and force
$\hat\Delta(\xi,\!Z)$, and in turn $\hat s(\xi,\!Z)$, to abruptly grow again to higher positive values. 
Note also that the Cauchy problem (\ref{heq1})-(\ref{heq2}) is equivalent to the integral one [analogous to (\ref{heq1rint})]
\be
\hD (\xi,Z)= \int_0^\xi\!\!d\eta\,\frac {1\!+\!v(\eta)}{2\hs ^2(\eta,Z)}-
\frac{\xi}2,\qquad \hs (\xi,Z)-1= \int_0^\xi\!\!\!d\eta\!\!\int^{\hze(\eta,Z)}_Z\!\!\!\!\!\!\!\!\!\!\!\!
dZ' \: K\widetilde{n_0}(Z').
  \label{sDelta}
\ee

As the rhs(\ref{inteq1}) is zero for $t\le 0$, 
within short time intervals $[0,t_s]$ [whose length we determine {\it a posteriori}, see (\ref{neglectChange})]  we can 
approximate \ $\bAp(t,z)\simeq \Bap(ct\!-\!z)$, \ i.e. neglect the change of the pump, and by (\ref{bupExpl}) \ $\bup_e(t,z)\simeq e\Bap(ct\!-\!z)/{mc^2}$.
Setting $\hat\bu^{\scriptscriptstyle \perp}_e(\xi)\!=\! e\Bap(\xi)/{mc^2}$, the forcing term $v\!=\!v(\xi)$ becomes a {\it known} function of $\xi$ (only), and 
 (\ref{heq1})  a family parametrized by $Z$ of {\it decoupled ODEs} of the same form as (\ref{reduced}).
Setting \ $q\!\equiv\!\hat\Delta$,   $p\!\equiv\!-\hat s$, \  one recognizes that (\ref{heq1}) for every fixed $Z$
 are the Hamilton equations $q'=\partial \check H/\partial p$, $p'=-\partial \check H/\partial q$ of a system with 1 degree of freedom and
Hamiltonian $\check H(q,p,\xi;Z)\!\equiv\!\hat H(q,-p,\xi;Z)$, where
\bea
\ba{l}
\hat H(\Delta,s,\xi;Z)\equiv \gamma(s,\xi)+ \U(\Delta;Z), \qquad \U(\Delta;Z)\!\equiv\!K\left[
\widetilde{{\cal N}}\!(Z \!+\!\Delta) \!-\!\widetilde{{\cal N}}\!(Z)\!-\! \widetilde{N}\!(Z)\Delta\right],\\[8pt]
 \gamma(s,\xi)\!\equiv\!\displaystyle \frac{s^2+\mu^2(\xi)}{2s}, \qquad \mu\!\equiv\!\sqrt{1\!+\!v},
\qquad \widetilde{{\cal N}}(Z)\equiv
\int^Z_0\!\!\!d\zeta\,\widetilde{N}(\zeta)\!=\!\int^{Z}_0\!\!\!d\zeta\,\widetilde{n_0}(\zeta)\, (Z\!-\!\zeta) .
\ea                                \label{hamiltonian}
\eea
In defining $\U$ we have fixed, for each $Z$, the free additive constant so that
 $\U(0,Z)\equiv 0$. 
Up to the factor $mc^2$, $\hat H,\U,\hat H\!-\!1\!-\!\U,$ are resp. the total, potential, 
kinetic energies of a single $Z$-electron. 
By their very definitions,  $\mu$ is the electrons' Lorentz factor when $u^z=0$ (i.e. due only to their transverse velocity), $\widetilde{N}(z)$ grows, $\widetilde{{\cal N}}(z)$ grows and is convex. Moreover,

\begin{prop}
 $\U\ge 0$,  and $\U=0$ iff $\Delta\!=\!0$.  
$\hat H\ge\gamma\ge\mu\ge 1$,  and $\hat H\!=\!\mu$  iff $(\Delta,s)\!=\!\big(0,\mu\big)$. 
\label{prop1}
\end{prop}

Given a solution 
$P(\xi;Z)\!\equiv\!\big(\hat \Delta(\xi;Z),\hat s(\xi;Z)\big)$ of (\ref{heq1})
 let $h(\xi,Z)\equiv \hat H[\hat\Delta(\xi,Z),\hat s(\xi,Z),\xi;Z]$.
At each $\xi$ the variables $\hat \Delta,\hat s$ are related by (\ref{hamiltonian}a), which we can rewrite in the form
\be
\frac 12\left[\hat s+\frac{\mu^2}{\hat s}\right]= \bar\gamma(\hat \Delta,\xi;Z)\equiv h(\xi,Z)-\U(\hat\Delta;Z),                            \label{H-level-cycle}
\ee
Solving this equation with respect to (w.r.t.) $\hat s$ we find the two solutions 
\be
\spm\big(\hat\Delta,\xi;Z\big)\equiv \bar\gamma\big(\hat\Delta,\xi;Z\big)\pm\sqrt{ \bar\gamma^2\big(\hat\Delta,\xi;Z\big)-\mu^2(\xi)}; 
   \label{spm}
\ee
whether $\hat s=\sp$ or $\hat s=\sm$ is determined by the continuity of $\hat s$ as a function of $\xi$. For $\xi>l$ all direct dependences on $\xi$ disappear from (\ref{H-level-cycle}-\ref{spm}).

Equations (\ref{heq1})-(\ref{heq2}) can 
be solved numerically, or by quadrature for  $\xi>l$ (by energy conservation). Then the whole $\hat\bx_e(\xi,\bX)$ can be recovered via (\ref{hatsol}), which now reads
\be
\hat\bx_e(\xi,\bX)=\bX+\hat\BD(\xi,Z),
\qquad \quad \hat \BD\!^{\scriptscriptstyle \perp}\!(\xi,Z)\!:=\!\!\displaystyle\int^\xi_0\!\!\! d\eta \,\frac{\hat\bu^{\scriptscriptstyle \perp}_e(\eta)}{\hat s(\eta,Z)}, \quad \hat \Delta\!^z(\xi,Z):=\hat \Delta (\xi,Z),        \label{hatsol'}
\ee
whence again one can recover $\bx_e(t,\bX)$. Since $\bx^{\scriptscriptstyle \perp}_e\!-\!\bXp$ and
 $\hbx^{\scriptscriptstyle \perp}_e\!-\!\bXp$ do  not depend on $\bXp$, the Jacobians  of the map $\bX\mapsto\hbx_e\equiv(x_e,y_e,z_e)$ at fixed $t$ and of the map $\bX\mapsto\hbx_e\equiv(\hxe,\hye,\hze)$ at fixed $\xi$
 from the Lagrangian  to the Eulerian coordinates   reduce to $J=\partial_Zz_e$ and to $\hJ$,
\bea
\hJ(\xi,Z)=\frac {\partial\hze(\xi,Z)}{\partial Z}=:1\!+\!\varepsilon(\xi,Z),\qquad\qquad
\hat\sigma(\xi,Z):=\frac {\partial\hat s(\xi,Z)}{\partial Z}, \label{DefhJhsigma}
\eea
where $\varepsilon:=\partial_Z \hat\Delta$; here we have also introduced  $\hat\sigma$ for later use.
Taking $z$ as the independent variable, i.e. inverting  the 3rd component of (\ref{hatsol'}),
we find that $\hat Z_e(\xi,z)$ fulfills  the equation $\hat Z_e(\xi,z)=z-\hat\Delta[\xi,\hat Z_e(\xi,z)]$;
replacing $\xi\mapsto ct\!-\!z$ and recalling (\ref{clear}) we find that $ Z_e(t,z)$  satisfies
\ $ Z_e(t,z)\!=\!z\!-\!\hat\Delta[ct\!-\!z,Z_e(t,z)]$;
\ deriving the latter equation and using (\ref{heq1}a) we find
\bea
\frac1{J(t,\!z)}
=\partial_z Z_e(t,\!z)=1\!+\!\hat\Delta'\!-\!(\partial_Z \hat\Delta)\partial_z Z_e(t,\!z)\quad\Rightarrow\quad
\frac1{J(t,\!z)}
=\frac{1\!+\!v\!+\!\hat s^2}{2\hat s^2 \hJ}\Big\vert_{(\xi,Z)=\big(ct\!-z,Z_e(t,\!z)\big)} .\label{invZtoz}
\eea
Since the evolution determined by (\ref{heq1}) prevents $\hat s$ to vanish (see section \ref{during}), 
$\hat J\!>\!0$ is a necessary  and sufficient condition for the invertibility of  the maps 
 $\bx_e\!:\!\bX\!\mapsto\!  \bx$  at fixed $t$ and
$\hat\bx_e(\xi,\cdot):\bX\mapsto \bx$ at fixed $\xi$, and thus for the self-consistency of 
this hydrodynamical model; it will be studied in section \ref{Wave-breakings}.
We can recover the Eulerian variables $\bu_e,\gamma_e,\bb_e$  from their hatted counterparts
by the replacement $(\xi,Z)\mapsto \big(ct\!-z,Z_e(t,\!z)\big)$, for instance
\bea
\bu_e(t,\!z )=\hat\bu\!\left[ct\!-\!z,\!Z_e(t,\!z)\right].        \label{sol'}
\eea
Replacing (\ref{invZtoz}) in (\ref{n_h}), and noting that by (\ref{u_es_e}) 
$(1\!+\!v\!+\!\hat s^2)/2\hat s^2\!=\!\hat\gamma/\hat s\!=\!1/(1\!-\!\hat\beta^z)$, 
we express the Eulerian electron density more explicitly in terms of the solution by the formula \cite{FioDeN16}
\bea
n_e(t,z)
=\left.\frac{\hat\gamma\:\widetilde{n_0}}{\hat s \,\hat J}
\right\vert_{(\xi,Z)=\big(ct\!-z,Z_e(t,z)\big)}
=\left.\frac{\widetilde{n_0}(Z)}
{(1\!-\!\hat\beta^z) \hat J}\right\vert_{(\xi,Z)=\big(ct\!-z,Z_e(t,z)\big)}.         \label{expln_e}
\eea
This formula allows us to derive  estimates and  bounds on $n_e$, $\delta\bA\!^{{\scriptscriptstyle\perp}}\equiv\bAp\!-\!\Bap$
even before solving  (\ref{heq1})-(\ref{heq2}).
The approximation \ $\bAp(t,z)\simeq \Bap(ct\!-\!z)$ \  is acceptable as long as the   motion determined by its use makes the \ $|\mbox{rhs}(\ref{inteq1})|$ small w.r.t. $|\Bap|$, what can be tested checking that the  first correction to $\bAp$, obtained replacing $\bAp(t,z)\mapsto\Bap(ct\!-\!z)$ in the rhs(\ref{inteq1}),
\bea
\delta\bA\!^{{\scriptscriptstyle\perp}(1)}(t,z)\equiv
 -  \frac{2\pi e^2}{mc^2} \! \int_{T\cap  D\!_{t,z}}\!\!\!\!\!\!\!\!\!\!\!\!  cdt' \,  dz'\,
 \left(\!\frac{n_e\Bap}{\gamma}\!\right)\!\left(t',z'\right),          \label{correction1}
\eea
is small. If so,  replacing $\bA\!^{{\scriptscriptstyle\perp}(1)}(t,z)\equiv\Bap(ct\!-\!z)\!+\!\delta\bA\!^{{\scriptscriptstyle\perp}(1)}(t,z)$ (the corrected $\bAp$) into (\ref{bupExpl}) and solving  (\ref{heq1})-(\ref{heq2}) with the new $\bup_e$  one can determine the motion more accurately; and so on. This iterative procedure will  also generate higher harmonics,  dispersion and depletion of the pump; we expect that after a while, well inside the bulk,  $\bAp$ will transform into a self-consistent  effective pulse travelling with a group velocity slightly smaller than $c$ and a reflected EM wave. This deserves separate and appropriate  investigation.

 The ratio $n_e/\gamma$ can be obtained from (\ref{expln_e}).
Using $\xi'= ct'\!-\!z'$ and $\xip'\equiv ct'\!+\!z'$ as integration variables the integral (\ref{correction1}) over the rectangle 
$T\cap  D\!_{t,z}$ simplifies as an integral over $\xi',\xip'$
\bea
\delta\bA\!^{{\scriptscriptstyle\perp}(1)}(t,z)= -\frac{2\pi e^2}{mc^2}
 \! \int\limits^{ct+z}_0\!\!\frac{d\xip'}2 \! \int\limits^{ct-z}_0\!\!\! d\xi' \, \Bap(\xi') 
\left.\frac{\widetilde{n_0}}{\hat s \,\hat J}
\right\vert_{Z=\hat Z_e\left[\xi',(\xi'+\xip')/2\right]}.                   \label{correction1'}
\eea
We can easily estimate $\delta\bA\!^{{\scriptscriptstyle\perp}(1)}$
if $\Bep$ is as in (\ref{modulate}), with slowly varying $\epsilon(\xi)$, 
$\widetilde{n_0}(Z)$ over the scale $\lambda=2\pi/k$. 
In fact, from $\Bap(\xi)\!\simeq\!  \epsilon(\xi) \,\Bep_o{}'(\xi)/k^2$
and  (\ref{oscillestimate})  we find at leading order in $1/k$
\bea
\delta\bA\!^{{\scriptscriptstyle\perp}(1)}(t,z) &\simeq &
 \! \int\limits^{ct+z}_0\!\!\! d\xip'  \!\! \int\limits^{ct-z}_0\!\!\! d\xi' \, 
\Bep_o{}'(\xi')\,\epsilon(\xi')\, f(\xi',\!\xip')\simeq  \! \int\limits^{ct+z}_0\!\!\! d\xip'  \, 
\left[\Bep_o(\xi)\,\epsilon(\xi)\, f(\xi,\xip')\right]_{\xi=ct-z}\nn
&=& -g(ct\!-\!z,ct\!+\!z)\,\Bep(ct\!-\!z)  \label{correction1''}   \
\eea
\bea
 f(\xi,\xip)\equiv \frac{\pi e^2}{mc^2k^2}\left.\frac{\widetilde{n_0}}{\hat s \,\hat J}
\right\vert_{Z=\hat Z_e\left[\xi,(\xi+\xip)/2\right]}, \qquad 
 g(\xi,\xip)\equiv \!\!  \int^{\xi}_0\!\! d\xip'  \, f(\xi,\xip').
 \label{deffg}
\eea
Consequently, at leading order in $1/k$ beside $\Bap$, also $\delta\bA\!^{{\scriptscriptstyle\perp}(1)},\bA\!^{{\scriptscriptstyle\perp}(1)}$ are proportional to $\epsilon$ and hence approximately
vanish  outside the stripe $0\le \xi\equiv ct\!-\!z\le l$. However,
$\delta\bA\!^{{\scriptscriptstyle\perp}(1)}(t,z)\propto\Bep_o(\xi)$, while $\Bap(\xi)\propto\Bep_o(\xi\!+\!\pi/2k)$.
In section \ref{Wave-breakings} we show that (\ref{Lncond}) or more generally  (\ref{Lncond'}) imply that in this stripe $\hat s(\xi,Z)>1$, $\hat J(\xi,Z)\simeq 1$, whence
$\widetilde{n_0}/\hat s \,\hat J\lesssim n_b$;
using $|\Bep_o|=1$, we find
\bea
\left|\bA\!^{{\scriptscriptstyle\perp}}(t,z)-\Bap(ct\!-\!z)\right| 
\lesssim (ct\!+\!z)\,\frac{\pi e^2 n_b}{mc^2k^2}\,\epsilon(ct\!-\!z) .
\eea
Hence $\delta\bA\!^{{\scriptscriptstyle\perp}(1)}$ is negligible w.r.t. $\Bap$, 
and $\bAp\!\simeq\!\Bap$ is a good approximation,  if 
\be
\frac{e^2 n_b\lambda}{2 mc^2}(ct+z)\ll 1. 
\label{neglectChange}
\ee

\section{Motion of the plasma electrons in the plane model}
\label{plasmaelectrons-in-plane-model}

Given {\it specific} input data $\Bep,\widetilde{n_0}$, one can numerically solve (\ref{heq1})-(\ref{heq2}) for a large number of values  of $Z$ (e.g. in a fine lattice over the interval of interest)  with relatively little computation power. This has been done e.g. to obtain the WLs of
fig. \ref{Worldlinescrossings}.
Rather, in this section we  analytically study such equations for {\it generic} input data.

\subsection{Motion during the laser-plasma interaction}
\label{during}

Since $\widetilde{N}(Z)$  grows with $Z$, by (\ref{heq1}b) the zeros of $\hD(\cdot,\!Z)$ 
are extrema of $\hs(\cdot,\!Z)$, and vice versa.
 Let us recall how $\hD(\cdot,\!Z),\hs(\cdot,\!Z)$ start evolving  from their initial values (\ref{heq2}).  
As said, for small $\xi\!>\!  0$ all electrons reached by the pulse start oscillating transversely and 
drifting forward  (pushed by the ponderomotive force); 
in fact, $v(\xi )$  becomes positive, implying in turn that so do  the right-hand side (rhs) of (\ref{heq1}a) and $\hD$; \  the $Z\!=\!0$ electrons leave behind themselves a  layer of ions $L_t$ of finite thickness 
completely deprived of electrons (this is depicted dark yellow in fig. \ref{Worldlinescrossings}).  The growth of $\hD$ implies also the growth 
both of the rhs of (\ref{heq1}b) (because the latter grows with $\hD$) and of $\hs$. \  
\ $ \hD(\xi,\!Z)$ keeps growing  as long as $1\!+\!v(\xi)\!=:\!\mu^2(\xi)>\hs^2(\xi,\!Z)$, reaches a maximum at $\tilde\xi_1(Z)\equiv$ the smallest  $\xi\!>\!0$ such that the rhs (\ref{heq1}a)  vanishes.
$ \hs(\xi,\!Z)$ keeps growing  as long as $\hD(\xi,\!Z)\!\ge\!0$ and reaches a maximum at the first zero $\tilde\xi_2>\tilde\xi_1$ of $ \hD(\xi,\!Z)$. For small  $\xi>\tilde\xi_2$, while  $\hD(\xi,\!Z)$ is negative, 
$ \hs(\xi,\!Z)$ starts decreasing if $Z\!>\!0$, remains   constant if $Z=0$.
If $Z>0$  both  $\hat\Delta,\hat s$ keep decreasing as long $\hat s\!\ge\!\mu$.  $ \hD(\xi,\!Z)$  reaches a negative minimum at 
$\check\xi_3(Z)\equiv$ the smallest  $\xi\!>\!\tilde\xi_2$ such that the rhs (\ref{heq1}a)  vanishes 
again. 
We also denote by $\tilde\xi_3(Z)$ the smallest  $\xi\!>\!\check\xi_3$ such that $ \hs(\xi,\!Z)=1$, and 
define $\tilde\xi_i':=\min\{\tilde\xi_i,l\}$. As noted after (\ref{alphaapprox}), if $\Bep$ is slowly modulated  then $v(l)\!\simeq\!0$, and $\check\xi_3\simeq\tilde\xi_3$ if in addition $l<\tilde\xi_3$. We shall say that 
\bea
\ba{ll}
\mbox{a pulse is {\it essentially short w.r.t. $\widetilde{n_0}$} if}\qquad & 
\hs(\xi,Z)\ge 1, \\[4pt]
\mbox{a pulse is {\it strictly short w.r.t. $\widetilde{n_0}$} if}\qquad & 
\hD(\xi,Z)\ge 0,
\ea 
\qquad  \forall\:\:Z\ge 0,\:\: 0\le\xi\le l;   \label{Lncond'}
\eea
clearly, a strictly  short pulse is also essentially short.
Equivalently, a pulse is essentially short if $l\le \tilde\xi_3(Z)$ (i.e.  $\tilde\xi_3'=l$).

For given  pulse energy, maximizing  the energy transfer from the pulse to the plasma wave   can be achieved \cite{SprEsaTin90PRL,FioFedDeA14} through a suitable $l\sim\tilde\xi_2$, what is compatible with  (\ref{Lncond'}a). Assuming (\ref{Lncond'}a) one can derive useful  apriori bounds on the solutions of  (\ref{heq1}),
(\ref{basic}), in particular on the Jacobian $\hat J$,  using that  $1/\hat s^2,1/\hat s^3\le 1$ for all electrons; moreover, each electron  is overcome by the pulse before (or at latest very shortly after) its longitudinal displacement $\Delta$  reaches a negative minimum for the first time.  Condition  (\ref{Lncond'}b) means that each electron has become overcome by the pulse before its $\Delta$ becomes negative  for the first time; it
implies that the ion layer $L_t$ is not refilled by electrons, and that no electron gets out of the plasma bulk, before the pulse has passed.

Let $\hDOu$, $\hDd$ be respectively an upper and a lower bound for $\hD$, i.e.  $\hDd(Z)\!\le\! \hD(\xi,Z)\!\le\! \hDOu(Z)$,  for all $\xi\in[0,\tilde\xi_3']$; moreover, let 
\bea
\ba{l}
\check n(\xi,Z)\!\equiv\!\widetilde{n_0}\big[\hze(\xi,Z)\big], \quad
n_u(Z)\!\equiv\!\!\!\max\limits_{z\in\big[Z+\hDd,Z+\hDOu\big]}\!\{\widetilde{n_0}(z)\},%=:n_u^1(Z), 
\quad n_d(Z)\!\equiv\!\!\!\min\limits_{z\in\big[Z+\hDd,Z+\hDOu\big]}\!\{\widetilde{n_0}(z)\},\\[10pt]
n_u''(Z)\!\equiv\!\max\limits_{z\in\big[Z,Z+\hDOu\big]}\{\widetilde{n_0}(z)\},%=:n_u^1(Z), 
\quad n_d''(Z)\!\equiv\!\min\limits_{z\in\big[Z,Z+\hDOu\big]}\{\widetilde{n_0}(z)\},%=:n_d^1(Z),
\quad n_u'(Z)\!\equiv\!\max\limits_{z\in\left[Z+\hDd,Z\right]}\{\widetilde{n_0}(z)\}. \quad 
\ea\label{n-bounds} 
\eea
Then \ $n_d\le \check n(\xi,Z)\le n_u\le n_b$
\ if \ $0\le\xi\le\tilde\xi_3'$, \ $n_d''\le \check n(\xi,Z)\le n_u''\le n_b$
\ if \ $0\le\xi\le\tilde\xi_2'$. \
Ref. \cite{FioDeAFedGueJov22}, section 2, determines   
 {\it a priori} bounds  on the solutions $\hs,\hD$, $\hat H$ and on $\tilde\xi_i$,  $\hDOu,\hDd$  valid for all $Z$
and $\xi\in[0,\tilde\xi_3']$, {\it without} solving  (\ref{heq1}-\ref{heq2}). 
We just recall a few bounds.  \
$\hDO(\xi)\equiv \int_0^\xi\!\!d\eta\,v(\eta)/2$ is the $z$-displacement for zero density, which is a growing function.
It is easy to show \cite{FioDeAFedGueJov22} that 
$\hD(\xi,\!Z)\le \hDO(\xi)$  for $\xi\in[0,\tilde\xi_3']$, whence %$0\le\xi\le\tilde\xi_3'$. 
the simplest choice is $\hDOu\equiv\hDO(l)$; whereas $\hDd(Z)=0$ if the pulse is strictly short, otherwise $\hDd(Z)$ is the negative solution of the equation $ \U( \Delta;\!Z)= K n_u''(Z)\hDOu^2/2$ in the unknown $\Delta$   (as a first estimate, $\hDd\simeq -\hDOu$). In section \ref{Wave-breakings} we shall use these bounds to formulate no-WBDLPI conditions.

Abbreviating  $M_u\equiv Kn_u$, \ $M_d\equiv Kn_d$, \  $M_u'\equiv Kn_u'$,
$M_u''\equiv Kn_u''$, \ $M_d''\equiv Kn_d''$, \ let
\bea
\ba{l}
\displaystyle 
s_u\equiv 1\!+\!\frac{M_u''}{2}\hDOu^2\!+\!\sqrt{\!\left(1\!+\!\frac{M_u''}{2}\hDOu^2\right)^2\!-\!1},\qquad
g(\xi,\!Z)\equiv\frac {M_u''(Z)}{2}\!\!\int_0^\xi\!\!\!\!d\eta\,(\xi\!-\!\eta)\,v(\eta),\\[12pt]
\hsU(\xi,\!Z)\equiv\min\left\{s_u(Z),1 +g(\xi,\!Z)\right\},  \qquad
\displaystyle f(\xi,Z)\equiv\!\!\int_0^\xi\!\!\! d\eta\,(\xi\!-\!\eta)\!\left\{\!\frac {1\!+\!v(\eta)}{\left[\hsU(\eta,Z)\right]^2}- 1\!\right\};
\ea          \label{sDelta1Def}       
\eea
sufficient conditions  for the pulse to be essentially  (resp. strictly) short are \cite{FioDeAFedGueJov22}:
\bea
\forall  Z\ge 0 \qquad 
f(l,Z) \ge\left(1\!-\!\frac{M_d''}{M_u'}\right) \max_\xi{f(\xi,Z)}  
 \qquad \big(\mbox{resp. $\:\: f'(l,Z)\ge 0$}\big) .   \label{ShortPulse1'}
\eea
Before checking whether (\ref{ShortPulse1'}b) is fulfilled, it is easier to check the stronger condition $M_u''(Z)\,  l^2(1\!+\!2\hDOu/l) \le 2$.

If $\Bep$ is as in (\ref{modulate}) with slow modulation $\epsilon(\xi)$, 
via (\ref{oscillestimate}) we can easily estimate  the transverse oscillation (\ref{hatsol'}) and its maximum as
\be
\hat \BD\!^{\scriptscriptstyle \perp}\!\simeq\!\frac{-e\Bep}{k^2mc^2\hat s}, \qquad
|\BD^{{\scriptscriptstyle \perp}}_{{\scriptscriptstyle M}}|\simeq \frac e{k^2mc^2}\,\max\left\{\frac{\epsilon}{\hat s}\right\}\le 
\frac{a_0s_u}{k},
\label{MaxTransvOscill}
\ee 
because $|f'|\!\ll\! |k f|$ holds also 
for $f\!=\!\epsilon/\hat s$, and  $e\epsilon/kmc^2\le a_0\!\equiv\! eE^{\scriptscriptstyle \perp}_{\scriptscriptstyle M}/kmc^2$, $\hs\ge s_d\equiv 1/s_u$.

\subsection{Motion after the laser-plasma interaction}
\label{after}

In the intervals of  $\xi$ where $v(\xi)$ is a constant so are $\mu$ and $h$, 
and the $Z$-trajectory in $(\Delta,s)$ phase space ($Z$-path) described by $P(\xi;Z)$  is a level curve of $H$,
$H(\Delta,s;Z)\!=$ const.  In particular, for  $\xi\!\le\!0$, 
we have $v\!=\!0$, $\mu\!=\!h\!=\!1$, and the $Z$-path reduces to  the point $(\Delta,s)\!=\!(0,1)$.
For  $\xi\!\!>\!\!l$, \ the $Z$-path fulfills $H(\Delta,s;Z)\!=\!h(Z)$, where $h(Z)\!\equiv\!H|_{\xi=l}\!=\!1\!+\!\!\int^l_0\!\!d\xi v'(\xi)/\hat s(\xi,Z)
$ by (\ref{EnergyGain}), $\mu\!\equiv\!\sqrt{1\!+\!v(l)}\!\ge\!1$ is the relativistic factor due 
to a constant quiver (i.e. transverse) velocity, $\bu=\bup$;  $ v(l)\simeq 0$
[see (\ref{alphaapprox})], and again \ $\mu\simeq 1$  if $\Bep$ is slowly modulated. 
For all $Z\!>\!0$, $\U\!\to\!\infty$ as $|\Delta|\!\to\!\infty$\footnote{In fact,
$\frac{mc^2}{4\pi e^2}\left|\frac{\partial\U}{\partial\Delta}\right|=\left|\widetilde{N}\!(Z \!+\!\Delta)\!-\! \widetilde{N}\!(Z)\right|=\left|\int^{Z \!+\!\Delta}_Z\!\! d\zeta\,\widetilde{n_0} (\zeta)\right|$. This diverges as $\Delta\to+\infty$, becomes 
$\int^{Z}_0\!\! d\zeta\,\widetilde{n_0} (\zeta)=$const$>0$ for $\Delta\le -Z$.
Hence $\left|\frac{\partial\U}{\partial\Delta}\right|\ge$const$>0$ for both $\Delta\to\pm\infty$,
and $\U\to\infty $ as $|\Delta|\to\infty$.
},
hence the $Z$-paths for all  $Z\!>\!0$ (see e.g. Fig.s \ref{graphsb}d-f)
are cycles around the {\it center} $C\!\equiv\!(\Delta,s)\!=\!(0,\mu)$ (the only critical point),
and the corresponding motions are periodic in $\xi$; the period $\xi_{{\scriptscriptstyle H}}$
 is related to the  period $t_{{\scriptscriptstyle H}}$ of the plasma
oscillation by $\xi_{{\scriptscriptstyle H}}=ct_{{\scriptscriptstyle H}}$, because the initial and final $z_e$ are the same.
 As $Z\to 0^+$ the cycle becomes more and more elongated towards negative $\Delta$'s, and
for $Z=0$ the path is  open on the left, because
$\U(\Delta;0)\!=\!0$ for $\Delta\!\le\!0$. This implies $\hat\Delta(\xi,0)\!\to\!-\!\infty$
as $\xi\!\to\!\infty$; in other words, in the plane wave idealization (only) the $Z\!=\!0$ electrons escape to $z_e\!=\!-\infty$ infinity.  
The two solutions $\spm(\Delta;Z)$, defined via (\ref{spm}) for $\Delta\in[\Dm,\DM]$ together make up the cycle; in  fig. \ref{graphsb}d-f $\sp$ is plotted half green and half red, while $\sm$ is plotted half orange and half blue. \
We denote as $P_0,P_1,P_2,P_3$ the points that respectively lie the most left, down, right and up along the generic $Z$-cycle. 
$P_0,P_2$ respectively  minimize, maximize $\Delta$ while  $P_1,P_3$ 
respectively  minimize, maximize  $s$.
$\DM(Z)>0,\Dm(Z)<0$, the  maximal and minimal $\Delta$, are the solutions of the equation  $H(\Delta,s)\!=\!h$
that make $\gamma$ minimal, i.e. are the solutions of the equation
$h\!-\!\mu\!=\!\U(\Delta;Z)$ in the unknown  $\Delta$; the curves 
$\big(\Delta,\sp(\Delta,Z)\big)$, $\big(\Delta,\sm(\Delta,Z)\big)$ share the endpoints $\big(\DM,\mu\big)$, $\big(\Dm,\mu\big)$ and together make up the cycle. There is a unique $Z_b\!>\!0$ such that $\Dm(Z_b)=-Z_b$.
The minimal $s$ is the minimum $\smm$ of $\sm$, and  the maximal $\sM$ 
of $s$ is the maximum $\sM$ of $\sp$; both are obtained at $\Delta=0$
(in fact $\partial \spm/\partial \Delta\propto \partial \U/\partial \Delta=0$ 
only for $\Delta=0$), i.e.
\be
\smm(Z)=\sm(0,Z)=h\!-\!\sqrt{h^2\!-\!\mu^2},\qquad \sM(Z)=\sp(0,Z)=h\!+\!\sqrt{h^2\!-\!\mu^2}; \label{smM}
\ee
also the maximum Lorentz factor $\gammaM=h=(\smm\!+\!\sM)/2$ is attained at $\Delta=0$.
Summarizing, 
\bea
\ba{ll}
P_0\equiv(\Dm,\mu), \qquad\qquad
& P_1\equiv(0,\smm), \\[6pt]
P_2\equiv(\DM,\mu), \qquad\qquad
& P_3\equiv(0,\sM).
\ea\eea
As $P(\xi,Z)$ moves anticlockwise along the generic cycle, it passes in the order through
$P_0,P_1,P_2,P_3$ at each turn; we shall denote as $\bar\xi^i$ ($i=1,2,3,4$) 
the $\xi$-lapse necessary to go from $P_{i-1}$ to $P_i$, and  as $\xi^i_n$ ($i=0,1,2,3$) the value  of $\xi$ such that  $P(\xi^i_n,Z)=P_i$ during the $n$-th turn, $n\in\NN_0$. Clearly
$\xiH=\bar\xi^1\!+\!\bar\xi^2\!+\!\bar\xi^3\!+\!\bar\xi^4$, \ and 
\bea
\xi_n^0=\xi_0^0\!+\!n\xiH,\qquad\xi_n^1=\xi_n^0\!+\!\bar\xi^1,\qquad\xi_n^2=\xi_n^1\!+\!\bar\xi^2,
\qquad\xi_n^3=\xi_n^2\!+\!\bar\xi^3.
\eea
On every piece of the path  where $\varepsilon\!\equiv\!\mbox{sign}(\Delta')$ is constant (\ref{heq1}) is integrable by quadrature: replacing (\ref{spm}) in (\ref{heq1}a) we find \
$d\xi\!=\!\mp\spm d\Delta/\sqrt{ \bar\gamma^2-\mu^2}$. \ 
Hence the  $\xi$-lapse $\Delta \xi\!>\!0$ that passes going (anticlockwise) from any origin $P_*\!=\!(\Delta_*,s_*)$
to any endpoint  $P_{\#}\!=\!(\Delta_{\#},s_{\#})$  is
\be
\Delta \xi=\int\limits^{\Delta_{\#}}_{\Delta_*}\!\!\frac{  \varepsilon \,s_{-\varepsilon}\: d\Delta}{\sqrt{ \bar\gamma^2(\Delta;Z)-\mu^2}}=\displaystyle\int\limits^{\Delta_{\#}}_{\Delta_*}\!\!d\Delta \left[\frac{  \varepsilon\bar\gamma(\Delta;Z)}{\sqrt{ \bar\gamma^2(\Delta;Z)-\mu^2}}-1
\right]
\ee
[in fact $\varepsilon\!=\!\mbox{sign}(\Delta_{\#}\!-\!\Delta_*)$]. 
As a result, the period\footnote{The NR plasma period $t_{{\scriptscriptstyle H}}^{{\scriptscriptstyle nr}}=\sqrt{\frac{\pi m}{\widetilde{n_0}(Z) e^2}}$ is  obtained in the limit $h\to\mu$ of vanishing oscillation amplitudes.
}  and the $\xi$-lapses  $\bar\xi^i$ are given by
\bea
&& c\, t_{{\scriptscriptstyle H}}=\xi_{{\scriptscriptstyle H}}=2\int\limits^{\DM}_{\Dm}\!\!\!d\Delta\, \frac{\bar\gamma(\Delta;Z)}{\sqrt{ \bar\gamma^2(\Delta;Z)-\mu^2}} , \label{period}\\
&&\bar\xi^4=\frac14 \xiH-\Dm,\quad
 \bar\xi^1=\frac14 \xiH+\Dm,\quad
\bar\xi^2=\frac14 \xiH-\DM,\quad
\bar\xi^3=\frac14 \xiH+\DM.
\eea

\subsection{Auxiliary problem: constant initial density}
\label{auxi}

\begin{figure}[hbtp]
\includegraphics[width=7.8cm]{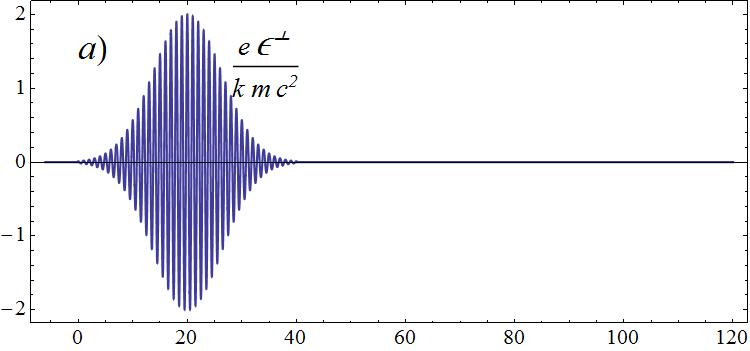} \hfill \includegraphics[width=8cm]{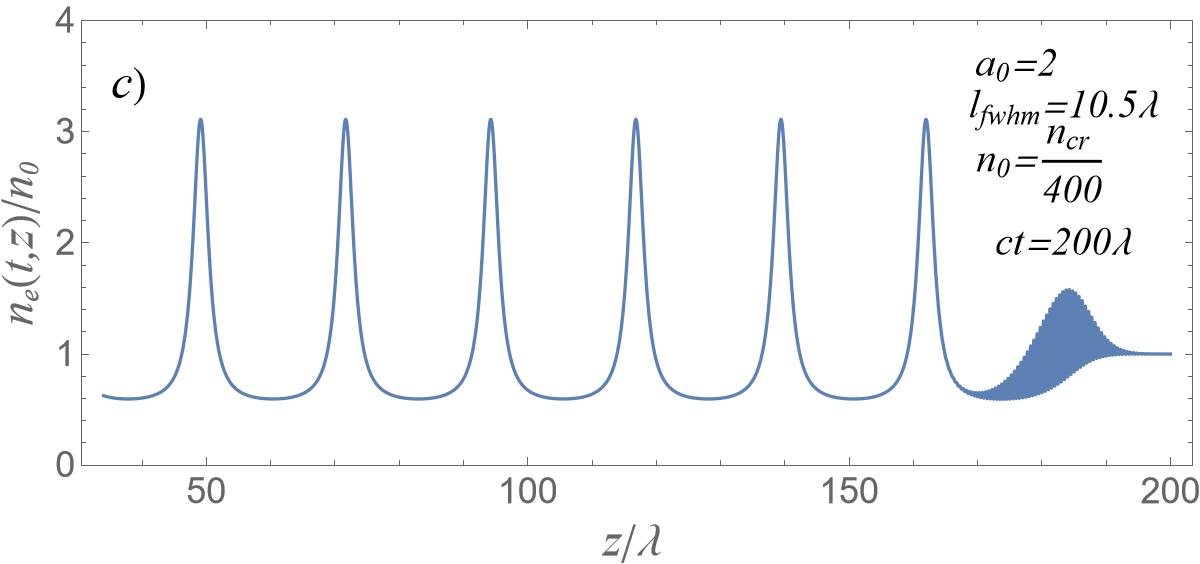}\\
\includegraphics[width=7.8cm]{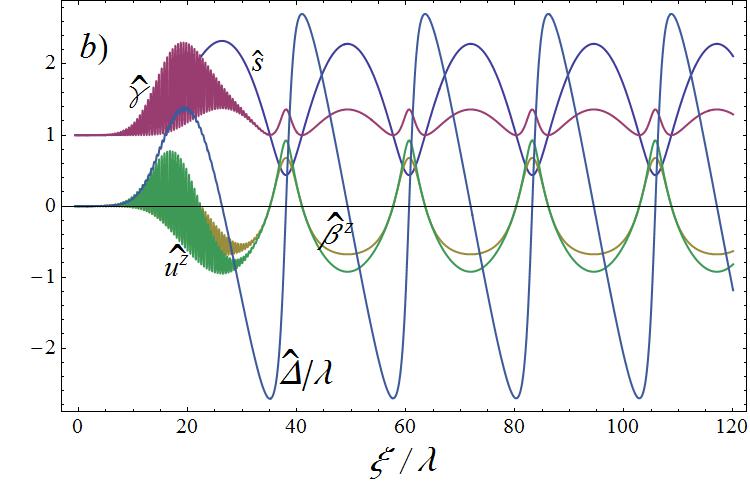} \hfill
\includegraphics[width=8.2cm]{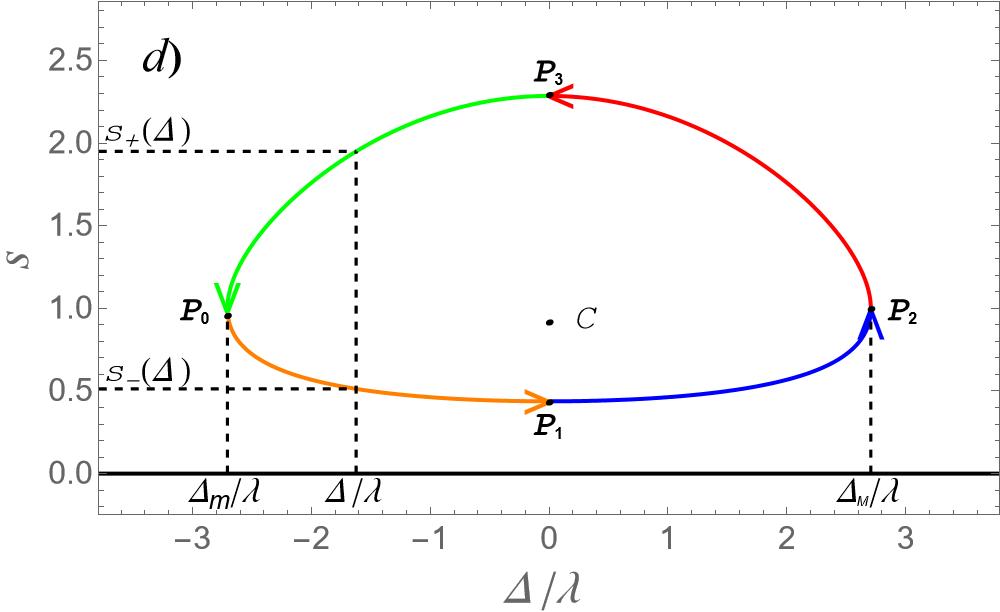}\\
\includegraphics[height=6cm]{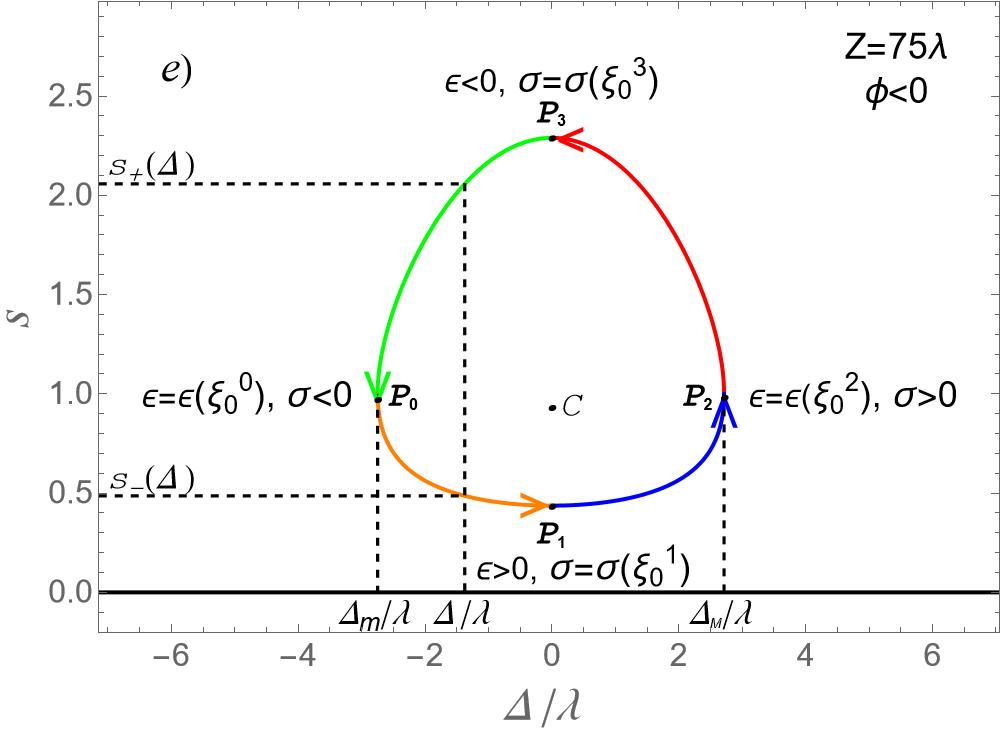}
\hfill \includegraphics[height=6cm]{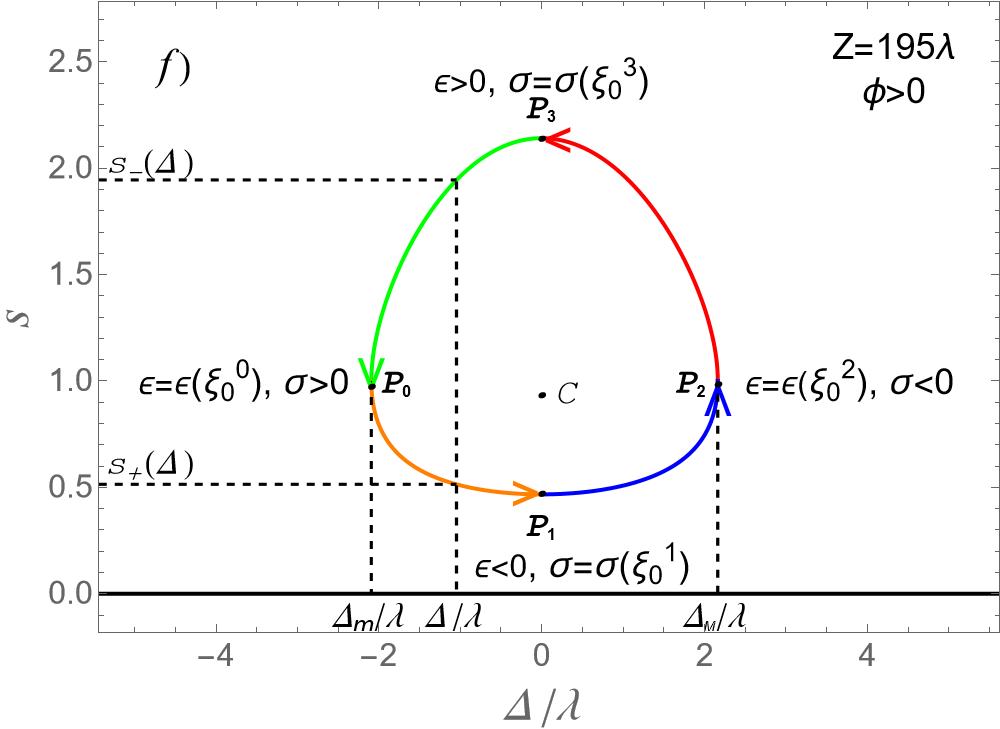}
\caption{\ (a)  Normalized Gaussian pump of length $l_{fwhm}\!=\!10.5\lambda$, linear polarization, 
peak  amplitude $a_0\!\equiv\!\lambda eE^{\scriptscriptstyle \perp}_{\scriptscriptstyle M}/mc^2\!=\!2$,
as in section III.B of \cite{BraEtAl08} (if $\lambda=0.8\mu$m this leads to a peak  intensity $I\!=\!1.7\!\times\!10^{19}$W/cm$^2$). 
\ \ (b) \ Corresponding  solution of (\ref{e1}-\ref{heq2})  if
$\widetilde{n_0}(Z)\!=\!n_0\!\equiv\! n_{cr}/400$, 
 where as usual $n_{cr}\!=\! \pi mc^2/e^2\lambda^2$; this value of 
$n_0$ is as in the right plateau of the density $\widetilde{n_0}(Z)$ plotted in
fig. \ref{Worldlinescrossings}, which we also  pick from \cite{BraEtAl08}, hence this solutions describes the $Z$-electrons for all $Z>z_p$.
 Correspondingly, $h\!=\!1.36$, and, as anticipated, $\hat s$ is indeed  insensitive to fast oscillations of $\Bep$. \
(c) \ Corresponding normalized electron density inside the bulk as a function of $z$ at \ $t=200\lambda/c$. \ \ 
(d) \  Corresponding  phase portrait (at $\xi>l$), with indication of the points $P_0,P_1,P_2,P_3$. 
Phase portraits (at $\xi\!>\!l$) induced by the same pump on  the 
$Z=75\lambda$ electrons (e) and the $Z=195\lambda$ electrons (f) of the  density\ $\widetilde{n_0}(Z)$ plotted in fig. \ref{Worldlinescrossings}.
In (e), (f) we have also pinpointed the signs of $\varepsilon,\sigma$ along different arcs of the portraits 
in the large $\xi$ limit (see section \ref{Jsigma-after}); they are opposite, because so are the signs of $\Phi(Z)$ at $Z=75\lambda,195\lambda$.}
\label{graphsb}
\end{figure}

If \ $\widetilde{n_0}( Z ) =n_0$ \ then 
the $Z$-dependence disappears completely from
(\ref{heq1}-\ref{heq2}), which  reduces to the  Cauchy problem of a forced, relativistic harmonic oscillator
with equilibrium at $\Delta\!=\!0$:
\bea
&& \Delta'=\displaystyle\frac {1\!+\!v}{2s^2}\!-\!\frac 12,\qquad\qquad\qquad
 s'=M\Delta,\label{e1}\\[10pt]
&&  \Delta(0)\!=\!0, \qquad\qquad\qquad\qquad\:\:   s(0)\!=\! 1,\label{e2}
\eea
where $ M \!\equiv\!4\pi e^2n_0/mc^2 \!=\!\frac{\omega_p^2}{c^2}$ ($\omega_p$ is the {\it plasma angular frequency}). 
The equation 
$s''=M [(1\!+\!v)/s^2\!-\!1]/2$, which follows from (\ref{e1}),  is essentially
formula (3)  in \cite{BerMur90}\footnote{The $x,y,k_p,\Bp,\gamma_\perp,...$ used there are our  $\xi,s,k,\bu,\mu=\sqrt{1\!+\!v},...$.}, which was studied there assuming a  pulse (\ref{modulate}) circularly polarized; the solution was found: in closed form for a  modulating amplitude $\epsilon$ constant in $[0,l]$ and zero outside; numerically for a few other $\epsilon$'s having Gaussian rise and fall with the same or different variances.
The potential energy in (\ref{hamiltonian}) takes the form 
$\U(\Delta,Z)\!\equiv\!M\Delta^2/2$. 
In fig. \ref{graphsb} we plot a typical Gaussian $\Bep(\xi)$ and the corresponding solution of (\ref{e1}-\ref{e2}) 
if $Ml_{fwhm}^2\!\simeq\!11$, $Ml^2\!\simeq\!158$. The abrupt rise of $\hat\Delta$ 
in the location where $\hat s$ reaches a minimum [see the comment after eq. (\ref{heq2})] is evident; the solution manifestly becomes periodic   for $\xi\!\ge\! l$, i.e.  after the pulse has passed.  Not only the solution $\big(\Delta(\xi),s(\xi)\big)$, but all of $h$, 
$\hat\bu_e,\hat\BD$, are $Z$-independent.  It follows that $\partial\hat\Delta/\partial Z\!\equiv\!0$ and by (\ref{invZtoz}) the automatic invertibility of $z_e(t,\!Z)$;
moreover, the inverse functions $Z_e(t,\!z)$ and $\hat Z_e(\xi,\!z)$ have the closed form  
\be
 Z_e(t,z)=z-\Delta(ct\!-\!z)\qquad\Leftrightarrow\qquad \hat Z_e(\xi,z)=z\!-\!\Delta(\xi),                                                \label{sol"}
\ee 
which makes all Eulerian fields - such as   (\ref{sol'}) or (\ref{expln_e}) - completely explicit and 
dependent on $t,z$ only through $ct\!-\!z$, i.e. propagating as travelling-waves; in particular 
(\ref{expln_e}) becomes
\bea
 &&  n_e(t,z)=\frac{n_0}2 \,
\left[1\!+\! \frac{1\!+\!v(ct\!-\!z)}{s^2(ct\!-\!z)}\right]=
\frac{n_0}{1-\beta^z(ct\!-\!z)},         \label{expln_e'}
\eea
implying the known remarkable  consequence \ $ n_e(t,z) > n_0/2$ \ \cite{AkhPol56}.
$H$ is conserved in $\xi$-intervals where $v(\xi)\!\equiv\!v_c\!\equiv$const; its value 
$\bar h$ is related to  $\DM\!=\!-\Dm$  by $\bar h\!-\!\mu\!=\!M \DM^2/2$. \ 
For all $\bar h$ the path $P(\xi;Z)$ is a cycle independent of $Z$ around  the center $C$ as in fig. \ref{graphsb}d, corresponding to a  periodic motion; the $Z$-independent period (\ref{period})  is\footnote{Setting $w:=\frac{\Delta}{\DM}=\frac{\Delta\sqrt{M}}{\sqrt{2(\bar h\!-\!\mu)}}$
we obtain (\ref{period0}) as follows:
\bea
\frac {\bar \xi_{{\scriptscriptstyle H}}}4
= \!\int\limits^{\DM}_{0}\!\!\!\! 
\frac{d\Delta}{\sqrt{\bar\gamma(\Delta)\!-\!\mu}}\frac{\bar\gamma(\Delta)}{\sqrt{\bar\gamma(\Delta)\!+\!\mu}}=
\sqrt{\frac{2}M}\!\int\limits^{1}_0\! \! \frac{dw}{\sqrt{1\!-\!w^2 }}
\frac{2\mu\!+\!(\bar h\!-\!\mu)(1\!-\!w^2)\!-\!\mu}{\sqrt{2\mu\!+\!(\bar h\!-\!\mu)(1\!-\!w^2)}}\nn=
\sqrt{\frac{2}M}\!\int\limits^{1}_0\! \! \frac{dw}{\sqrt{1\!-\!w^2 }}
\left[\sqrt{\bar h\!+\!\mu}\sqrt{1\!-\!\alpha^2 w^2}-\frac{\mu}{\sqrt{\bar h\!+\!\mu}\sqrt{1\!-\!\alpha^2 w^2}}\right]=\sqrt{\frac{2(\bar h\!+\!\mu)}M}\left[E(\alpha)-\frac
{\mu}{\bar h\!+\!\mu}K(\alpha)\right].\nonumber
\eea}
\be
c\, \bar t_{{\scriptscriptstyle H}}= \bar \xi_{{\scriptscriptstyle H}}
=4\sqrt{\frac{2(\bar h\!+\!\mu)}M}\left[E(\alpha)-\frac
{\mu}{\bar h\!+\!\mu}K(\alpha)\right], 
\qquad \alpha:=\sqrt{\frac{\bar h\!-\!\mu}{\bar h\!+\!\mu}} , \label{period0}
\ee
where $K,E$ are the complete elliptic integrals of the first and second kind.
This respectively reduces to $ct_{{\scriptscriptstyle H}}^{{\scriptscriptstyle nr}}$,
$ct_{{\scriptscriptstyle H}}^{{\scriptscriptstyle ur}}\simeq \frac{15\pi}{8}\sqrt{\frac{2\bar h}M}$ in the nonrelativistic, relativistic limits $\bar h\to\mu$, $\bar h\to\infty$.

\section{Motion of a test electron in the plasma wave}
\label{test-electron}

A generic test electron injected at $\xi=\xi_0>l$ in the PW behind the pulse is subject itself to the
longitudinal electric field (\ref{explE}), which enters its equation of motion (\ref{reduced}). 
Therefore the evolution of its dynamical variables $\hat z_i(\xi),\hat s_i(\xi)$ (the suffix $i$  stands for injected) is ruled by a Cauchy problem with the same ODEs (\ref{heq1}), but  initial conditions different from (\ref{heq2}), %of the form
\bea
\ba{llllll}
\hat z_i'(\xi) &=& \displaystyle\frac {\mu_i^2}{2\hat s_i^2(\xi)}\!-\!\frac 12, \qquad 
&\hat s_i'(\xi) &=& K\left\{\!
\widetilde{N}\left[\hat z_i(\xi) \right] \!-\! \widetilde{N}\left[\hat Z_e\left(\xi,\hat z_i(\xi) \right) \right] \!\right\}, \\[12pt] 
\hat z_i(\xi_0) &=& z_{i0}, \qquad\quad &\hat s_i(\xi_0) &=& s_{i0};
\ea                    \label{heq-test}
\eea
here \ $\mu_i^2\!=\!1\!+\!\bup_i{}^2$,   $\bup_i$ is the transverse component of its 4-velocity, a constant for $\xi>\xi_0$.

Now assume that $\widetilde{n_0}(z)\!=\!n_0$ if $z\!\ge\! z_s$, for some
$n_0\!\le\!n_b$,  $z_s\!\ge\! 0$ (see fig. \ref{fig1}).
Consequently,  
\ $\widetilde{N}(z)-n_0(z\!-\!z_s)=N_s\equiv\int^{z_s}_0 \widetilde{n_0}(Z)dZ=$ const \ if $z\ge z_s$. \
Let $Z_p$ be the smallest $Z>z_s$ such that $\hze(\xi,Z_p)\ge z_s$ for all $\xi$; in other words 
it follows that $\Dm(Z_p)=z_s\!-\!Z_p$. For all $Z\ge Z_p$
the solution of (\ref{heq1})-(\ref{heq2}) is the one of (\ref{e1}-\ref{e2}) obtained with
 uniform initial electron (and proton) density 
$\widetilde{n_0}(Z)=n_0$; correspondingly, 
$\hat \Delta(\xi,Z)\equiv\hze(\xi,Z)\!-\!Z=\Delta(\xi)$, and $\hat Z_e$ is given by (\ref{sol"}b).
If the initial conditions (\ref{heq-test}) of  the test electron fulfill $\hat Z_e(\xi_0,z_{i0})=z_{i0}\!-\!\Delta(\xi_0)\ge Z_p$ 
(what implies $z_{i0}>z_s$), then (\ref{heq-test}b)  boils down to $\hat s_i'(\xi)=M\Delta(\xi)$, which is
an equation  decoupled from $\hat z_i$ that can be integrated first;  then also  the equation for $\hat z_i$  can be integrated. Note that these equations contain no longer any back-reaction mechanism  [cf. the discussion in section \ref{during} applied to eq. (\ref{e1})] preventing $\hat s_i(\xi)$ to vanish at any $\xi>\xi_0$. Introducing
the constant $\delta s\equiv s_{i0}\!-\!s(\xi_0)$ and integrating  (\ref{heq-test}) we thus find
\bea
\hat s_i(\xi)=s_{i0}+M\!\!\int^\xi_{\xi_0}\!\!\!\! dy\,\Delta(y)=\delta s+s(\xi),\qquad
\hat z_i(\xi)=z_{i0}+\!\!\int^\xi_{\xi_0}\!\!\!\! dy\left[\frac {\mu_i^2}{2\hat s_i^2(y)}
\!-\!\frac 12\right].  \label{test-motion}
\eea
This highlights another reason why it is convenient to use $s$ and $s_i$ as fundamental dynamical variables: \ $\hat s_i\!-\!s$ \ is constant, while the same nice property is not shared by
the differences \ $\hat \gamma_i\!-\!\gamma$, $\hat u^z_i\!-\!u^z$. \ Assuming
for simplicity  $\mu_i=\mu$ [what is fulfilled e.g. if the test electron is self-injected by WB
of a PW induced by a slowly modulated monochromatic pulse: $\bup_i(l)\simeq 0\simeq\bup(l)$], we need to distinguish the following cases:

\begin{enumerate}

\item $\delta s=0$. \ Then $s_i(\xi)=s(\xi)$, 
$\hat z_i(\xi)=z_{i0}\!-\!\Delta(\xi_0)\!+\!\Delta(\xi)$, i.e. the test electron has the same periodic motion of the $Z=z_{i0}\!-\!\Delta(\xi_0)$ electrons of the PW.

\item $\delta s>0$. \ It follows that $s_i(\xi)>s(\xi)$,
and\footnote{In fact, since $s_i(\xi)>s(\xi)$ the following difference is negative:
\bea
\hat z_i(\xi)\!-\!\hze\!\left[\xi,\hat Z_e(\xi_0,z_{i0})\right]=\hat z_i(\xi)\!-\!\hat Z_e(\xi_0,z_{i0}) \!-\!\Delta(\xi)=\hat z_i(\xi)\!-\!z_{i0}\!+\! \Delta(\xi_0)\!-\!\Delta(\xi)=\!\!\int^\xi_{\xi_0}\!\!\!\! dy\left[\frac {\mu^2}{2\hat s_i^2(y)}\!-\!\frac 12\right]\nn
+\Delta(\xi_0)\!-\!\Delta(\xi)=\!\!\int^\xi_{\xi_0}\!\!\!\! dy\left[\frac {\mu_i^2}{2\hat s_i^2(y)}\!-\!\frac 12\right]\!-\!\!\int^\xi_{\xi_0}\!\!\!\! dy\left[\frac {\mu^2}{2\hat s^2(y)}\!-\!\frac 12\right] =
\!\!\int^\xi_{\xi_0}\!\!\!\! dy\left[\frac {\mu_i^2}{2\hat s_i^2(y)}\!-\!\frac {\mu^2}{2s^2(y)}\right] \label{diffzZ}
\eea
}
$\hat z_i(\xi)<\hze\!\big[\xi,\hat Z_e(\xi_0,z_{i0})\big]$ for all  $\xi>\xi_0$.  The variation $\Delta z_i$ over a period $\xiH$
is negative (since it is zero when $\hat s_i$ is replaced by $s<\hat s_i$), and  
$\hat z_i(\xi)$ decreases   by $\Delta z_i$  after each period $\xiH$. 
This yields negative average velocities w.r.t. to `times' $\xi,t$ respectively equal to $\Delta z_i/\xiH>0$ and 
$c\Delta z_i/(\xiH\!+\!\Delta z_i)$. 

\item $\delta s<0$. \ It follows that $s_i(\xi)<s(\xi)$ 
and [(replacing in (\ref{diffzZ})]
$\hat z_i(\xi)>\hze\!\big[\xi,\hat Z_e(\xi_0,z_{i0})\big]$ for all  $\xi>\xi_0$. If $\xi_0\in]\xi^1_{l-1},\xi^1_l]$ 
(with some $l\in\NN$) then
$\hat s_i$ attains its minimum \ $s_i^m\equiv \smm\!+\!\delta s$ \ at $\xi=\xi^1_h$, $h=l,l\!+\!1,...$; \ in the vicinity of $\xi^1_h$ we can approximate
\be
\hat s_i(\xi)\simeq  s_i^m+ s''(\xi^1_h)\frac{(\xi\!-\!\xi^1_h)^2}2= 
s_i^m+ \frac M4\!\left(\frac{\sM^2}{\mu^2}\!-\!1\right)(\xi\!-\!\xi^1_h)^2.\label{approx-test-s}
\ee

\begin{enumerate}
\item
\label{s_i^m>0}
 $s_i^m>0$. \  Then the path of the test electron in the $(z,s)$ space 
is an open curve as illustrated in fig. \ref{selfinjected-untrapped}, where the 
test electrons  originate from self-injection as described in section \ref{self-injected-electrons}.  
The maximum (normalized) energy, attained  at the minimum  $s_i^m$ of $\hat s_i(\xi)$, is
\be
\gammaM_i=\frac{\mu_i^2}{2s_i^m }+\frac{s_i^m}2;
\label{gammaM}
\ee
while as the integrand of (\ref{test-motion}b)  is $\xiH$-periodic ($\xiH$ is the period (\ref{period0}) 
associated to $n_0=\widetilde{n_0}(Z)$, for $Z>Z_p$),   its integral $\Delta z_i$ over a period $\xiH$ is positive 
, and $\hat z_i(\xi)$ grows   by $\Delta z_i$  after each period $\xiH$. 
This yields positive average velocities w.r.t. to `times' $\xi,t$ respectively equal to $\Delta z_i/\xiH>0$ and 
$c\Delta z_i/(\xiH\!+\!\Delta z_i)$. The test electron in the averages moves forward, but {\it does not remain in the same plasma wave trough trailing the pump} where it was at $\xi=\xi_0$; it slides back to other ones.
Choosing $[\xi^0_n,\xi^0_{n+1}]$ as the integration interval,
in the limit $s_i^m\downarrow 0$  the leading contribution to $\Delta z_i$ comes from around
$\xi=\xi^1_n$ via (\ref{approx-test-s}) and reads  [with $\sM$ given in (\ref{smM})]
\bea
\Delta z_i=\!\!\int\limits^{\xi^0_{n+1}}_{\xi^0_n}\!\! d\xi\left[\frac {\mu_i^2}{2\hat s_i^2(\xi)}\!-\!\frac 12\right]
\stackrel{s_i^m\downarrow 0}{\sim}
\frac {\mu_i^2\pi}{(s_i^m)^{3/2}\sqrt{2M\!\left(\frac{\sM^2}{\mu^2}\!-\!1\right)}}\simeq
\frac {2\pi \,(\gammaM_i)^{3/2}}{\mu_i\sqrt{M\!\left(\frac{\sM^2}{\mu^2}\!-\!1\right)}}.
\eea

\item 
\label{s_i^m<0}
$s_i^m< 0$. Then the path of the test electron in the $(z,s)$ space 
is again an open curve; but there is  $\xi_f\in]\xi_0,\xi^1_l[$ such that $\hat s_i(\xi)>0$ for all $\xi_0\le\xi<\xi_f$, whereas $\hat s_i(\xi_f)=0$. In terms of $\xi$ the evolution is well-defined only for $\xi<\xi_f$, but, as 
noted after (\ref{hatt}), this corresponds to all $t<t_f=\infty$.
As $\xi\to\xi_f$ $\hat s_i(\xi)$ vanishes, and $\hat z_i(\xi)$ diverges; the test electron  
{\bf is trapped, i.e. remains, in the same plasma wave trough trailing the  pulse}  where it was at $\xi=\xi_0$, and its momentum grows with $\xi$ (or equivalently $t, \hat z_i$).  
This is illustrated e.g. in fig. \ref{selfinjected-trapped}, where the trapped 
test electrons  originate from self-injection as described in section \ref{self-injected-electrons}.
More precisely, near $\xi_f$ we can approximate
\be
\hat s_i(\xi)\simeq   - s' (\xi_f)\,(\xi_f\!-\!\xi)=
\left|s' (\xi_f)\right|\,(\xi_f\!-\!\xi).    \label{approx-test-s'}
\ee
 Replacing this in (\ref{test-motion}) we obtain as the leading term in the limit $\xi\to \xi_f$
\bea
\hat z_i(\xi)\, \simeq  \int^\xi_{\xi_0}\!\!  \frac {dy \,\mu_i^2}{2\hat s_i^2(y)}
\simeq  \int^\xi_{\xi_0}\!\!  \frac {dy \,\mu_i^2}{2\left[  s'(\xi_f) \,(\xi_f\!-\!y)\right]^2}
\simeq \frac {\mu_i^2}{2\left[ s'(\xi_f)\right]^2 (\xi_f\!-\!\xi)}. \label{approx}
\eea
We can express $\hat s_i$ in terms of the longitudinal coordinate $\hat z_i$ by solving (\ref{test-motion}b) for $\xi$ and replacing the result in  (\ref{test-motion}a).  If  $\hat z_i$  is large the approximation (\ref{approx}) is good and can be solved for $\xi_f\!-\!\xi$ in the closed form $\xi_f\!-\!\xi\simeq  \mu_i^2/2 \hat z_i\left[ s'(\xi_f)\right]^2$,  
which replaced in (\ref{approx-test-s'}) yields 
\bea
\hat s_i\simeq\frac {\mu_i^2}{2 \left|s' (\xi_f)\right|\,\hat z_i},\qquad\qquad
\hat \gamma_i=\frac{\mu_i^2}{2\hat s_i}+\frac{\hat s_i}2\simeq\frac{\mu_i^2}{2\hat s_i}
\:\simeq \: \left|s' (\xi_f)\right|\,\hat z_i,           \label{s_i^m<0|s_iz_i}
\eea
i.e. {\bf the Lorentz factor (and energy) grow linearly  with the distance travelled}.
The results become more explicit by replacing $\left|s' (\xi_f)\right|\!=\!M\Delta(\xi_f) $.

\item 
\label{s_i^m=0}
 $s_i^m=0$. Then  $\hat s_i(\xi_f)=0$ occurs with $\xi_f=\xi^1_l$,  and at leading order
\bea
\hat z_i(\xi)\simeq \frac {2\mu_i^2}{ M^2\!\left(\frac{\sM^2}{\mu^2}\!-\!1\right)^2 }
\frac 1{ (\xi\!-\!\xi^1_l)^3}, \qquad \hat s_i\simeq 
\frac {\mu_i^{4/3}}{2^{4/3} \hat z_i^{2/3} \left[M\!\left(\frac{\sM^2}{\mu^2}\!-\!1\right)\right]^{1/3}},\\
\hat \gamma_i=\frac{\mu_i^2}{2\hat s_i}+\frac{\hat s_i}2\simeq\frac{\mu_i^2}{2\hat s_i}
\simeq\hat z_i^{2/3} \: \left[2\mu_i^2 M\!\left(\frac{\sM^2}{\mu^2}\!-\!1\right)\right]^{1/3}.
\label{approx'}
\eea
Again, as $t\to\infty$ the test electron  is trapped (i.e.
remains) in the same plasma wave trough where it was at $t=t_0$, and its $ \gamma_i$ diverges with $t$ (or equivalently, with $z_i$), but at a slower rate.  

\end{enumerate}
\end{enumerate}

{\bf Remarks.} \ From the above discussion we see that:

\renewcommand{\labelenumi}{\theenumi}
\renewcommand{\theenumi}{\roman{enumi})}%
\begin{enumerate}
\item The inequality  \ $s_i^m\le 0$ \
is the {\it trapping condition for a test electron in a PW trough}. 

\item In the present schematization the pump  and the PW trailing it have the phase velocity {\it exactly} equal to $c$ (no less!); hence  the (slower of light) test electrons cannot overshoot the bottom of any plasma trough. In particular, 
in cases 3.2, 3.3 they are {\it phase-locked} all the time to the  accelerating part of the trough where they were injected. 

\end{enumerate}

\section{Wave-breakings and their localization}
\label{Wave-breakings}

The hydrodynamic description (HD) used so far is justified for all $\xi$ such that  the map $\bX\mapsto\hbx_e$
is invertible, i.e. such that $Z'<Z''$ implies $\hat z(\xi,Z')<\hat z(\xi,Z'')$,
or, equivalently, as long as the $Z',Z''$  worldlines  (in reduced 2-dimensional spacetime $\RR^2$)  do not intersect. 
Physically,  this means as long as any two distinct electrons layers (which are identified by 
distinct values $Z',Z''$ of $Z$) remain longitudinally separated. 
Wave-breaking (WB) (i.e. going of two or more electrons layers into each other) at ``time" $\xi$ is characterized by the intersection of different worldlines at that value of $\xi$, i.e. by the existence of $Z',Z'',...$ such that $\hat z(\xi,Z')=\hat z(\xi,Z'')=...$.
WB is excluded as long as  $\hat J>0$  everywhere,  otherwise it occurs in a
suitable spacetime region including the one where $\hJ$ becomes zero or negative.  

Differentiating the equations of motion (\ref{heq1}) w.r.t. $Z$ we immediately find that the $\hat J,\hat\sigma$
defined in (\ref{DefhJhsigma}) fulfill the nonlinear ordinary Cauchy problem 
\bea
&&\ba{ll}
\hat J'=-\kappa\hat\sigma, \qquad\quad  & \hat\sigma'=K\left(\check n\, \hat J\!-\! \widetilde{n_0}\right), \\[8pt]
\hat J(0,Z)=1, \qquad\quad &\hat\sigma(0,Z)=0,\ea                \label{basic} 
\eea
\bea
\mbox{where  }\qquad
\kappa:=\frac {\mu^2}{\hat s^3},\quad \check n(\xi,Z):=\widetilde{n_0}\left[\hze(\xi,Z)\right],
\quad \hat\omega^2:=\frac{K \check n\hs^3}{\mu^2}.
\nonumber
\eea
In fig.s \ref{graphs2}, \ref{graphs2'}, we have plotted some typical solutions of (\ref{basic}), under the conditions described  there.  To study  the evolution of $\hJ,\hs$
we first recast (\ref{basic}) in the equivalent form
\bea
\bchi'=A\bchi+\blambda,\qquad\qquad\bchi(0)=\0,    \label{basic''} \\[12pt]
\mbox{where }\qquad\bchi:=\left(\!\!\ba{l} \hat\varepsilon \\
\hat\sigma\ea\!\!\right)\!,\quad A:=
\left(\!\ba{cc} 0 &-\kappa \\ K\check n & 0 \ea\!\right)\!,\quad
\blambda:=K[\check n\!-\!\widetilde{n_0}]\left(\!\!\ba{l} 0\\ 1\ea\!\!\right)\!,    \label{basic''-2} 
 \eea
$\hat\varepsilon=\hJ\!-\!1$, and express the general solution of (\ref{basic''})  as
\be
\bchi(\xi)=\int^{\xi}_0 \!\!d\eta\: \G(\xi,\eta)\blambda(\eta).
\label{gensol} 
\ee
Here we have made   the notation temporarily lighter by hiding 
the dependence of $A,\G,\bchi,\blambda,...$
on $Z$; the `Green function' $\G(\xi,\eta)$ is the family (parametrized by $\eta$)
of $2\times 2$-matrices  solving the homogeneous Cauchy problems
\ $\G'=A\G, \quad \G(\eta,\eta)=I$. \
Since $\lambda^1=0$, only the second column of $\G$, which we denote as
$\left(\D, \LL\right)^T$, 
appears in (\ref{gensol}); this fulfills the Cauchy problem
\be
 \LL'=K \check n\D, \qquad \D'=-\kappa\LL, \qquad \LL(\eta,\eta)=1, \quad \D(\eta,\eta)=0.  \label{hDeq}
\ee
In the appendix we study this system  reducing it  to 
a `time'-dependent harmonic oscillator with angular frequency $\hat\omega$ by a change of `time' (i.e. independent) variable.
Re-exhibiting the $Z$-dependence, we can rewrite  (\ref{gensol})   in the form
\bea
\left(\!\!\ba{l} \hat\varepsilon(\xi,\! Z)  \\
\hat\sigma(\xi,\! Z) \ea\!\!\right)\! &=& \int^{\xi}_0 \!\!d\eta\: K[\check n(\eta,\! Z)\!-\!\widetilde{n_0}(Z)] 
\left(\!\!\ba{l} \D(\xi,\!\eta;\!Z)\\
\LL(\xi,\!\eta;\!Z)\ea\!\!\right)
 \label{gensol'} 
\eea

In the next two sections we study the solution (\ref{gensol'}) of the
system  (\ref{basic}), or equivalently  (\ref{basic''}), separately in the  $\xi$-lapses  $[0,l]$
and $[l,\infty[$; \ clearly,  
 in the latter \ $A(\xi,Z)$ \ is \ $\xiH(Z)$-periodic.

\subsection{Bounds on $\hat J,\hat\sigma$ for  $0\le\xi\le l$, and no WBDLPI conditions}
\label{Bounds0l}

 \bea
\mbox{Let} \qquad \qquad 
\phi(\xi,\xi_0;Z)\equiv
\!\!\int^{\xi}_{\xi_0}\!\!\!\!   d\eta\,  \sqrt{ \frac{K \check n\,\mu^2}{\hat s^3}(\eta,Z)}, \qquad  
\Gamma(\xi,\!Z)\equiv \left\{\!\frac{\widetilde{n_0}(Z)\,\mu^2(\xi)}{[\check n \hat s^3](\xi,\!Z)}\right\}^{1/4}
 \label{defphi}\\[8pt]
\hat\sigma_a(\xi,\!Z)\equiv
 \!\!\int^{\xi}_{0} \!\! \!\!   d\eta  \,   K\,\left[\check n(\eta,\! Z)\!-\!\widetilde{n_0}(Z)\right] 
\frac{\Gamma(\eta,\!Z) }{\Gamma(\xi,\!Z)} 
\cos[\phi(\xi,\!\eta;Z)],  \label{hatsigma-approx} \\[8pt]
\hat J_a(\xi,\!Z)\equiv 1+\Gamma (\xi,\!Z)\, \sqrt{\!K\widetilde{n_0}(Z)}
\!\int^{\xi}_{0} \!\! \!\!     d\eta\,\left[1\!-\!
\frac{\check n(\eta,\! Z)}{\widetilde{n_0}(Z)}\right]  \Gamma (\eta,\!Z) \, \sin \left[\phi(\xi,\!\eta;\!Z)\right] .
\label{hatJapprox} 
\eea

\begin{figure}[hbtp]
\includegraphics[width=16.5cm]{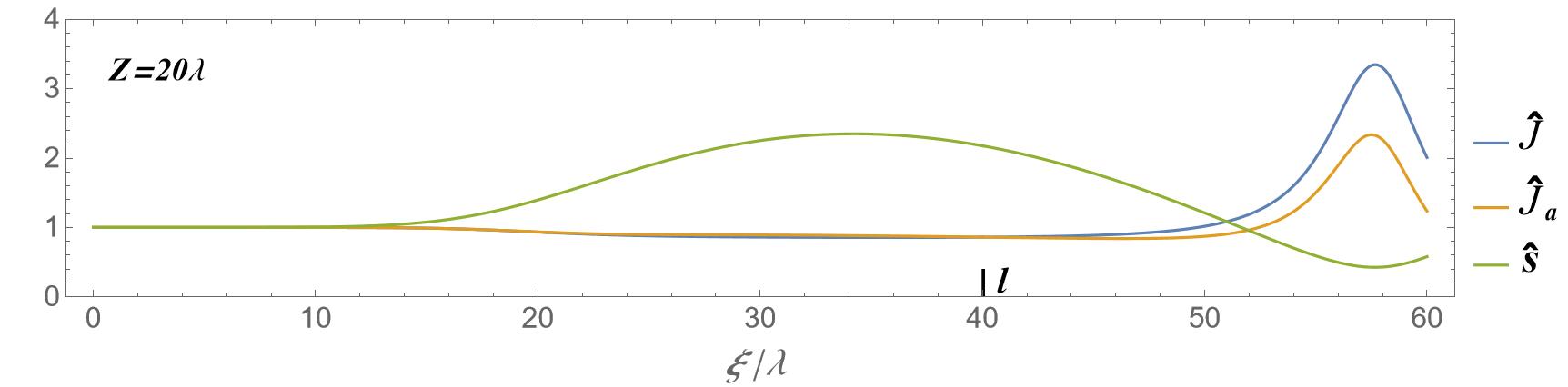}\\
\includegraphics[width=16.5cm]{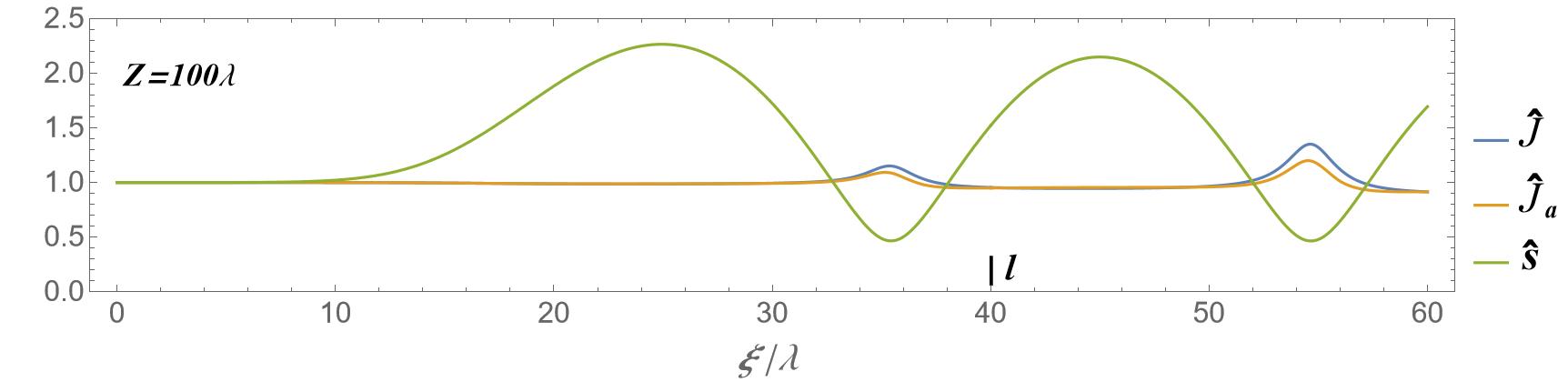}\\
\includegraphics[width=16.5cm]{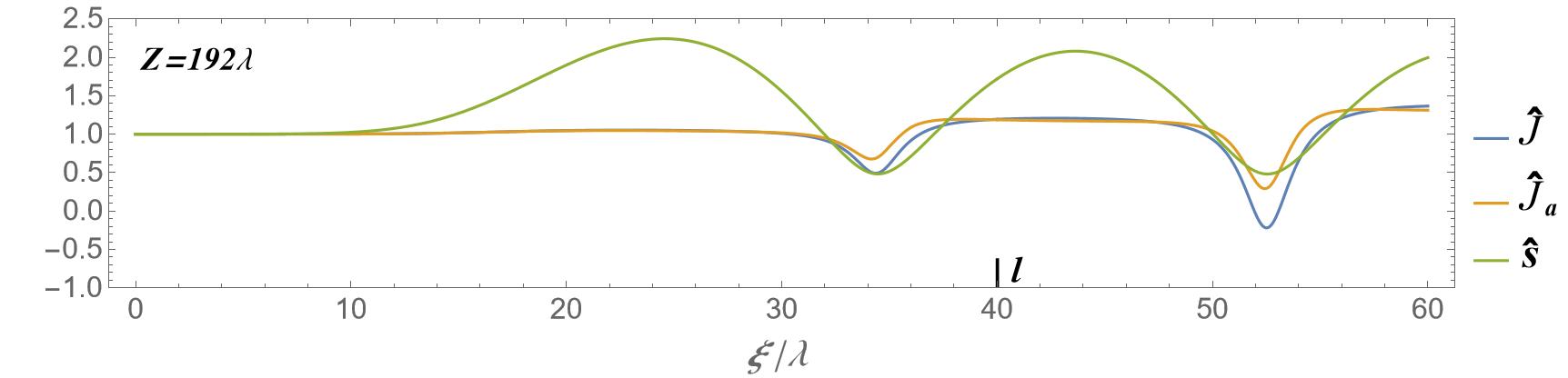}\\
\includegraphics[width=16.5cm]{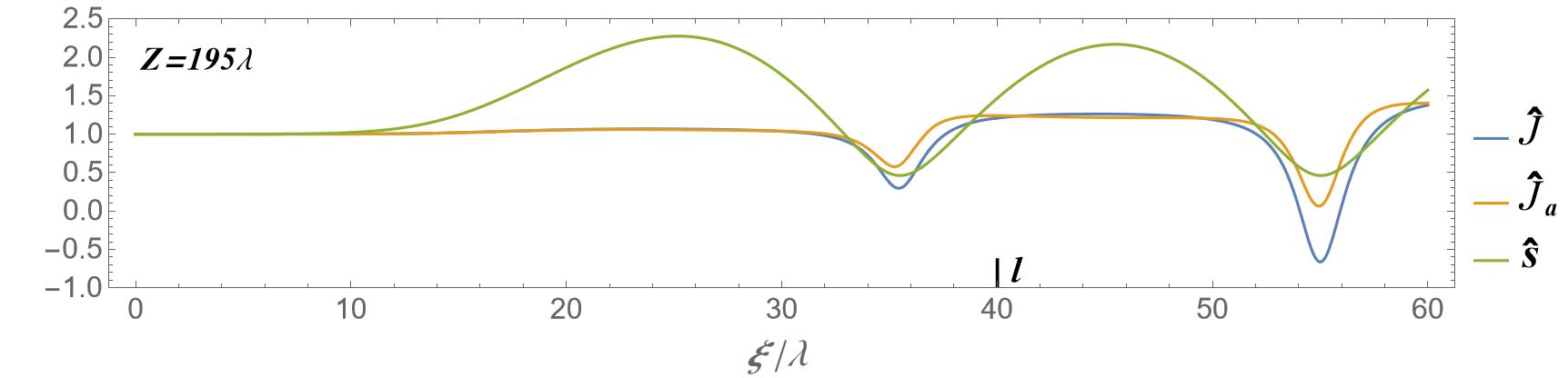}\\
\includegraphics[width=16.5cm]{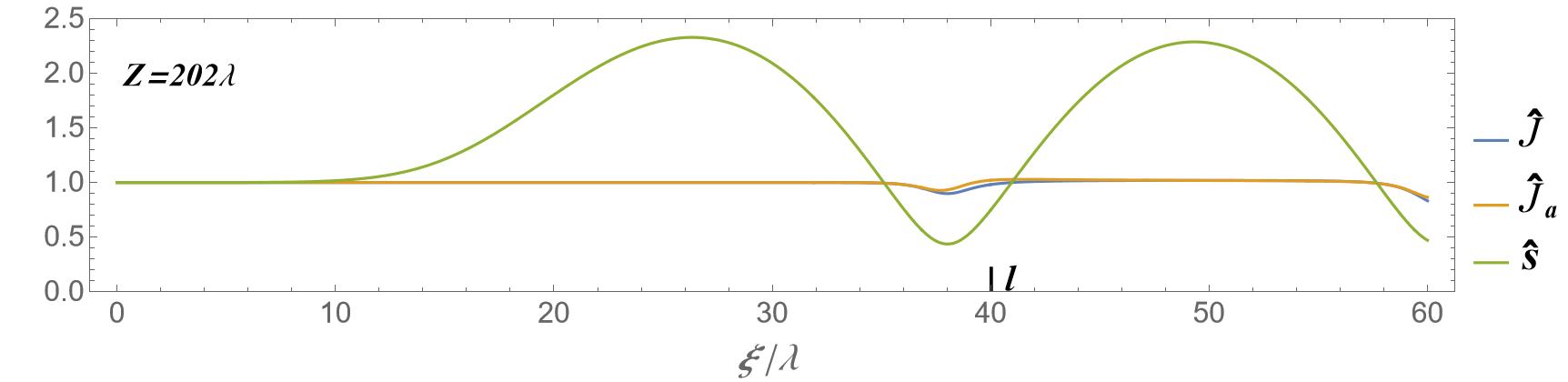}
\caption{Plots of  $\hat J,\hat J_a$  for few values of $Z$
under the conditions considered in fig. \ref{Worldlinescrossings}. Correspondingly $\phi_f$ ranges from about $2\pi$ to about $4.5\pi$. As we can see, $\hat J,\hat J_a$ agree very well except where $\hat s$ approaches 0. Hence, $\hat J(\xi,Z)\simeq\hat J_a(\xi,Z)$ for $\xi\in[0,l]$ if the pulse is essentially short.}
\label{J_vs_Ja}
\end{figure}

\begin{prop} If $\widetilde{n_0}(z)$ is continuous, $d\widetilde{n_0}/dz$ is defined piecewise and bounded, and 
\bea
\left|\int_{0}^{\xi}\!\!\!   d\eta\,\left[\frac{\hat\omega'}{2\hat\omega}(\eta,\!Z)
 \,\sin[2\phi(\eta,\!0,\!Z)\right]\right|\ll \,\phi(\xi,\!0,\!Z),
\label{good-approx-xi}
\eea
then   we can approximate 
\be
\hat J(\xi,\!Z)\simeq \hat J_{a}(\xi,\!Z), \qquad \hat\sigma(\xi,\!Z)\simeq \hat\sigma_a(\xi,\!Z). \label{hatJsigmaapprox} 
\ee
\label{hatJsigma-approx}
\end{prop}
The proof is in appendix \ref{Proofs-0l}. For small $\xi$ $(\hat J_a,\hat\sigma_a)$ is a good approximation of $(\hat J,\hat\sigma)$ expressed in closed form in terms of the family of solutions of (\ref{heq1}-\ref{heq2}); actually, it is the first element of a sequence of approximations of $(\hat J,\hat\sigma)$ that can be determined using the method of \cite{Fiotdho}.
If $\Bep$ is a slowly modulated wave (\ref{modulate}) one can estimate oscillatory integrals
via (\ref{oscillestimate}). This entails replacing $v$ by $v_a$ (see  (\ref{v_a})) in all integrals. The corresponding transformed (\ref{good-approx-xi}) is implied by the stronger condition
\bea
\hat \Lambda_a(\xi,Z):=\int_{0}^{\xi}\!\!\!   d\eta\,\left|\frac{\hat\omega_a'}{2\hat\omega_a} \right|(\eta,\!Z)\ll\,\phi_a(\xi,\!0,\!Z),
\label{good-approx-xi'}
\eea
which is easier to check. 
$2\hat \Lambda_a$ is the  {\it total variation} of $\log\hat\omega_a(\cdot,Z)$ in $[0,\xi]$
($2\hat \Lambda_a(\xi,\!Z)=\left|\log\frac{\hat\omega_a(\xi,Z)}{\hat\omega_a(0,Z)}\right|$ if  $\hat\omega_a(\cdot,Z)$ is monotone).
Since $\phi(\xi,\xi_0;Z)$  strictly grows (resp. decreases) with $\xi$ (resp. $\xi_0$), 
the first term of (\ref{hatJapprox}) and the integrands of (\ref{hatJapprox}), (\ref{hatsigma-approx})
oscillate between positive and negative values as $\xi$ grows.
$\hat J_a(\xi,\!Z)$ is a very good approximation of $\hat J(\xi,\!Z)$ at least as long as $\hat s\ge 1$, see e.g. fig. \ref{J_vs_Ja}; if the pulse is essentially short, this means for $0\!\le\!\xi\!\le\! l$. 

\medskip
The numerical resolution of (\ref{basic}), or evaluation of the rhs(\ref{hatJapprox}), for values of $Z$ in a sufficiently fine lattice  allows to  
understand with sufficient confidence  whether and where WB takes place for a specific set of input data  (the laser pulse and the initial electron density) and $\xi$ not too large. 
We now derive apriori bounds  on  $\hat J,\hat\sigma$ as functionals of the
input data, more precisely of $v$ and of the bounds\footnote{We stress that, by the definition (\ref{n-bounds}), the range of $\check n$  coincides with that of $\widetilde{n_0}$, 
and therefore it keeps bounded (whereas $\hat n_e,n_e$ diverge where $\hat J=0$); $n_u(Z),n_d(Z)$ are upper and lower bounds for $\widetilde{n_0}(Z)$
in the interval of $Z$'s occupied by the $Z$-electrons during their interaction with the pulse.
If $\widetilde{n_0}(Z)\equiv n_0=$ const then   $n_u= n_u'=n_d=n_0$ as well.
} $n_u(Z),n_d(Z)$ defined in
(\ref{n-bounds}), {\it without explicitly solving} %the $Z$-electrons' equations of motion 
(\ref{heq1})-(\ref{heq2}) \& (\ref{basic}),  via  (\ref{hatJsigmaapprox}) and  apriori
bounds  on $\hat\Delta,\hat\sigma$ (as presented in section \ref{during}). These bounds  on  $\hat J,\hat\sigma$ allow to  derive sufficient conditions on these data ensuring that $\hat J$ 
remains positive in a  given spacetime region, so that  WBDLPI %respectively takes or 
does not take place. 

In section \ref{Proofs-est-phi} we show that for $\xi\!\in\![0,\tilde\xi_3(Z)]$
the function $\phi(\xi,\!0;\!Z)$ is bounded by the inequalities 
\be
\phi(\xi,0;Z)
\le \left[\xi+\hat\Delta(\xi,Z)\right]\,\sqrt{K n_u(Z)} \le\left[\xi+\hDO(\xi)\right]\,\sqrt{K n_u(Z)}.
\label{est-phi}
\ee
In particular, since 
$\hat\Delta\le\hDOu$ for all $\xi\in[0,\tilde\xi_3]$,  the inequality
\bea
\phi_f(Z):= \phi(l,0;Z)\le 
 \sqrt{ K n_u} \, \left(l\!+\!\hDOu\right)=:\rho_u\,\pi
 \label{phi(l)bound}
\eea
holds if (\ref{Lncond'}a) does. 
In appendix \ref{Proofs-0l} we prove that if  $\phi_f\in[0,\pi]$ then for all $\xi\in[0,l]$
\bea
\frac{\widetilde{n_0}}{n_u}\!+\!  \left(\!1\!-\!\frac{\widetilde{n_0}}{n_u}\! \right) 
\Gamma \,
 \cos\phi \: \lesssim  \:  \hat J  \: \lesssim  \: \frac{\widetilde{n_0}}{n_d}- 
\left(\! \frac{\widetilde{n_0}}{n_d} \!-\!1\! \right)  \Gamma \,
\cos\phi  ,                    \label{cond0}
\eea
[here $\phi\equiv\phi(\xi,\!0;\!Z)$]. 
Hence, $J>0$, and no WBDLPI involves the $Z$ electrons, if %$0\le\phi_f\le\pi$ and
\be
0\le\phi_f(Z)\le\pi\quad\mbox{and} \quad \left(1\!-\!\frac{n_u(Z)}{\widetilde{n_0}(Z)} \right)\Gamma
(l,Z)\: \cos \phi_f(Z)  \:<\: 1.
 \label{condNoWB}
\ee
 If $\rho_u\le 1/2$, then  rhs(\ref{condNoWB}b)$\le 0$, and this condition is  automatically satisfied. 
For a given pump, as $n_b$ goes to zero  so do $n_u,\rho_u,\phi_f$, 
and by (\ref{est-phi}) $\rho_u\le 1/2$; hence
\begin{prop} 
For any fixed pump (\ref{pump}) there is no WBDLPI anywhere if 
\ \ $n_b\le\frac{\pi^2}{4K(l\!+\!\hDOu)^2}$.
\end{prop}
If $ \rho_u\le 1$, 
and the stronger, but simpler, additional assumption
\be
\left(\frac{n_u(Z)}{\widetilde{n_0}(Z)}\!-\! 1\! \right) %\hat s^{3/4} (l,Z)
\left(\frac{\widetilde{n_0}(Z)\, \big[\hsU(l,Z)\big]^3}{n_d(Z)}\right)^{1/4}<1,
  \label{condNoWB'}
\ee
is fulfilled, then no WBDLPI involves the $Z$ electrons
[in fact, (\ref{condNoWB'}) implies (\ref{condNoWB})  because $\mu^2\ge 1$, $\check n\ge n_d$, $\hsU\ge \hs$ \cite{FioDeAFedGueJov22}]. 
\ In  appendix  \ref{Proofs-0l} we also show that if  $\phi_f\in[0,\pi/2]$ then 
\bea
-\left(\!\frac{\widetilde{n_0}}{n_d}\!-\!1\!\right) \:\sin\phi
\:\lesssim\: \frac{\Gamma}{\omega_0}
\: \hat\sigma
\:\lesssim\: \left(\!1\!-\!\frac{\widetilde{n_0}}{n_u}\!\right)  \:\sin\phi
\label{barsigma-bounds0}
\eea
for all $\xi\in[0,l]$;
moreover, we prove the following Proposition, valid for all $\phi_f$:
\begin{prop} 
Under the assumptions of Proposition \ref{hatJsigma-approx},
for all $\xi\in[0,\tilde\xi_3']$, $\tilde\xi_3'\equiv\min\{l,\tilde\xi_3\}$ 
\bea
\frac{\widetilde{n_0}}{n_u} -
\left(\!\frac{\widetilde{n_0}}{n_d} \!-\!  \frac{\widetilde{n_0}}{n_u}\!\right)\! 
\MM_u-\tilde\delta\,\br\!_u \: \lesssim \: \hat J(\xi,Z)  \: \lesssim \:  
 \frac{\widetilde{n_0}}{n_d} + \left(\!\frac{\widetilde{n_0}}{n_d} \!-\!  \frac{\widetilde{n_0}}{n_u}\!\right)\! 
\MM_u+ \tilde\delta \, \br\!_u,\qquad 
\label{barepsilon-long'''hat} 
\eea
\bea
\frac{\Gamma(\xi,Z)}{\omega_0(Z)}\,
\left|\hat\sigma(\xi,Z)\right|\: \lesssim \:   
\left(\!\frac{\widetilde{n_0}}{n_d} \!-\!  \frac{\widetilde{n_0}}{n_u}\!\right)\! 
\MN_u+  \tilde\delta
\label{barsigma-long'''hat}
\eea
where \ \ $\br\!_u\equiv\left(\frac{\widetilde{n_0}\, s_u^3}{n_d}\right)^{1/4}\! \!,$  
\  $\omega_0= l\sqrt{\!K\widetilde{n_0}},$  \      $\mu^2_{{\scriptscriptstyle M}}\equiv1+\max v\simeq 1+a_0^2$, \ 
$\MM_u\equiv \left[\frac{\phi_f}{\pi}\right] 
\!\left(\!\frac{n_u\, s_u^6\, \mu^2_{{\scriptscriptstyle M}}}{n_d }\!\right)^{1/4}$,
\bea
\tilde\delta\equiv
 \max\!\left\{\!\left(\!\frac{\widetilde{n_0}}{n_d}\!-\! 1\!\right) \theta\!\left(\!\frac{\phi_f}{\pi}\!-\!1\!\right) \!,\! \left(\!1\!-\! \frac{\widetilde{n_0}}{n_u}\!\right)  \theta\!\left(\!\frac{\phi_f}{\pi}\!-\!\frac 12\!\right) \!\right\}\! , \quad
\MN_u\equiv  \left[\frac{\phi_f}{\pi}\!+\!\frac 12\right] \left(\!\frac{n_u s_u^3\mu^2_{{\scriptscriptstyle M}}}{\widetilde{n_0}}\!\right)^{1/4}\!\!; 
\label{defs}
\eea
here $\phi_f\equiv\phi(l,\!0;\!Z)$, $\theta(x)$ is Heaviside function, and $[a]$ stands for the integer part of $a\ge 0$. 
\label{Jsigmabounds}
\end{prop}
Here we have not displayed the
$Z$-dependences of $\tilde\xi_3,\tilde\xi_3',\widetilde{n_0},n_u,n_d, \tilde\delta, s_u,\MM_u,\MN_u,\br\!_u,\phi_f$
to shorten the formulae. Note that, by definition,  $\tilde\delta,\MN_u$ \ vanish if $\phi_f <\frac\pi 2$, $\MM_u$
 vanishes if $\phi_f <\pi$.
Hence,  $\hat J(\xi,Z)>0$ for all  $\xi\in[0,l]$, i.e.no WBDLPI involves the $Z$ electrons, if 
\bea
\left(\!\frac{n_u}{n_d} \!-\! 1\!\right) \MM_u
+ \tilde\delta\,\frac{n_u}{\widetilde{n_0}}\,\br_u    \: < \: 1
\label{nowavebreaking}
\eea
and (\ref{Lncond'}a) hold; if  $\widetilde{n_0}(z)$ grows with $z$ for $z\in[Z,Z\!+\!\hDOu]$, then
$\widetilde{n_0}=n_d$, and (\ref{nowavebreaking}) simplifies to
\bea
\left(\!\frac{n_u}{n_d} \!-\! 1\!\right) \left(\MM_u+ \br_u \right)   \: < \: 1.
\label{nowavebreaking1}
\eea
If for all $Z\ge 0$ one of these conditions is fulfilled, then WBDLPI takes place nowhere.

\medskip
In the NR regime  $a_0^2\simeq0$, $\hDOu/l\ll 1$, $s_u\simeq 1$ [see (\ref{sDelta1Def})], 
\bea
\phi_f\le \rho_u\pi\simeq l\sqrt{\!Kn_u},\quad
\br_u\simeq  \left(\!\frac{\widetilde{n_0}}{n_d}\!\right)^{1/4}\! \! ,\quad
\displaystyle\MM_u \lesssim \left[\rho_u\right]   \left(\!\frac{n_u}{n_d}\!\right)^{1/4},
  \label{useful}
\eea
and satisfying (\ref{nowavebreaking})  amounts to satisfying  one of the following conditions:
\bea
\ba{ll}
\displaystyle \mbox{a)}\quad & \rho_u\le \frac 1 2, \\
\displaystyle  \mbox{b)}\quad & \rho_u< 1 \quad \mbox{and} \quad   \displaystyle\left(\!\frac{n_u}{\widetilde{n_0}}\!-\!1\! \right)\left(\!\frac{\widetilde{n_0}}{n_d}\!\right)^{1/4}\: < \: 1,\\[10pt]
\mbox{c)}\quad & \displaystyle\left(\!\frac{n_u}{n_d}\!-\!1\! \right)  \left(\!\frac{n_u}{n_d}\!\right)^{1/4}\!\left[\rho_u \right]  + \! \left(\!\frac{\widetilde{n_0}}{n_d}\!\right)^{\!\!1/4}  \!
\max\!\left\{\frac{n_u}{n_d}\!-\!\frac{n_u}{\widetilde{n_0}} , \frac{n_u}{\widetilde{n_0}}\!-\! 1 \right\} \: < \: 1;
\ea
\label{NRnowavebreaking}
\eea
if $\widetilde{n_0}(z)$ grows with $z$ for $z\in[Z,Z\!+\!\hDOu]$, then
$\widetilde{n_0}=n_d$, $\br_u\simeq 1$,
and  (\ref{NRnowavebreaking}b,c) simplify to
\bea
\ba{ll}
\displaystyle \mbox{b)}\quad &  \rho_u< 1\quad \mbox{and} \quad  \displaystyle\left(\!\frac{n_u}{\widetilde{n_0}}\!-\!1\! \right) \: < \: 1,\\[10pt]
\mbox{c)}\quad &  \displaystyle\left(\!\frac{n_u}{\widetilde{n_0}}\!-\!1\! \right)   \left\{1+\left[\rho_u \right] \left(\!\frac{n_u}{\widetilde{n_0}}\!\right)^{1/4} \right\} \: < \: 1.
\ea
\label{NRnowavebreaking2}
\eea
The no-WBDLPI  condition (\ref{NRnowavebreaking}) is substantially weaker\footnote{
Inequality (\ref{NRnowavebreaking3}) implies (\ref{NRnowavebreaking}) unless $0.34\le\delta'\le 0.46$. \ In fact: 
\begin{enumerate}
\item If $\delta'>1$, i.e.  $\sqrt{M_u(Z)}l< 0.81$, then (\ref{NRnowavebreaking3}) implies (\ref{NRnowavebreaking}a), independently of the variation $\delta$. Note that (\ref{NRnowavebreaking}a) is the much weaker condition $\sqrt{M_u(Z)}l\le \pi/2\simeq 1.57$.

\item One easily shows that \ $\left(\!\frac{n_u}{\widetilde{n_0}}\!-\!1\! \right)\left(\!\frac{\widetilde{n_0}}{n_d}\!\right)^{1/4}\le g(\delta)\equiv\frac{\delta}{(1\!-\!\delta)^{5/4}}< g(\delta')$; \
$g(x)$ is strictly growing for $x\in[0,1[$ and equals 1 for $x=0.46$. Therefore
 if $1\ge\delta'>0.46$ then  (\ref{NRnowavebreaking3}) implies (\ref{NRnowavebreaking}b).

\item One easily shows that \  lhs(\ref{NRnowavebreaking}c)$\le g(\delta)(1\!+\!\sqrt{M_u}l/\pi)
\le g(\delta')(1\!+\!0.81/\pi\delta')\equiv \tilde g(\delta')$; \
$\tilde g(x)$ is strictly growing for $x\in[0,1[$ and equals 1 for $x=0.34$. \
Therefore
 if $\delta'<0.34$ then  (\ref{NRnowavebreaking3}) implies  (\ref{NRnowavebreaking}c).
\end{enumerate}
} than   (48) of \cite{FioDeAFedGueJov22}, which reads
\be
\delta< \delta' \equiv \frac{0.81}{\sqrt{M_u(Z)}\,l}, \qquad\qquad \mbox{where} \quad \delta\equiv 1\!-\!\frac{n_d}{n_u}\le 1, \:\: M_u\equiv Kn_u  
\label{NRnowavebreaking3}
\ee
and, on the contrary, is applicable also to discontinuous $\widetilde{n_0}$. 
If $\widetilde{n_0}(Z)$ varies `slowly', i.e. $\delta\ll 1$, then $\phi_f\simeq\omega_0$, 
$\tilde\delta \simeq \delta/2$, and (\ref{NRnowavebreaking}) 
essentially simplifies to \ $\left(\left[\frac{\omega_0}\pi\right] +\frac 12\theta\left(\!\frac{\omega_0}\pi\!-\!\frac12\!\right)\right)  \delta \: < \: 1$; \
this is automatically satisfied if $\omega_0<\pi$.
As noted in \cite{FioDeAFedGueJov22}, the rate of growth of  $\widetilde{n_0}(Z)$ at
$Z\!\simeq\!0^+$  is more critical than at larger $Z$ for the timing of the first WB: if $\widetilde{n_0}(Z)= O(Z)$ (discontinuous $\widetilde{n_0}'(Z)$ at $Z=0$) WB occurs for very small $z$'s earlier  than if $\widetilde{n_0}(Z)=O(Z^2)$ at least (continuous $\widetilde{n_0}'(Z)$ at $Z=0$).
Since in  LWFA  the density of the plasma obtained from a supersonic gas jet hit by the laser pulse usually fulfills $\widetilde{n_0}(z)=o(z^2)$, our results imply that under rather broad conditions there is no WBDLPI for very small $z$'s.

If there is some $\Upsilon>0$ such that $|d\log\widetilde{n_0}/dZ|\le\Upsilon$  for all $Z$, then we can replace 
$\tilde\delta$ by $\Upsilon\max\{\Delta_u,\!-\!\Delta_d\}$ and
$\delta$ by $\Upsilon(\Delta_u\!-\!\Delta_d)$. In the moderate relativistic regime, the above no-WBDLPI are only slightly stronger than  (\ref{NRnowavebreaking}), (\ref{NRnowavebreaking3})
because $1/\hat s\le 1$, and the fourth roots in the definitions of Proposition \ref{Jsigmabounds}
map positive numbers to other ones much closer to 1.

\subsection{Behaviour of $\hat J,\hat\sigma$ for $\xi>l$}
\label{Jsigma-after}

\begin{figure}%[hbtp]
\includegraphics[width=16.5cm]{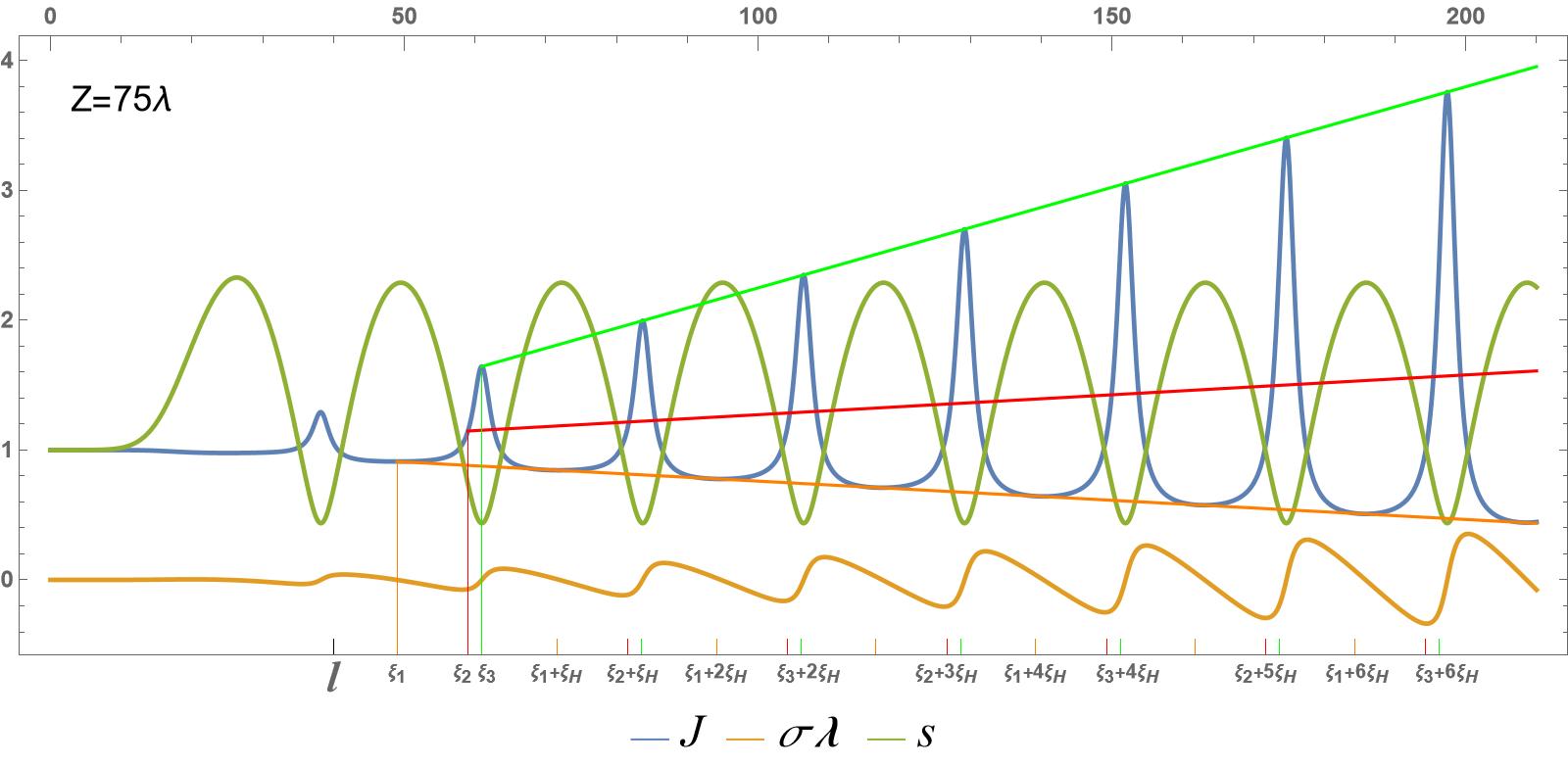}
\caption{Evolution of $\hat J,\hat\sigma$ vs. that of  $\hat s$ for $Z=75\lambda$ 
under the conditions  considered 
in Ref. \cite{BraEtAl08} and described in the caption of fig.   \ref{Worldlinescrossings}; at $Z=75\lambda$ \ $\widetilde{n_0}'>0$, $\Phi<0$, $\vartheta<0$.    
}
\label{graphs2}
\end{figure}

Since  the matrix $A(\xi,Z)$ defined in (\ref{basic''-2}) is periodic  for $\xi>l$,
we could determine the qualitative behaviour of $\G$ and hence of $\hat J,\hat\sigma$
applying Floquet theory (see e.g. \cite{Chicone99}) to the equation $\G'=A\G$.
Actually, we will find such  behaviours more directly as follows.

Differentiating both sides of the periodicity identity $\hze\left[\xi\!+\!n\xiH(Z),Z\right]=\hze(\xi,Z)$ 
(valid for all $n\in\NN$ and $Z$) w.r.t. $Z$ and using the one 
$\hze'\!\left[\xi\!+\!n\xiH(Z),Z\right]=\hze'\!\left[\xi,Z\right]$ 
we find
\bea
\frac{\partial \hze}{\partial Z}\!\left[\xi\!+\!n\xiH(Z),Z\right]=\frac{\partial \hze}{\partial Z}(\xi,Z)-\hat\Delta'\!\left[\xi,Z\right]n\frac{\partial\xiH}{\partial Z}. 
\nonumber
\eea
A similar relation can be obtained from %the periodicity identity 
$\hat s\left[\xi\!+\!n\xiH(Z),Z\right]=\hat   s(\xi,Z)$.
Thus, we obtain
\bea
\ba{l}
\displaystyle  \hat J \!\left[\xi\!+\!n\xiH(Z),Z\right]= \hat J (\xi,Z)- n\,\hat\Delta' (\xi,Z)\,\Phi(Z), \\[8pt]
\displaystyle  \hat\sigma\!\left[\xi\!+\!n\xiH(Z),Z\right]=\hat\sigma(\xi,Z)- n\, \hat s' (\xi,Z)\, \Phi(Z).
\ea\qquad \Phi \equiv\frac{\partial\xiH}{\partial Z}=c\frac{\partial\tH}{\partial Z}\label{pseudoper}
\eea
Consequently\footnote{In fact, $a\equiv \hat J(\xi,Z)-\xi b(\xi,Z)$ is $\xi$-periodic iff (\ref{pseudoper}) is fulfilled. Similarly for $\hat\sigma$.}, both $\hat J, \hat\sigma$ are LQP, i.e. fulfill
(\ref{lin-pseudoper}), in $\xi$ with period $\xiH$ and $b$ respectively given by
 $b(\xi,Z)=-\hat\Delta' (\xi,Z)\,\frac{\partial \log\xiH}{\partial Z}$,  $b(\xi,Z)=-\hat s' (\xi,Z)\,\frac{\partial \log\xiH}{\partial Z}$. 

Similarly, taking the (laboratory) time $t$ as the independent
parameter in place of $\xi$, differentiating the relation $z_e\left[t\!+\!n\tH(Z),Z\right]=z_e(t,Z)$ w.r.t. $Z$ 
we find 
\bea
\displaystyle  \frac{\partial z_e}{\partial Z} \!\left[t\!+\!n\tH(Z),Z\right]=\frac{\partial z_e}{\partial Z} (t,Z)- n\,\dot\Delta (t,Z)\,\frac{\partial\tH(Z)}{\partial Z}.
 \label{pseudoper'}
\eea
This implies that $J(t,Z)\equiv \frac{\partial z_e}{\partial Z}$ is LQP in $t$ with period $\tH$, in
agreement with formulae (4-6) of \cite{BraEtAl08}, which were obtained by a Fourier series expansion of $z_e(\cdot,Z)$.

\begin{figure}
\includegraphics[width=16.5cm]{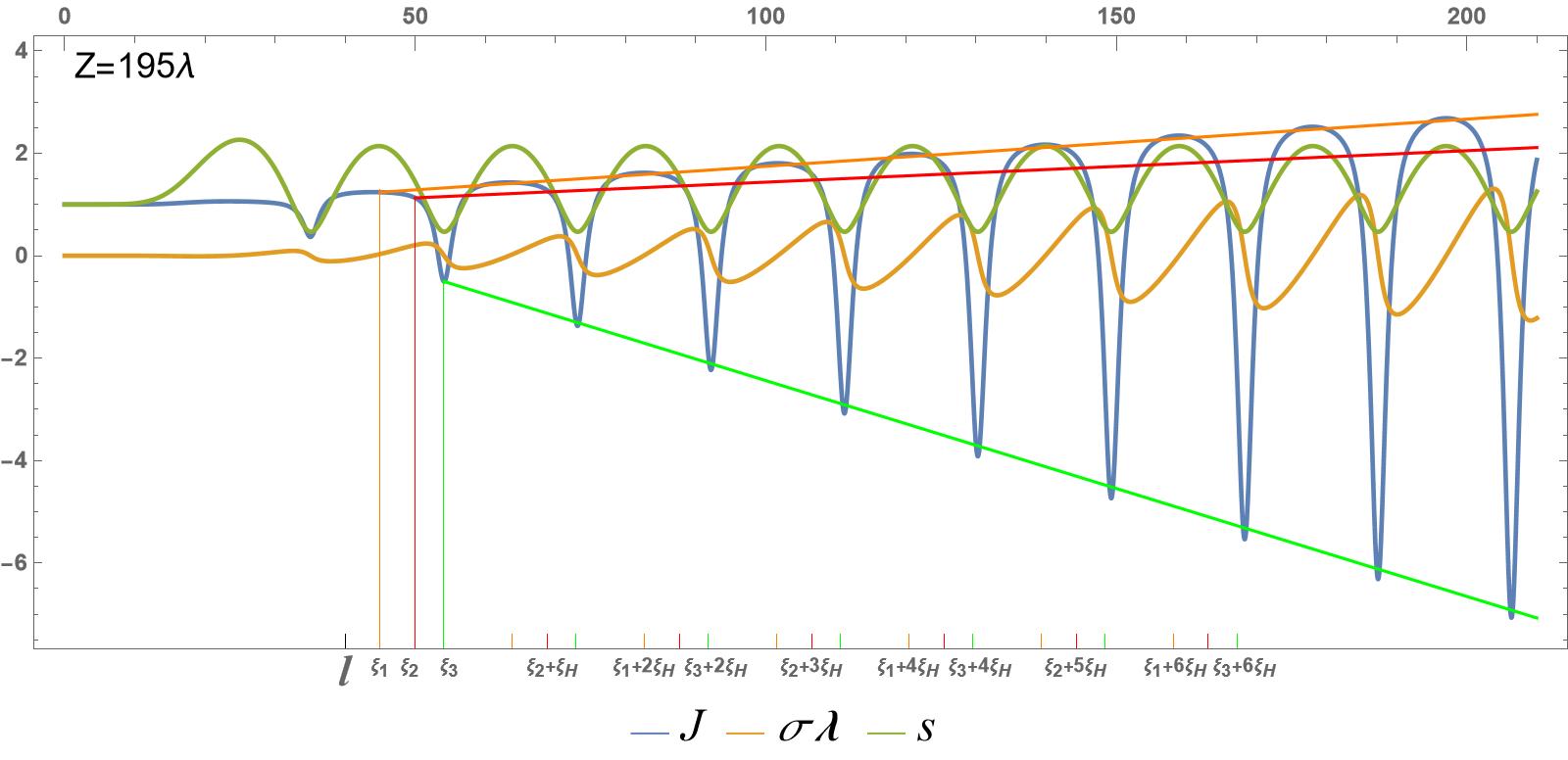} 
\caption{Evolution of $\hat J,\hat\sigma$ vs. that of  $\hat s$  for $Z=195\lambda$, under the conditions  considered 
in Ref. \cite{BraEtAl08} and described in the caption of fig.  \ref{Worldlinescrossings}; there 
\ $\widetilde{n_0}'(Z)<0$, $\Phi(Z)>0$, 
$\vartheta(Z)>0$.  }
\label{graphs2'}
\end{figure}

As illustrations, in fig. \ref{graphs2}, \ref{graphs2'} we have plotted the
graphs  of $\hat s,\hat\Delta,\hat J,\hat\sigma$  under the conditions of Ref. \cite{BraEtAl08} described in  fig.  \ref{graphsb}, at two values of $Z$ where $\widetilde{n_0}(Z)$ resp. grows, decreases.

Relations (\ref{pseudoper}) can be used to extend the knowledge of  $ \hat J ,\hat\sigma$ from the interval $\I\equiv[l, l\!+\!\xiH[$ to all $\xi\ge l$ [and similarly for (\ref{pseudoper'})]; as a first approximation of
$\hat J ,\hat\sigma$ in $\I$ one can adopt (\ref{hatJsigmaapprox}).
Fixed any $\xi\in\I$, (\ref{pseudoper}) defines a pair of {\it linear} discrete 
maps $n\in\NN\mapsto ( J_n, \hat\sigma_n)$, where
$ J_n=(\ref{pseudoper}a)$, $\hat\sigma_n=(\ref{pseudoper}b)$; 
all the $ J_n$ lie on a straight line with slope $- \hat\Delta' (\xi,Z)\,\Phi(Z)$,
while all the $\hat\sigma_n$ lie on a straight line with slope $- \hat s' (\xi,Z)\,\Phi(Z)$, see fig. \ref{graphs2}, \ref{graphs2'}.
If $\Phi(Z)=0$ 
both lines are horizontal, and $ \hat J (\xi,Z),\hat\sigma(\xi,Z)$ are $\xiH$-periodic. If $\Phi(Z)\neq 0$
the signs of the slopes are determined by that of $\Phi$ and by the quadrant
where $P(\xi,Z)$ is. Typically, \ sign$(\Phi)=-$sign$(d\widetilde{n_0}/dZ)$,  \ by the weak dependence of $h$ on $Z$, see (\ref{Phivartheta}).
 Since $\hat\Delta',\hat s'$ oscillate about zero, so do the slopes. Hence for sufficiently large $n$ also $ \hat J (\xi,Z),\hat\sigma(\xi,Z)$ oscillate
between positive and negative values, with amplitude linearly growing with $n$; WB occurs as soon as $\hat J\le 0$.  More precisely, if in $\I$ it is $\hat J>0$ (no WB), then:

\begin{enumerate}
\item If $\widetilde{n_0}(Z)=n_0=$ const,  then $\Phi(Z)=0$  for all $Z$, and WB cannot occur anywhere. 
 
\item If $\Phi(Z)<0$, then there is $n\in\NN$ such that  $ \hat J (\xi,Z)\le 0$
at some $\underline{\xi}\in]\xi_n^2,\xi_{n+1}^0[$  (see e.g. fig. \ref{graphs2}a,b),  i.e. in the upper part of the orbit, because  there $\hat\Delta'(\xi,Z)<0$.

\item  If $\Phi(Z)>0$, then there is $n\in\NN$   such that  $\hat J (\xi,Z)\le 0$
at some $\underline{\xi}\in]\xi_n^0,\xi_n^2[$  (see e.g. fig. \ref{graphs2'}a,b), i.e. in the lower part of the orbit, because there 
$\hat\Delta'(\xi,Z)>0$.

\end{enumerate}
We can  determine %or estimate 
the number  $n_{br}$ of oscillations leading to the first WB by  (\ref{pseudoper}); if 
$|\hat J\!-\!1|\ll 1$ in $\I$,  approximating the first term at rhs(\ref{pseudoper}a) by 1, noting that the second term is minimum when -sign$(\Phi)\hat\Delta'$ is (i.e. either for 
$\xi\simeq\xi_{n_{br}}^1$, or for $\xi\simeq\xi_{n_{br}}^3$), using (\ref{heq1}a),  (\ref{smM})  we find 
\bea
n_{br}\simeq
 \min_Z\, \left[\left(\frac {h(Z)}{\sqrt{h^2(Z)\!-\!\mu^2}}-\mbox{sign}\big(\Phi(Z)\big)\right)\frac1{|\Phi(Z)|}\right];
\label{nwb}
\eea
here $[a]$ stands for the integer part of $a\in\RR^+$, and $\mu^2=1\!+\!v(l)\!=$const. As expected, this diverges as $h\to\mu$, goes to zero as $h\to\infty$; in other words, the nonrelativistic, ultrarelativistic regimes lead to late, early WB respectively.

In spite of their oscillations, $\hat J(\xi,Z),\hat\sigma(\xi,Z)$ are constrained at all $\xi$ by the relation 
\bea
\ba{l}
\displaystyle\hat\sigma f- \hat J\,\U_1+ \hat\Delta K\widetilde{n_0}+\vartheta =0,  \qquad \qquad 
\mbox{where }\\[10pt]
f(s)\equiv\frac{\mu^2}{2\hat  s^2}-\frac 12,\qquad \U_1\equiv\frac{\partial \U}{\partial \hat\Delta},\qquad \vartheta(Z)\equiv\frac{dh(Z)}{dZ}, 
\ea     \label{identity}
\eea
which can be obtained
differentiating the identity \ $\frac {\hat s^2(\xi,Z)\!+\!\mu^2}{2\hat s(\xi,Z)}+\U[\hat\Delta(\xi,Z);Z]\!=\!h(Z)$ \ w.r.t. $Z$. Applying (\ref{identity}) where $\hat s'$ or $\hat\Delta'$ vanish we find in particular, for all $n$, 
\bea
\ba{ll}
\displaystyle \hat J(\xi_n^0,Z)=\frac {\vartheta(Z) +K\widetilde{n_0}(Z) \Dm(Z)}{\Um(Z)},
\qquad &
\displaystyle \hat J(\xi_n^2,Z)=\frac {\vartheta(Z) +K\widetilde{n_0}(Z) \DM(Z)}{\UM(Z)},\\[10pt]
\displaystyle\hat\sigma(\xi_n^1,Z)=\vartheta(Z)\!\left[1-\frac{h(Z)}{\sqrt{h^2(Z)\!-\!\mu^2}}\right],\qquad
&\displaystyle\hat\sigma(\xi_n^3,Z)=\vartheta(Z)\!\left[1+\frac{h(Z)}{\sqrt{h^2(Z)\!-\!\mu^2}}\right],
\ea                           \label{specialvalues}
\eea
where we have abbreviated $\UM=\left.\frac{\partial U}{\partial \hat\Delta}\right|_{\hat\Delta=\DM}>0$, $\Um=\left.\frac{\partial U}{\partial \hat\Delta}\right|_{\hat\Delta=\Dm}<0$. 
Hence: $\hat\sigma(\xi_n^3,Z), \hat\sigma(\xi_n^1,Z)$ respectively have the same, opposite sign as $\vartheta(Z)$; moreover, if $\vartheta\ge 0$ (resp. $\vartheta\le 0$) then automatically $\hat J(\xi_n^2,Z)$ (resp.  $\hat J(\xi_n^0,Z)$) is positive.
Typically, $\vartheta(Z)$ is negligible w.r.t. $K\widetilde{n_0}\Dm,K\widetilde{n_0}\DM$.
Note that the values (\ref{specialvalues}) of $ \hat J,\hat\sigma$ are {\it independent} of $n$, while
\bea
\ba{ll}
\displaystyle\varepsilon(\xi_n^1,Z)
=\Phi_n^1\left[1-\frac{h}{\mu^2}\big(h+\sqrt{h^2\!-\!\mu^2}\big)\right],
\quad & \displaystyle\varepsilon(\xi_n^3,Z)
= \Phi_n^3\left[1-\frac{h}{\mu^2}\big(h-\sqrt{h^2\!-\!\mu^2}\big)\right],\\[14pt]
\displaystyle\hat\sigma(\xi_n^0,Z)
=-\Phi_n^0\,\Um,\qquad\quad  & \hat\sigma(\xi_n^2,Z)= -\Phi_n^2\,\UM  
\ea                      \label{specialvalues2}
\eea
{\it depend} on $n$\footnote{$\hat\Delta'(\xi_n^1,Z)=-\frac{\partial \hat H}{\partial s}(P_1; Z)$ 
and the Taylor formula around $\xi=\xi_n^1$ imply \ $\hat\Delta(\xi,Z)=-(\xi\!-\!\xi_n^1)\frac{\partial H}{\partial s}(P_1; Z)+O\big[(\xi\!-\!\xi_n^1)^2\big]$, \ whence, differentiating w.r.t. $Z$ and setting $\xi=\xi_n^1$, \
$\varepsilon(\xi_n^1,Z)=\frac{\partial \xi_n^1}{\partial Z}\frac{\partial H}{\partial s}(P_1; Z)=\Phi_n^1\left[\frac 12-\frac{\mu^2}{2\smm^2}\right]=\Phi_n^1\left[\frac 12-\frac{\sM^2}{2\mu^2}\right]=\Phi_n^1\left[1-\frac{h}{\mu^2}(h+\sqrt{h^2\!-\!\mu^2})\right]$; \ similarly one proves the results for $\varepsilon(\xi_n^3,Z)$,
$\hat\sigma(\xi_n^0,Z),\hat\sigma(\xi_n^2,Z)$.}; here  $\Phi_n^i(Z)\equiv \frac{d  \xi_n^i}{d Z}$.  The signs of the factors of the $\Phi_n^i$  in the first, second column are resp. negative, positive.
At least for  large $n$, the $\Phi_n^i\simeq n\,\Phi$ have the same sign as $\Phi$, implying that $\varepsilon(\xi_n^1,Z),\varepsilon(\xi_n^3,Z)$
have opposite sign, and $\hat\sigma(\xi_n^0,Z), \hat\sigma(\xi_n^2,Z)$ as well.

\medskip
If the relative variation of $\widetilde{n_0}$ in the interval $[Z\!+\!\Dm,Z\!+\!\DM]$ are small, we can easily find very good approximations
of $h(Z),\xiH(Z),\vartheta(Z),\Phi(Z),...$: We first compute
the functions  $\bar h(n_0),\bxiH(n_0)$ yielding the dependences of $h,\xiH$ on $n_0$ in the case of a constant initial density $\widetilde{n_0}(Z)=n_0$ (section \ref{auxi}).  Then we approximate 
$h(Z)\simeq \bar h\big[\widetilde{n_0}(Z)\big]$, $\xiH(Z)\simeq \bxiH\big[\widetilde{n_0}(Z)\big]$, and
\be
\Phi(Z)\equiv \frac{\partial\xiH}{\partial Z} \simeq\frac{d \widetilde{n_0}}{d Z}  \left.\frac{\partial\bxiH}{\partial n_0}\right\vert_{n_0=\widetilde{n_0}(Z)}, \qquad
\vartheta(Z)\equiv \frac{\partial h}{\partial Z} \simeq\frac{d \widetilde{n_0}}{dZ}  \left.\frac{\partial\bar h}{\partial n_0}\right\vert_{n_0=\widetilde{n_0}(Z)}    \label{Phivartheta}
\ee
The dependence $\bxiH(n_0)$ is obtained from formula (\ref{period0}); since the dependence   $\bar h(n_0)$
is much slower than that of  $1/\sqrt {n_0}$, one finds \ $\frac{\partial\bxiH}{\partial n_0}\simeq -\bxiH/2n_0$, \ which replaced in (\ref{Phivartheta}) gives 
\bea
\Phi & \simeq & -  \frac{\bxiH}{2}\,\frac{d \log\widetilde{n_0}}{d Z}, \label{Phi-estimate}
\eea
which in turn allows to estimate $n_{br}$ when replaced in (\ref{nwb}).

\section{Wave-breaking and WFA of self-injected electrons}
\label{self-injected-electrons}

Assume that some $Z, Z'$-electron layers cross each other at  $\xi=\xi_c$, i.e.
$\hat z(\xi_c,Z)=\hat z(\xi_c,Z')\equiv z_c$. A $Z$-electron will scatter off
a $Z'$-electron  if at the scattering time also their  transverse coordinates are very close to each other.
Using kinetic theory
one could predict the distribution of scattered electrons in phase space. However, due to the low density of our plasma, we  expect that significant scatterings will be rare, and each 
$Z$-electron will feel the averaged force generated by the whole layer of $Z'$-electrons. Therefore, we continue to treat the plasma as collisionless. Consequently, the problem remains 1D, and the transverse %(or {\it quiver}) 
velocity $\bup$ of all electrons remains the same as before the collisions. For simplicity, we  assume that $\bup$ is negligible; as noted after (\ref{alphaapprox}),  this is well justified if the laser pulse is slowly modulated.
If only `few' electron layers cross each other (more precisely, 
if equalities of the form $\hat z_e(\xi_c,Z)=\hat z_e(\xi_c,Z')$ take place only for $Z,Z'$ in a short interval $[Z_l,Z_r]$ and $\xi_c$ in a short interval $[\xi_{br},\xi_{br}']$),  then  these  electrons 
will not modify the plasma wave substantially (as they cause a small WB)
and can be treated as test particles.
Hence to determine the motion of one of these test electrons in the PW
we can  apply the approach of section \ref{test-electron}.
In particular, assuming that $\widetilde{n_0}(z)$ is of the type depicted in fig. \ref{fig1},
 if for some $Z<Z_p$ these test $Z$-electrons  reach\footnote{$z_q$ is the minimal longitudinal coordinate which  in the hydrodynamic regime can be reached only by $Z$ electrons with $Z\ge z_s$, i.e. of the plateau in the initial density.} a chosen  $z_{i0}\ge z_q\equiv z_s\!+\!\DM(z_s)$
for the first time at $\xi=\xi_0(Z)$, then  their motion for $\xi>\xi_0 $  will be given by (\ref{test-motion}), as long as $\hat z_i(\xi)\ge z_q$. They will: be boosted backwards
(to $z<Z$) if $\delta s\equiv s_{i0}- \hat s\big(\xi_0,\hat Z_e(\xi_0,z_{i0})\big)>0$; oscillate around $z_{i0}$ if $\delta s=0$; undergo a WFA if $\delta s<0$, i.e. if they are injected with a positive $p^z$
in the accelerating part of the PW. In particular, the latter will be trapped in a trough of the PW if $s_i^m\le 0$. The initial value $s_{i0}$ is the final value of $\hat s_i(\xi)$ in the short interval $[\xi_c,\xi_0]$; to find it one needs to
 determine the motion of the $Z$-electrons there. How to do this will be discussed elsewhere \cite{FioAAC22,Fionew}.

\begin{figure}%[ht]
\begin{center}
\includegraphics[width=17cm]{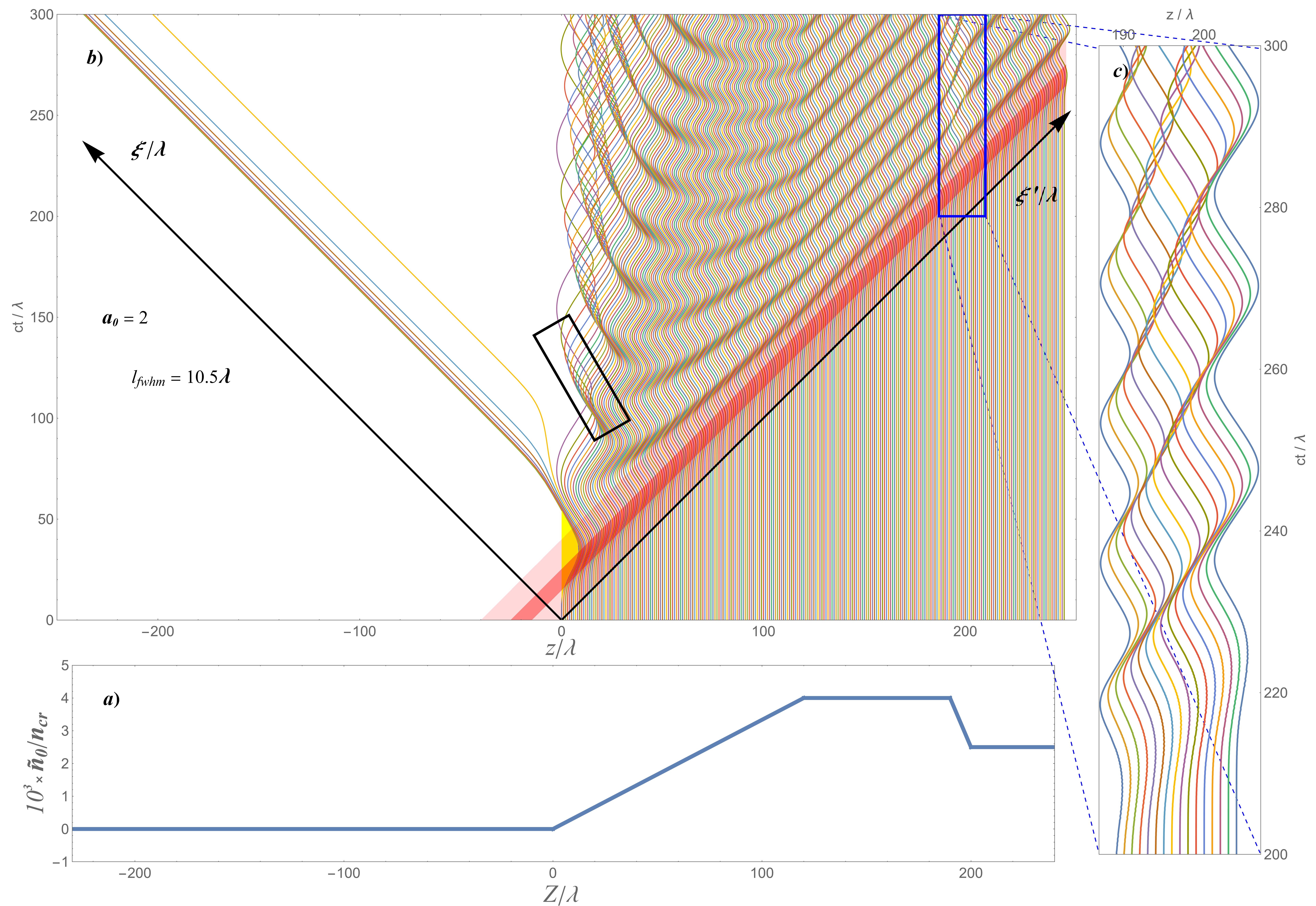}
\end{center}
\caption{a) Initial plasma density $\widetilde{n_0}(Z)$ with a linear up-ramp, a first plateau, a steep linear down-ramp and a final lower plateau,  as in \cite{BraEtAl08}, section III.B. \ \ b) Projections onto the $z,ct$ plane of the electron worldlines (WLs) in Minkowski space induced by the
impact of the pulse of fig. \ref{graphsb}a on  $\widetilde{n_0}(Z)$: WLs of $Z\!\sim\!0$ electrons stray left away ({\it slingshot effect}),
WLs of other up-ramp electrons first intersect after about $5/4$ oscillations (black box), WLs of down-ramp electrons first intersect after about  $7/4$ oscillations 
(blue box). \ \ c)   \ Zoom of the blue box of b). 
Here: $\xi'\equiv ct\!+\!z$; the dark yellow region is the layer $L_t$ (cf. section \ref{during}) containing ions and no electrons; we have resp. painted pink, red the support of $\Bep(ct\!-\!z)$  (considering $\Bep(\xi)\!=\!0$ outside $0\!<\!\xi\!<\!40\lambda$) and the region where the modulating intensity $\epsilon^2$ is above the FWHM, i.e. $-l'/2\!<\!\xi\!-\!20\lambda\!<\!l'/2$, with $l'\equiv l_{fwhm}=10.5\lambda$. 
The pulse of fig. \ref{graphsb} can be considered essentially short [cf. (\ref{Lncond'}a)] and fulfilling the no-WBDLPI conditions of section \ref{Bounds0l} only if we consider the pulse length as some $l''\le 28\lambda$, instead of the $l=40\lambda$ (conventionally) adopted here; nevertheless, it induces no WBDLPI.
}
\label{Worldlinescrossings}       % Give a unique label
\end{figure}

\begin{figure}
\includegraphics[width=16cm]{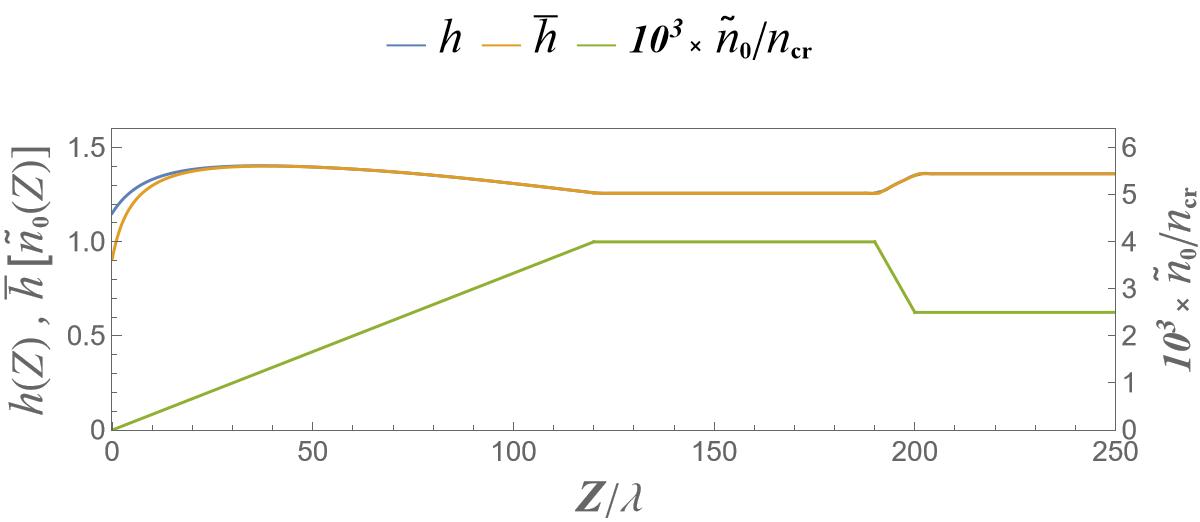}
\caption{Conditions as in fig. \ref{Worldlinescrossings}.
Blue is plot the corresponding normalized final energy $h$ of the $Z$-electrons  after interacting with the  pulse described in fig. \ref{graphsb}; we plot green  the
normalized initial electron density $\widetilde{n_0}$ for comparison.
Except for small values of  $\widetilde{n_0}(Z)$, $h(Z)$ is practically indistinguishable (cf. the end of section \ref{Jsigma-after}) from the energy $\bar h(n_0)$  (orange plot)  attained with a constant density $n_0=\widetilde{n_0}(Z)$; $\bar h(n_0)$ has been introduced in section \ref{auxi}.
}
\label{h_vs_Z}
\end{figure}

\begin{figure}[b]
\includegraphics[width=16.8cm]{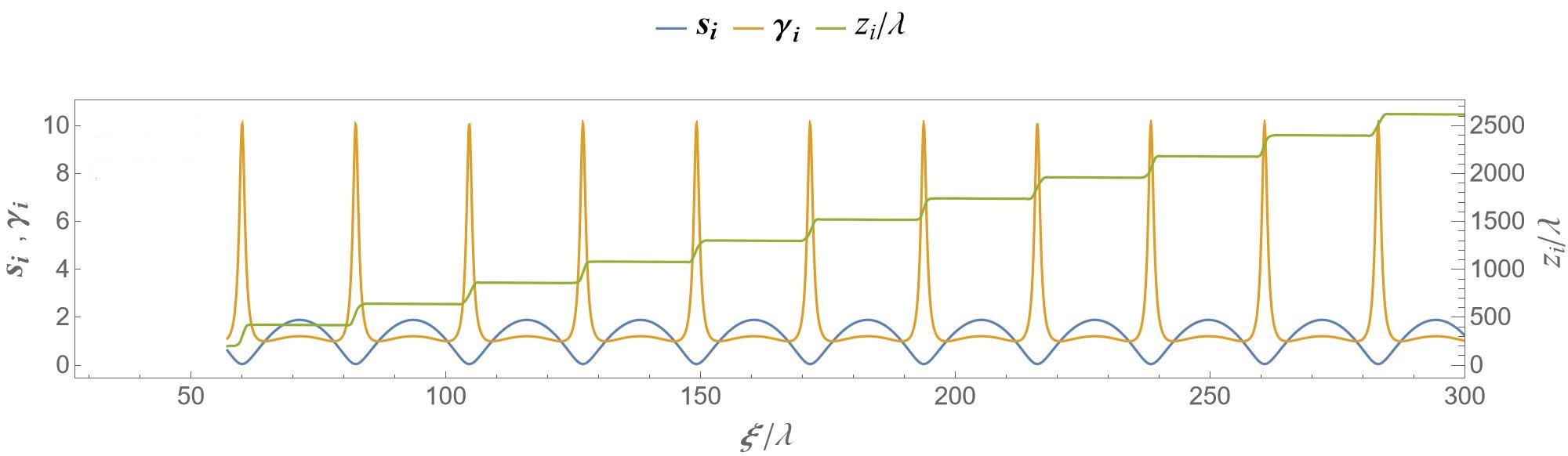} 
\caption{Conditions as in fig. \ref{Worldlinescrossings}. Plots of $s_i,\gamma_i,z_i$ vs. $\xi$ for sample electrons having $s_i^m>0$  
(see section \ref{test-electron}) self-injected just after the  $1^{st}$ PW trough trailing the pulse. 
}
\label{selfinjected-untrapped}
\end{figure}
To illustrate the validity and the effectivity of our approach, we have considered the LWFA
induced by the pulse of fig.  \ref{graphsb}.a onto the density $\widetilde{n_0}$ of fig. \ref{Worldlinescrossings}.a; the latter decreases linearly over a distance $10\lambda$ and was the best one for LWFA out of the three  considered in the section III.B of Ref.  \cite{BraEtAl08}.
Solving numerically the corresponding Cauchy problems (\ref{heq1})-(\ref{heq2}) we have determined and plotted (fig. \ref{Worldlinescrossings}.b)
the WLs of the $Z$-electrons, for $Z/\lambda=1,2,...,250$; the black, blue boxes encircle the first collisions among electron layers in the up-ramp and down-ramp.  In  fig. \ref{Worldlinescrossings}.c we have zoomed the blue box.  In fig. \ref{h_vs_Z} we plot (blue graph) the corresponding normalized energy $h(Z)$. 
We denote by $\xi_{br}^1,\xi_{br}^2,...$ the starting `instants' for the 1st, 2nd,... WB, i.e. the (locally) minimal $\xi$
for which there is a $Z$ (which  we shall denote  by $Z_{br}^1,Z_{br}^2,...$) such that $\hat J(\xi,Z)=0$ and $\hat J'(\xi,Z)<0$, so that $\hat J(\xi_{br}^a,Z_{br}^a)=0$ for $a=1,2,...$.
For small $\xi>\xi_{br}^a$ there is
an interval  $\I_\xi\equiv [Z_l^a(\xi),Z_r^a(\xi)]\supset Z_{br}^a$
such that if $Z\in\I_\xi$ then $\hat z_e(\xi,Z)=\hat z_e(\xi,Z')$ for at least another $Z'\in \I_\xi$, i.e. any such $Z$ electron layer collides with at least another one. (By continuity,
$Z_l^a(\xi_{br}^a)=Z_{br}^a=Z_r^a(\xi_{br}^a)$.)
From fig. \ref{fig1} we can read off that $(\xi_{br}^1,Z_{br}^1)\simeq (51.7\lambda,191.8\lambda)$ and $(\xi_{br}^2,Z_{br}^2)\simeq (68.7\lambda,190.6\lambda)$ respectively.
As a test of the approximation scheme
introduced at the end of section \ref{Jsigma-after}, we also plot  (orange) the normalized energy  induced by the same pulse onto a constant density $n_0=\widetilde{n_0}(Z)$; the approximation is excellent if $Z\ge 20\lambda$, not only for  $h(Z),\xiH(Z),\vartheta(Z),\Phi(Z),...$,
but also for the solutions $\big(\hat z_e(\xi,Z),\hat s(\xi,Z)\big)$   of (\ref{heq1}-\ref{heq2}) themselves.

\begin{figure}[t]
\includegraphics[width=16cm]{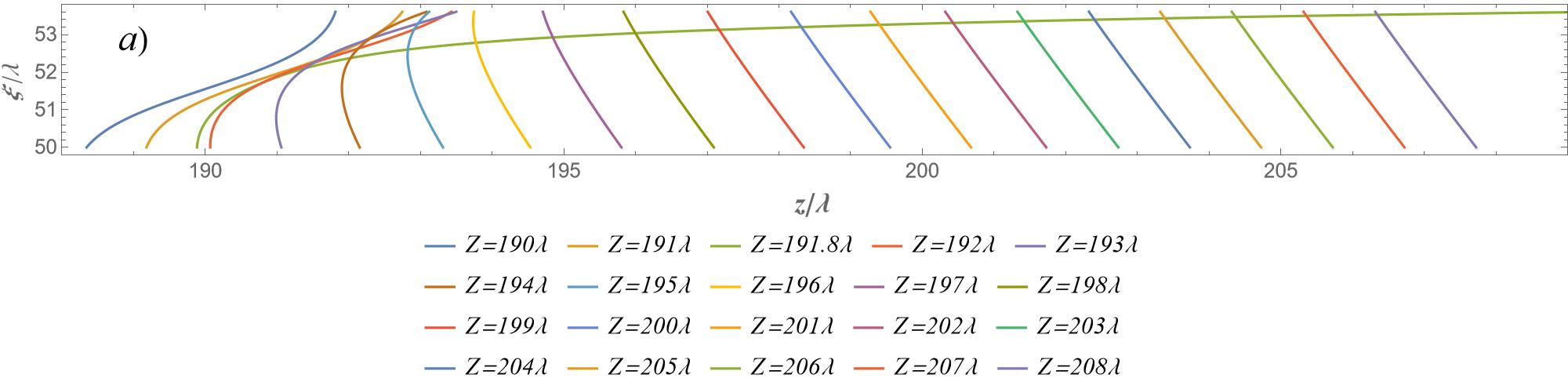}\\
\includegraphics[width=1.9cm]{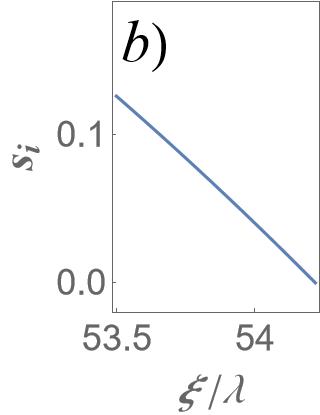}\hfill\includegraphics[width=14.4cm
]{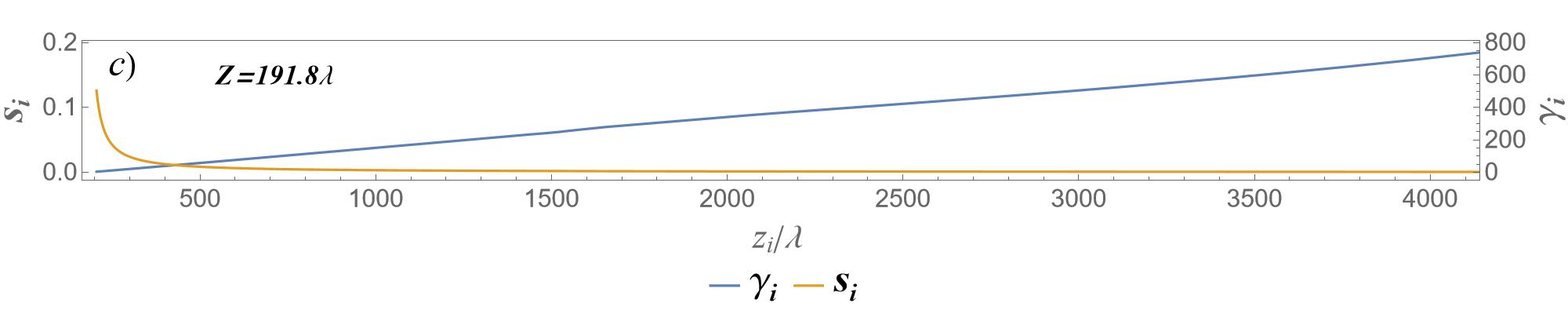}  
\caption{Conditions as in fig. \ref{Worldlinescrossings}. a) The WL (in $z,\xi$ space) of the $Z_{br}^1\!=\!191.8\lambda$ electrons (self-injected by the  $1^{st}$ WB and trapped in the next PW trough trailing the pulse) while crossing the WLs of other $Z$ electrons of the density down-ramp. \ b) Plot of  $s_{i}(\xi)$, \ c) plot of $\gamma_i(z_i)$ and path in the $z\!-\!s$ plane  for the $Z_{br}^1$  electrons along the density plateau region. 
}
\label{selfinjected-trapped}
\end{figure}

Then we have studied the latter for the  $Z$-electrons involved in the first or second WB more in detail; we have chosen  $Z$ belonging to equidistant meshes of step $ \delta Z=0.1\lambda$. 
The $Z$ electrons emerging from either WB range over all types:
boosted backward ($\delta s>0$); oscillating ($\delta s=0$); boosted forward ($\delta s<0$). In  fig. \ref{selfinjected-untrapped}, \ref{selfinjected-trapped} 
we have  plotted the paths in phase space of two examples of the latter: in the first $s_i^m>0$, the longitudinal momentum of
the injected electron has a positive average, leading to a mean 
drift forward, with periodic oscillations synchronized with the sliding of the electron from each PW trough to its left neighbour; 
in the second $s_i^m <0$, the injected electron is trapped in a trough and its  longitudinal momentum grows indefinitely with $\xi$ (or $\hat z_i$).
For the latter  we can use the approximation (\ref{s_i^m<0|s_iz_i}b), with 
 $\left|s' (\xi_f)\right|=M\left|\Delta(\xi_f) \right|$, if  $\hat z_i$ is large. 
The maximal possible $\gammaM_i$ for fixed $\hat z_i,M$ and variable $s_{i0}$ is obtained if $\Delta(\xi_f)=\Dm$, i.e. if $s(\xi_f)=\mu$, and reads
\bea
\gammaM_i(z_i)\:\simeq \: M\left|\Dm \right|\, z_i
\:= \: \sqrt{2M\,\big(\bar h-\mu\big)}\:z_i.         \label{gamma^M(z_i)}
\eea
The optimal plateau value $n_0$ of $\widetilde{n_0}$ maximizes the argument of the square root in (\ref{gamma^M(z_i)}), i.e. fulfills
\be
\frac d {dn_0}\left\{n_0 \left[\bar  h(n_0)-\mu\right]\right\}=0.
\ee
Eq. (\ref{gamma^M(z_i)}) is reliable as long as depletion of the pump can be neglected.

In the present case for $Z\ge z_q$ it is $ \Dm\simeq- 2.7\,\lambda$, $\bar h\simeq 1.36$, $\mu\simeq 1$, while $M\lambda^2=\pi^2/100$, whence\ $\gammaM_i(z_i) = M |\Dm|  z_i \simeq  0.27\,z_i/\lambda$. \ If there are trapped $Z$-electrons for which 
$\Delta(\xi_f,Z)=\Dm$, after $t=2950\lambda/c$ seconds these electrons have travelled a longitudinal distance $\Delta z^i\simeq(2950\!-\!205\!-\!55)\lambda$ beyond $z_q\simeq 205\lambda$, and their energy is approximately $ \gammaM_i mc^2\simeq 366$MeV, which is (only) a little larger than the
result $\gammaM_i mc^2\simeq 320$MeV
obtained in \cite{BraEtAl08} by means of 2D  PIC simulations where the  pulse is the present plane one of fig. \ref{graphsb}.a, multiplied by
a Gaussian $\chi\!_{{\scriptscriptstyle R}}(\rho)$ ($\rho\equiv\sqrt{x^2\!+\!y^2}$) in the transverse directions with  FWHM $R=20\lambda$.
Actually, using the equations of Ref. \cite{FioAAC22,Fionew} one can determine the motion of the $Z^1_{br}=191.8\lambda$ electrons, whose layer  earliest crosses and overshootes other ones and hence are the fastest electrons emerging from the first WB; their motion across the  $\widetilde{n_0}\neq$ const transient region is plotted in fig. \ref{selfinjected-trapped} up,  across the $\widetilde{n_0}=n_0$ plateau one in fig. \ref{selfinjected-trapped} down.
One finds $\gamma_i(z_i)\simeq   0.19\,z_i/\lambda$ and after $t=2950\lambda/c$ seconds
an energy $ \gamma_i mc^2\simeq 248$MeV, which is 
 a little smaller than  $320$MeV. Electrons that will reach via WFA an energy of about $320$MeV are expected to  be self-injected by the second WB and to collide with part of the electrons self-injected by the first WB. Anyhow,
we interpret the compatibility of our preliminary results with those of \cite{BraEtAl08} as a  strong indication of the effectiveness of  our simple 1D model.

\section{Corrections due to the finite laser spot radius $R$}
\label{finiteR}

 Now we  estimate the effects of a pulse with a finite spot size $R$ at $t=0$.
As often done, we model the pulse as the free plane transverse wave  (\ref{pump}-\ref{modulate})  multiplied by a
`cutoff' function  $\chi\!_{{\scriptscriptstyle R}}(\rho)$, where  $\chi\!_{{\scriptscriptstyle R}}(\rho)$ is 1 if $\rho\!\equiv\!\sqrt{\!x^2\!+\!y^2}\!\le\! R$  and rapidly goes to zero for $\rho\!>\! R$.
We assume $l,\xiH\ll z_R\!\equiv\!kR^2/2$, because
in vacuum $\chi(\rho)\Bep(\xi)$ is close to a solution of  Maxwell equations
for $\xi$-lapses shorter than the Rayleigh length $z_R$. (Actually,  self-focusing strongly reduces diffraction in the plasma, and $\chi(\rho)\Bep(\xi)$ approximates the pulse for a  $\xi$-lapse much larger than $z_R$.)
Moreover, we assume that  $R\!>\!|\BD^{{\scriptscriptstyle \perp}}_{{\scriptscriptstyle M}}|$ (otherwise the solutions of section \ref{plasmaelectrons-in-plane-model} are unreliable even for the $\bX\!=\!(0,0,Z)$ electrons); by (\ref{MaxTransvOscill}), this inequality is fulfilled if  $a_0s_u\lesssim kR$.

By causality  \cite{FioDeN16},
at each instant $t$ all dynamical variables, in particular $n_e$, are strictly the same 
as in the plane model of section \ref{plasmaelectrons-in-plane-model} within the causal cone \
$\C^{\scriptscriptstyle R}_t\!\equiv\!\{(\rho'\!,\varphi'\!,z') \:\:|\:\: 0\!\le\!\rho'\!+\!ct\!-\!z'\!\le\!R\}$ \ (in cylindrical coordinates), 
and approximately the same in a neighbourhood of it (note that $\C^{\scriptscriptstyle R}_t$ trails the pulse with the speed of light). A first consequence is that for $t\ge 0$ the part of the PW trailing the pulse 
and contained inside $\C^{\scriptscriptstyle R}_t$ is exactly as in the plane model. 
Let us now denote by $\bar t,t_e,t_{br}$ the times of the first maximal penetration,  first backward expulsion, and  first WB for the $Z\!=\!0$ electrons in the plane model; clearly 
$0\!<\!\bar t\!<\!t_e\!<\!t_{br}$. 
A second consequence is that for $0< t < t_e$ there is a  ``hole'' $h_t$ in the electron
distribution including at least $\C^{\scriptscriptstyle R}_t\cap L_t$, where $L_t$ is the pure ion layer discussed in section \ref{plasmaelectrons-in-plane-model} and depicted
in fig. \ref{Worldlinescrossings}b; the size of $h_t$ reaches its maximum at  $t=\bar t$ and then decreases.

We denote as lateral electrons (LE) the ones initially located outside the surface $\rho\!=\!R$; 
they are attracted by the positively charged $h_t$ towards the $\vec{z}$-axis.
If the spot radius fulfills 
\be
(t_e\!-\!l/c)|v^\rho_a|>R> |\BD^{{\scriptscriptstyle \perp}}_{{\scriptscriptstyle M}}|,    \label{bubble}
\ee 
then the $(\rho,z)\sim (R,0)$ electrons (i.e. the first LE that reach $\vec{z}$)
collide with each other near the $\vec{z}$-axis and thus close part of $L_t$ into a (possibly temporary)  electron cavity (where $n_e\!=\!0$, and only ions are present)  \cite{RosBreKat91,PukMey2002}
 before  any electrons may be expelled backwards; this cavity may contribute to the formation 
 of a ion {\it bubble}, which then may disappear or trail the pulse
({\it bubble regime}).  In (\ref{bubble}) $v^\rho_a$ stands for the average inward
$\rho$-component of the velocity of LEs,
while $t_e\!-\!l/c$ is the time lapse between the passage of the end of the pulse
across the $z=0$ surface and the first expulsion of $Z=0$ electrons.
By geometric reasons  $|v^\rho_a|\!<\!|v^z_a|$, where $v^z_a$ is the
average $z$-component of the  $\bX\!=\!0$ electrons   velocity
in their backward trip within the bulk; 
a  rough estimate is $|v^\rho_a|\!\simeq\! |v^z_a|/2\!=\!\Delta\!\left(\bar t,0\right) /\,2(t_e\!-\bar t)$.

Whereas for larger $R$ fulfilling \cite{FioDeN16,FioDeN16b} 
\be
R\!\gg\! |\BD^{{\scriptscriptstyle \perp}}_{{\scriptscriptstyle M}}|, \qquad t_e \!-\!\bar t \sim R/c,   
\label{req}
\ee
the $0\!\le\!Z\!\le\!Z_b$, $\rho\!\lesssim\! r \equiv R-(t_e\!-\!l/c)|v^\rho_a|$ electrons
exit  the bulk shortly after $\C^{\scriptscriptstyle R}_t$ has completely entered it.
Conditions (\ref{req})  ensure that these electrons: i)  are expelled before LE collide with them and thus obstruct them the way out; ii)  move approximately as in the plane model 
before the expulsion. This follows from the inequalities
$r>0$, which in turn is implied by (\ref{req}b), $\bar t< l/c$, and $|v^\rho_a|<c$.
The expelled electrons are slightly decelerated by the electric force $F_e^z\sim 1/z_e^2$ generated by the net positive charge located at $z\gtrsim 0$ within $\rho\!<\!r$, but a bunch of them escape backward ($z_e \!\to\! -\infty$) with remarkable energy.
This was named {\it slingshot effect} in \cite{FioFedDeA14,FioDeN16,FioDeN16b} and is manifest in  the  WLs of the $Z\gtrsim 0$ electrons of 
fig. \ref{Worldlinescrossings}  straying left away (although in the plane model $F_e^z$ and the associated deceleration are larger).  If $\Bep$ is slowly modulated, the bunch
is well collimated (because $u^{{\scriptscriptstyle \perp}}_f\!\simeq\!0$).
For more details and quantitative predictions in some realistic examples see \cite{FioDeN16}.

Let $(\xi_{br},Z_{br})$ be the pair $(\xi,\!Z)$ with the smallest $\xi$ such that  $\hat J(\xi,\!Z)\!=\!0$.
For larger $R$,  more precisely if
\be
R\!\gg\! |\BD^{{\scriptscriptstyle \perp}}_{{\scriptscriptstyle M}}|, \qquad R>\xi_{br}, 
\label{collisions}
\ee
then  the $Z\sim Z_{br}$ electrons within $\C^{\scriptscriptstyle R}_t$
will move exactly as in  the plane model.
In particular, if $\widetilde{n_0}(Z)$ decreases in some interval $[z_d,z_s]$ and keeps constant for $Z>z_s$, then  some of the  down-ramp $Z$ electrons within $\C^{\scriptscriptstyle R}_t$ will be injected in the PW and
accelerated (see section  \ref{self-injected-electrons}) exactly as in the  $R=\infty$ %(i.e. plane) 
case.
Since the PW trails the pulse, the self-injected electrons captured in the intersection of a PW trough with $\C^{\scriptscriptstyle R}_t$ at some instant $t_0$ will remain in such an intersection and will experience the same acceleration also for all $t>t_0$, thus justifying the good agreement mentioned at the end of that section. Actually,  if the pulse is sharply peaked at some $\xi_p\in[0,l]$,
e.g. typically $\xi_p=l/2$, then a more effective (although not rigorous) definition of the causal cone is $\C^{\scriptscriptstyle R}_t\!\equiv\!\{(\rho'\!,\varphi'\!,z') \:\:|\:\: 0\!\le\!\rho'\!+\!ct\!-\!\xi_p\!-\!z'\!\le\!R\}$, and the right condition in (\ref{collisions}) can be replaced by the weaker one \
$ R>\xi_{br}- \xi_p$.

\section{Discussion, outlook and conclusions}
\label{conclu}

In this work we have analysed in detail 
the impact of a rather generic short laser pulse (the pump) normally onto a rather generic cold diluted plasma by elaborating (sections \ref{plasmaelectrons-in-plane-model}-\ref{self-injected-electrons})  on the improved plane hydrodynamic model \cite{Fio14JPA,Fio18JPA} (reviewed in section \ref{Setup}) 
and including qualitative corrections (section \ref{finiteR}) due to the finiteness of the radius $R$ of the real pulse spot 
(assumed circular).
We have assumed: the initial plasma density and the pulse in the plane model to be respectively of the type (\ref{n_0bounds}),
  (\ref{pump}) and {\it essential short}, i.e. to fulfill  (\ref{Lncond'}a) (here we have chosen the $\vec{z}$-axis in the direction of propagation of the pulse and as symmetry axis of the pulse spot, $xy$ as the transverse symmetry plane);
$R\gg|\BD^{{\scriptscriptstyle \perp}}_{{\scriptscriptstyle M}}|$
[see  (\ref{MaxTransvOscill})], and the real  initial plasma density to agree with  (\ref{n_0bounds}) at least
in the  cylinder $C_{\scriptscriptstyle R}$ of axis $\vec{z}$ and radius $R$.
 Condition (\ref{Lncond'}a) is a simplifying technical requirement, compatible 
with maximum pulse-to-plasma energy transfer (the latter requires a $l$ about half the PW length \cite{SprEsaTin90PRL,FioFedDeA14}); a sufficient condition for it is (\ref{ShortPulse1'}a).   
Predictions are valid strictly within (and approximately in a neighbourhood of) the causal cone \ $\C^{\scriptscriptstyle R}_t$ trailing the pulse as  long as the pump is not significantly changed, 
i.e. as  long as the transverse EM fields keep close to   (\ref{pump}); if the
pulse is a slowly modulated one  (\ref{modulate}), then $kR>a_0s_u$ 
guarantees $R>|\BD^{{\scriptscriptstyle \perp}}_{{\scriptscriptstyle M}}|$, and the change can be safely neglected  in the spacetime region  (\ref{neglectChange}); this region can be essentially enlarged replacing $n_b\mapsto n_0$ in  (\ref{neglectChange}), if $\widetilde{n_0}(z)=n_0$ except on a short interval $[0,z_s]$.

Several universal features emerge.

In particular, as explained in sections \ref{during} and  \ref{finiteR}, all electrons within the causal cone 
$\C^{\scriptscriptstyle R}_t$, when reached by the laser pulse, begin to move
forward, pushed by the ponderomotive force; consequently,  the $Z\!\sim\!0$ electrons 
leave behind themselves a region $h_t\subset C_{\scriptscriptstyle R}$ completely filled with ions  and deprived of electrons. For sufficiently small $R$, see (\ref{bubble}),
the rear part of this region is occupied by electrons coming from outside $C_{\scriptscriptstyle R}$ (LE), and
$h_t$ closes  into a (possibly temporary)  ion {\it bubble} behind the pulse. It would be worth investigating 
conditions for this bubble to disappear or trail the pulse; in the latter case the {\it bubble regime} would start
already at the vacuum-plasma interface. For larger  $R$, see (\ref{req}), a bunch of these $Z\sim 0$ electrons,
attracted and accelerated backwards by the positively charged $h_t$, are expelled from the
bulk   before LE obstruct their way out, and then escape to $z=-\infty$ with remarkable energy 
 ({\it slingshot effect}) \cite{FioFedDeA14,FioDeN16,FioDeN16b}, while deeper electrons perform periodical longitudinal oscillations; the latter have been
analyzed in detail in section \ref{after}.
As $R$ further increases, the energy of the backward expelled electrons decreases, and the slingshot effect becomes less important.
The dephased periodical longitudinal oscillations of the different $Z$ electron layers yield the PW (plasma wave),
as evident after passing from the Lagrangian to the Eulerian description. 
All this holds within $\C^{\scriptscriptstyle R}_t$ as long as 
the Jacobian $\hat J\equiv \det(\partial \hat\bx_e/\partial \bX)$ of the plane model  keeps positive.

In section \ref{Wave-breakings}
we have shown that $\hat J$ and $\hat\sigma\equiv 
\partial \hat s/\partial Z$ 
fulfill the ordinary Cauchy problem (\ref{basic}). Moreover, for $\xi>l$ they fulfill
 (\ref{pseudoper}), what allows to extend their knowledge from the first period $\xiH$ to all $\xi>l$.
Consequently, $\hat J,\hat\sigma$ are LQP w.r.t. $\xi$, in the sense (\ref{lin-pseudoper}), with common period $\xiH$. Similarly, $J(t,\bX)\equiv \det(\partial \bx_e/\partial \bX)$ fulfills (\ref{pseudoper'}) and is LQP
w.r.t. $t$ with  period $\tH$, as found in eqs (4-6) of \cite{BraEtAl08}.
We have also  shown that the  formulas in closed form (\ref{hatJapprox}),
 (\ref{hatsigma-approx}) approximate well $\hat J,\hat\sigma$ for small $\xi$, in particular $\xi< l$,
and thus allow us to determine the apriori bounds (\ref{cond0}), (\ref{barepsilon-long'''hat})  for $\hJ$ and (\ref{barsigma-bounds0}), (\ref{barsigma-long'''hat})  for $\hat\sigma$ based on the input data.
As a consequence,   (\ref{Lncond'}a) and either
 (\ref{condNoWB}), or  (\ref{condNoWB'}), or (\ref{nowavebreaking}) for all $Z>0$
imply that  $\hat J>0$, so that no  WBDLPI takes place, and  the hydrodynamic description
is self-consistent everywhere  during the laser-plasma interaction. 
In the NR regime these conditions reduce to  (\ref{Lncond}b) and 
  (\ref{NRnowavebreaking}). The latter further simplifies to  (\ref{NRnowavebreaking2}) 
if $\widetilde{n_0}(z)$ grows with $z$.
These inequalities, which involve only the input data $\widetilde{n_0}(z),\Bep(\xi)$, make  the qualitative expectations that small and/or slowly varying densities delay WB into quantitative preliminary conditions; the latter can be checked in few seconds by running on a common notebook some program that can be easily designed using an “off the shelf,” 
general-purpose numerical package (like {\it Mathematica}). In particular, in most LWFA 
experimental situations a supersonic gas (e.g. hydrogen or helium) jet orthogonal to $\vec{z}$ is hit by the laser pulse and locally converted into a plasma by the front of the pulse itself; even if the pulse hits the jet  just outside the nozzle, 
the corresponding $\widetilde{n_0}(z)$ is smooth, in particular has a continuous first derivative, with 
logarithmic derivative $d(\log\widetilde{n_0})/dz$ typically bounded by $\Upsilon=100$cm$^{-1}$, 
(see e.g. fig. 2 in \cite{HosEtAl02}, or fig. 5 in  \cite{VeiEtAl11}); moreover,
$\hDOu$  is a few micron, so that $\delta$  is very small, and (\ref{nowavebreaking}) is satisfied.
Hence there is no WBDLPI, and one can successfully predict the evolution of the plasma 
up to the first WB via a hydrodynamic description, either analytical, or numerical, by fluid simulation codes; the latter are less computationally demanding than PIC codes.

The spacetime location of the first WB (due to collisions  of different $Z$ electrons  layers) 
{\it after} the laser-plasma interaction is determined via (\ref{pseudoper}), and can be estimated via  (\ref{nwb}).
After the time $t_{br}$ of the first WB one needs to replaces the hydrodynamic description (HD) by a kinetic theory (including collisional terms); the above HD allows to compute the
 positions and momenta of the electron fluid elements at $t_{br}^-$, and therefore the associated distribution in
phase space,  to be adopted as initial condition for the application of the kinetic theory.
Nevertheless, for low density plasmas significant 2-body collisions are rare, hence one can neglect such collisions (i.e. continue to treat the plasma as collisionless) and consider only the effect of the mutual average forces generated by pairs of  $Z$ electron layers crossing each other \cite{Fionew}. If only `few' electron layers do, they do not  damage the PW significantly, and their electrons can be treated as test particles subject to the electric field associated to the PW. The equations of motion of  test particles injected in the PW, with $\xi$ as the `time' variable,  and the qualitative behaviour of their solutions  have been studied in section \ref{test-electron} (for more details see \cite{FioAAC22,Fionew}); in our model test particles cannot dephase, because the phase velocity of the PW is $c$. 
 In particular, we find  (section \ref{self-injected-electrons}) that the {\it maximal energy}
of the electrons  self-injected (via WB) in the PW and trapped by a PW single trough {\it grows approximately linearly} with the distance gone, cf. (\ref{s_i^m<0|s_iz_i}).
We have tested and illustrated our approach applying it in particular  to the (somewhat border-line) conditions
considered by  \cite{BraEtAl08} and reported in fig. \ref{Worldlinescrossings}, finding an encouraging agreement.

\section{Appendix}
\label{App}

\subsection{Proof of (\ref{est-phi})}
\label{Proofs-est-phi}

\begin{lemma}
\bea
\int\limits^\xi_0\!\!\!  d\eta\,\frac{\mu^2(\eta)}{\hat s^2(\eta,\!Z)} = \xi\!+\!2\hat\Delta(\xi,\!Z),\:\:
\int\limits^\xi_0\!\!\!  d\eta\,\frac{\mu(\eta)}{\hat s(\eta,\!Z)} \le \xi\!+\!\hat\Delta(\xi,\!Z),\:\:
\int\limits^\xi_0\!\!\!  d\eta\,\sqrt{\!\frac{\mu(\eta)}{\hat s(\eta,\!Z)}} \le \xi\!+\!\frac 12 \hat\Delta(\xi,\!Z).
\quad\label{nice-ineq}
\eea
\label{lemma'}
\end{lemma}
{\bf Proof} \ \ (\ref{nice-ineq}a) follows from (\ref{heq1}a).
On the other hand,
\bea
\left(\frac{\mu}{\hat s}\!-\!1\!\right)^2\!\ge 0 \quad\Rightarrow\quad 2\frac{\mu}{\hat s}\le 1\!+\!
\frac{\mu^2}{\hat s^2} \quad\Rightarrow\quad 
2\!\!\int^\xi_0\!\!\!  d\eta\,\frac{\mu(\eta)}{\hat s(\eta,Z)} \le \int^\xi_0\!\!\!   d\eta \left[\frac{\mu^2(\eta)}{\hat s^2(\eta,Z)}\!+\!1 \right] = 2\xi+2\Delta(\xi,Z), \nn
\left(\!\sqrt{\frac{\mu}{\hat s}}\!-\!1\!\right)^2\!\ge 0 \quad\Rightarrow\quad 2\sqrt{\frac{\mu}{\hat s}}\le 1\!+\!\frac{\mu}{\hat s} \quad\Rightarrow\quad 
2\!\!\int^\xi_0\!\!\!  d\eta\,\sqrt{\frac{\mu(\eta)}{\hat s(\eta,Z)}} 
\le 2\xi\!+\!\Delta(\xi,Z).\qquad \Box \nonumber
\eea
For $\xi\le\tilde\xi_3(Z)$, i.e. as long as $\hat s(\xi,Z)\!\ge\!  1$,  it is \
$\hat \Delta(\xi,Z) \le\hDO(\xi)$, \ and by (\ref{nice-ineq})  we find the  bounds \ $\int^\xi_0\!  d\eta\,\frac{\mu(\eta)}{\hat s^{3/2}(\eta,Z)} \le \xi+\Delta(\xi,Z)\le \xi+\hDO(\xi)$. \ 
Hence (\ref{defphi}) implies (\ref{est-phi}), as claimed.

\subsection{Properties of the `time-dependent' harmonic oscillator}
\label{tdep-osc}

We briefly report here some properties  of the `time-dependent' harmonic oscillator studied in \cite{Fiotdho}. By Proposition 1 in \cite{Fiotdho},  every nontrivial solution of the equation 
\be
\frac{d^2 q}{dx^2}=-\bom^2(x) q(x)                        \label{harmosceq}
\ee
(here $x$ plays the role of 'time') with $\inf_x\bom^2>0$ admits a strictly increasing sequence 
$\{x_h\}_{h\in\ZZ}\subset\RR$ of consecutive, interlacing zeros of $q,\frac{d q}{dx}$; \ more precisely,
 for all \ $j\in\ZZ$   \  $q$ vanishes and  $(-1)^j\frac{d q}{dx}$ has a positive maximum at $x=x_{2j}$, \
while $\frac{d q}{dx}$ vanishes and  $(-1)^jq$ has a positive maximum at $x=x_{2j+1}$.
(Indexing is defined up to $h\mapsto h\!+\!4j$.)
Taking $x_0$ as an independent variable, every $x_h$ is a strictly increasing function of $x_0$.
In  the case  $\bom\!=$const this reduces to \  $q(x)=A\sin[\bom(x\!-\!x_0)]$, 
and $x_h(x_0)\!-\!x_0=h\pi/4\bom$. If $\bom(x)$ is continuous and has bounded derivative defined almost everywhere, then 
$q,\frac{d q}{dx}$ can be written in the form
\be
 q =\sqrt{\frac{2I}{\bom}}\,\sin\psi, \qquad \frac{d q}{dx} =\sqrt{2I\bom}\,\cos\psi,
 \label{PolarCoord}
\ee
where the angle variable $\psi(x)$ is the solution of the Cauchy problem \ $\frac{d \psi}{dx}=\bom +\frac{1}{2\bom}\frac{d \bom}{dx}\sin(2\psi)$ \ with \ $\psi(x_0)=0$, \ 
or equivalently of the integral equation
\be
\psi(x)=\int_{x_0}^x\!\!\!  dz\,\bom (z)
+\int_{x_0}^x\!\!\!   dz\,\frac{1}{2\bom}\frac{d \bom}{dz} \,\sin\left[2\psi(z)\right],
\label{Inteqpsi}
\ee
while the action variable $I$ is given by 
\be
I(x)=I(x_0)
\exp\left\{-\!\!\int^x_{x_0}\!\!dz \left[\frac{1}{\bom}\frac{d \bom}{dz}\cos(2\psi)\right]\!(z)\right\}. \label{closedform-I}
\ee
The $x_h$  mentioned above  are related to $\psi$ by the relation  \ $\psi(x_h)=h\frac\pi 2$. \
If $\frac{1}{\bom}\frac{d \bom}{dz}$ is small, or   oscillates 'fast' about zero, then the
integrals containing it in (\ref{Inteqpsi}), (\ref{closedform-I}) can be neglected, 
$$
I(x)\simeq I(x_0)\qquad \psi(x)\simeq \phi(x):= \int_{x_0}^x\!\!\!  dz\,\bom (z)
$$ 
are good approximations, and the $x_h$ can be approximately characterized by 
\ $\phi(x_h)\simeq h\frac\pi 2$. More precisely, this is justified as long as 
one of the two following conditions is satisfied:
\bea
\left|\int_{x_0}^x\!\!\!   dz\,\frac{1}{2\bom}\frac{d \bom}{dz} \,\sin\left[2\phi(z)\right]\right|\ll|\phi(x)|,
\qquad
\Lambda(x):=\int_{x_0}^x\!\!\!   dz\,\left|\frac{1}{2\bom}\frac{d \bom}{dz} \right|\ll|\phi(x)|,
\label{good-approx}
\eea
Clearly the second implies the first.
If  $\bom(x)$ is monotone in $[x_0,x]$
then $2\Phi(x)=\left|\log\frac{\bom(x)}{\bom(x_0)}\right|$; otherwise $2\Phi$ is the
 {\it total variation} of $\log\bom(x)$ in such an interval,
  i.e.  the sum of a term of this kind for each  monotonicity interval contained  in  $[x_0,x]$.

\subsection{Proofs of the bounds and propositions of section \ref{Bounds0l}}
\label{Proofs-0l}

The function defined by \ $y(\xi,Z):=\int^\xi_0\!\!d\eta\,\kappa(\eta,Z)$ \
strictly grows w.r.t. $\xi$ (for each $Z$). We shall abbreviate  \ $y(l)\!\equiv\! y(l,Z)$. \
When adopting $y$ as a new `time' (i.e. independent) variable we shall  put a bar over 
dynamical variables seen as functions of $y$, e.g.
\be
\bar n(y,Z)\equiv \check n[\xi(y,Z),Z]=\widetilde{n_0}\left\{\hze\left[\xi(y,Z),Z\right]\right\}, 
\qquad \bom^2(y,Z)\equiv \frac{K \bar n}{\bar\kappa}(y,Z)  \label{def-bar-n-omega} 
\ee
($\bom$  has dimensions of an inverse   length, as $\hat\omega$) 
and abbreviate  \ $\dot {\bar f}\equiv \frac{\partial {\bar f}}{\partial y}$. \
The regularity assumptions of Propositon \ref{hatJsigma-approx}  about $\widetilde{n_0}(Z)$ imply that
 $\bar\omega(\cdot,Z)$ is continuous, $\bar\omega'(\cdot,Z)$ is defined at least piecewise and bounded everywhere. 
We find $\dot\bD=-\bL$,  $\dot\bL= \bom^2\bD$, and can rephrase the Cauchy problem  (\ref{hDeq}) also as that of a time-dependent harmonic oscillator
\be
 \ddot \bD(y,y_0)=-\bom^2(y)\:\bD(y,y_0), \qquad %\dot\bD=-\bL, 
\qquad 
\bD(y_0,y_0)=0, \quad \dot\bD(y_0,y_0)=-1.    \label{Deq}
\ee
This is solved by \ $\bD(\ti,\ti_0)=\q_1(\ti)\q_2(\ti_0)\!-\!\q_2(\ti)  \q_1(\ti_0)$, \ where
$\q_1,\q_2$ solve (\ref{Deq}a) with conditions $\q_1(0)=\dot \q_2(0)=1$,
$\q_2(0)=\dot \q_1(0)=0$\footnote{(\ref{Deq}c) holds because $-\dot\bD(y_0,y_0)$ is the Wronskian 
$W(y_0)$ of  $\q_1,\q_2$ evaluated at $y_0$; but $W(y_0)\!=\!W(0)\!=\!1$.}.
As a consequence,  $\bD,\dot\bD$ fulfill also the Cauchy problems associated to the 
time-dependent (with `time' variable $y_0$) harmonic oscillator
\bea
\bD"(y,\!y_0)=-\bom^2(y_0)\bD(y,\!y_0),\qquad &\bD(y,\!y)=0,\quad
&\bD\mrq(y,\!y)=1, \label{Deq'}\\[6pt]
\dot\bD"(y,\!y_0)=-\bom^2(y_0)\dot\bD(y,\!y_0),\qquad &\dot\bD(y,\!y)=-1,\quad
&\dot\bD\mrq(y,\!y)=0                \label{dotDeq}
\eea
(we abbreviate ${\bar f}\mrq\!\equiv\! \frac{d{\bar f}}{dy_0}$).
Re-exhibiting the $Z$-dependence, we can rewrite  (\ref{gensol'})   in the form
\bea
\left(\!\!\ba{l} \hat\varepsilon(\xi,\! Z)  \\
\hat\sigma(\xi,\! Z) \ea\!\!\right)\! &=& \!\!\int^{y(\xi, Z)}_0 \!\!\!\!dy_0
\!\! \ba{l} \left[1\!-\!\frac{\widetilde{n_0}(Z)}{\bar n(y_0,\! Z)}\right] \ea\!\! \bom^2(y_0,\! Z)
\left(\!\!\ba{l} \bD(y,\!y_0;\!Z)\\
- \dot\bD(y,\!y_0;\!Z)\ea\!\!\!\right) , \label{gensol''}  \nn
\left(\!\!\ba{l} \bar\varepsilon(y,\! Z)  \\
 \bar\sigma(y,\! Z) \ea\!\!\right)\! 
&=& \!\!\int^y_{0} \!\! dy_0
\!\! \ba{l} \left[1\!-\!\frac{\widetilde{n_0}}{\bar n(y_0,\! Z)}\right] \ea\!\! %\bD"(y,\!y_0;\!Z)
\left(\!\!\ba{l} -\bD"(y,\!y_0;\!Z)\\
\dot\bD"(y,\!y_0;\!Z)\ea\!\!\!\right) .
\label{barepsilon}
\eea
In the second line we have adopted $y$ as the independent variable and used (\ref{dotDeq}).

\subsubsection*{Proof of   (\ref{hatJsigmaapprox}a) in Proposition \ref{hatJsigma-approx},  of   (\ref{cond0}) 
and of the no WBDLPI condition (\ref{condNoWB})}

We apply the results \cite{Fiotdho} recalled in section  \ref{tdep-osc} to the solution 
$\bD(y,\!y_0)\equiv q(y_0)$ of (\ref{Deq'}), with $y_0$ playing the role of `time' $x$.
We denote by $\{\cy_n(y)\}_{n\in\ZZ}$ the  sequence of zeros 
associated via Proposition 1 of \cite{Fiotdho},  and
by $\check I,\cpsi$ the corresponding action, angle variables.
The  $\cy_n(y)$ fulfill   $\bD\big[y,\cy_{2h}(y)\big]\!=\!0$, \
 $(-)^h\bD\mrq\big[y,\cy_{2h}(y)\big]\!>\!0$, \
$(-)^{h}\bD\big[y,\cy_{2h+1}(y)\big]\!>\!0$, \ $\bD\mrq\big[y,\cy_{2h+1}(y)\big]\!=\!0$ 
\ (if $\bom=$const  it is \ $\bD(y,\!y_0)=\frac 1{\bom}\sin[\bom(y_0\!-\!y)]$
and $\cy_j(y)=y+j\pi/2\bom$). \ 
Consequently, \ $\bD(y,y_0)$, $-\bD"(y,y_0)$ are negative (resp. positive) if $h$ is
even (resp. odd) and  $y_0\in]\cy_{-2h-2}(y), \cy_{-2h}(y)[$.
Moreover, $\cy_j[\cy_{h}(y)]=\cy_{h+j}(y)$ for all $j,h\in\ZZ$; in particular,
$\cy_j[\cy_{-j}(y)]=\cy_0(y)=y$.

\bigskip
\noindent
By (\ref{good-approx-xi}) we can neglect the integrals containing $d(\log\bom)/dx$ 
in eq. (\ref{Inteqpsi}-\ref{closedform-I}); thus  we find 
 \bea
2\check I(y,\!y_0)\simeq 1/\bom(y), \qquad -\cpsi(y,\!y_0)\simeq \bar\phi(y,\!y_0)
\equiv\int^{y}_{y_0}\!\!\!\!   dx\: \bar\omega(x), \label{ESTcpsi}\\[6pt]
\bD(y,\!y_0) = \sqrt{\frac{2\I(y,\!y_0)}{\bom(y_0)}}\,\sin\left[\cpsi(y,\!y_0)\right]
\simeq -\frac{\sin\left[\bar\phi(y,\!y_0)\right]}{\sqrt{\bom(y)\bom(y_0)}}=:\bD_a(y,\!y_0) , \label{PolarCoordD}\\
\bD\mrq(y,y_0) = \sqrt{2\check\I(y,\!y_0)\,\bom(y_0)}\,\cos\left[\cpsi(y,\!y_0)\right]
\simeq \sqrt{\frac{\bom(y_0)}{\bom(y)}}
\cos \left[\bar\phi(y,y_0)\right], \label{PolarCoordDd}\\[6pt]
\bD\mrq\left[\cy_{2h}(y_0),y_0\right]\simeq    (-1)^h\sqrt{\!\frac{\bom(y_0)}{\bom\left[\cy_{2h}(y_0)\right]}},\qquad \bD\mrq\left[y, \cy_{-2h}(y)\right] \simeq    (-1)^{h}\sqrt{\frac{\bom[\cy_{-2h}(y)]}{\bom(y)}}.
\label{PolarCoordD'}
\eea
$\bar\phi$ is $\phi$ expressed in terms of $y$, cf. (\ref{defphi}). 
From (\ref{barepsilon}) and \ $\bar J(y)=\bar\varepsilon(y)+1$ \ we obtain
\bea
\bar J(y)= 1-\left[\bD\mrq(y,\!y_0)\right]_{y_0=0}^{y_0=y}+
\!\!\int^{y}_{0} \!\!\!\!\!  dy_0  \, \frac{\widetilde{n_0}}{\bar n(y_0)}  \bD"(y,\!y_0)
=\bD\mrq(y,\!0)+\!\!\int^{y}_{0} \!\!\!\!\!  dy_0  \, \frac{\widetilde{n_0}}{\bar n(y_0)}  \bD"(y,\!y_0)
  \label{barepsilon'}
\eea
Since $\frac{1}{\bom}\frac{d \bom}{dy_0}$   oscillates fast about zero, then using 
 \ $\bD"(y,\!y_0)=-\bom^2(y_0)\bD(y,\!y_0)$,  (\ref{PolarCoordD}) we obtain
\bea
\bar J(y)\simeq 1-\!\!\int^y_{0} \!\! dy_0\,\bom^2(y_0)
 \left[1\!-\!\frac{\widetilde{n_0}}{\bar n(y_0)}\right] 
 \frac{ \sin \left[\bar\phi(y,y_0)\right]}{\sqrt{\bom(y)\bom(y_0)}}=:\bar J_a(y);
 \label{hatJapprox'}
\eea
putting $y=y(\xi)$, $y_0=y(\eta)$, whence $ dy_0 \bom^2(y_0) =Kl\bar n(y_0)d\eta$, this yields  (\ref{hatJapprox}). 

By (\ref{ESTcpsi}), the $\cy_j(y)$  approximately solve the equations \ $\bar\phi[x,y]=j \pi/2$ in the unknown $x$.
If $y\in[0, \cy_2(0)]$, i.e. if $\bar\phi(y,0)\simeq -\cpsi[y,0] $ belongs to $[0,\pi]$, then $\sin \left[\bar\phi(y,y_0)\right]\ge 0$ in all the integration interval, so that  by (\ref{barepsilon'}) 
\bea
\frac{\widetilde{n_0}}{n_u}\big[1\!-\! \bD\mrq\!(y,\!0)  \big]=
\frac{\widetilde{n_0}}{n_u}\!\!\int^{y}_{0} \!\!\!\!\!  dy_0 
 \bD"\!(y,\!y_0) \le \bar J(y)-\bD\mrq\!(y,\!0)\le     \frac{\widetilde{n_0}}{n_d}\!\!\int^{y}_{0} \!\!\!\!\!  dy_0 
 \bD"\!(y,\!y_0)=\frac{\widetilde{n_0}}{n_d}\big[1\!-\! \bD\mrq\!(y,\!0) \big] \nn
\Rightarrow\qquad  
\frac{\widetilde{n_0}}{n_u}+  \left(\!1- \frac{\widetilde{n_0}}{n_u}\! \right)\! \bD\mrq(y,\!0)
\le  \bar J(y) \le \frac{\widetilde{n_0}}{n_d}+  
\left(\!1- \frac{\widetilde{n_0}}{n_d}\! \right)\! \bD\mrq(y,\!0)\qquad \qquad \qquad 
\label{barepsilon-lon}
\eea
Since $\bD\mrq(y,\!0)$ decreases with $y\in[0, \cy_2(0)]$, if $y(l)\in[0, \cy_2(0)]$ 
the minimum of lhs(\ref{barepsilon-lon}) during the laser-plasma interaction is
attained at $y= y(l)$, i.e.  at $\xi=l$. 
By (\ref{PolarCoordDd}) eq. (\ref{barepsilon-lon}) becomes 
\bea
\frac{\widetilde{n_0}}{n_u}\!+\!  \left(\!1\!-\!\frac{\widetilde{n_0}}{n_u}\! \right)\! \sqrt{\!\frac{\bom_0}{\bom(y)}} \cos\left[\bar\phi(y,0)\right] \lesssim  \bar J(y) \lesssim \frac{\widetilde{n_0}}{n_d}- 
\left(\! \frac{\widetilde{n_0}}{n_d} \!-\!1\! \right)\! \sqrt{\!\frac{\bom_0}{\bom(y)}}  \cos\left[\bar\phi(y,0)\right],                    \label{cond0'}
\eea
The inequalities (\ref{cond0}) follow observing that, by (\ref{defphi}), \ 
$\sqrt{\bom_0/\bom[y(\xi)]}=\Gamma(\xi)$.
Moreover,  setting $y=y(l)$  we find that $\hat J>0$, and there is no WBDLPI, iff (\ref{condNoWB}) holds, as claimed.

\subsubsection*{Proof of formula (\ref{hatJsigmaapprox}b) in Proposition \ref{hatJsigma-approx} and of  (\ref{barsigma-bounds0})}

Similarly, we  apply the results \cite{Fiotdho} to the solution $\dot\D(y,\!y_0)\equiv q(y_0)$ of (\ref{dotDeq}),
with $y_0$ playing the role of `time' $x$, and  as $\widetilde{I},\tpsi$
the corresponding action, angle variables. 
We denote by $\{\ty_h(y)\}_{h\in\ZZ} $ the sequence of zeros 
associated via Proposition 1 of \cite{Fiotdho}.
The $\ty_h(y)$ satisfy \  $\dot\bD\big[y,\ty_{2h}(y)\big]=0$, 
 $(-1)^h\dot\bD\mrq\big[y,\ty_{2h}(y)\big]>0$, 
$(-1)^{h}\dot\bD\big[y,\ty_{2h+1}(y)\big]>0$,  $\dot\bD\mrq\big[y,\ty_{2h+1}(y)\big]=0$. \
For all $j,h\in\ZZ$ it is $\ty_j[\ty_{h}(y)]=\ty_{j+h+1}(y)$; in particular,
$\ty_{-h-2}[\ty_{h}(y)]=\ty_{-1}(y)=y$. \
$\dot\bD(y,y_0)$, $-\dot\bD"(y,y_0)$ are negative (resp. positive) if $h$ is
even (resp. odd) and $y_0\in]\ty_{-2h-2}(y), \ty_{-2h}(y)[$. 
When $\bom\!=$const we have 
$\dot\bD(y,\!y_0)=-\cos[\bom(y_0\!-\!y)]$ and \ $\ty_j(y)=y+(j\!+\!1)\pi/2\bom$. \ 

\noindent
By (\ref{good-approx-xi}) we can neglect the integrals containing $d(\log\bom)/dx$ 
in eq. (\ref{Inteqpsi}-\ref{closedform-I}); thus  we find 
\bea
2\widetilde{I}(y,\!y_0)\simeq 1, \qquad \tpsi(y,\!y_0)\simeq -\bar\phi(y,\!y_0)-\pi/2, \label{ESTtpsi}\\[6pt]
\dot\bD(y,\!y_0) =\sqrt{\frac{2\I(y,\!y_0)}{\bom(y_0)}}\sin\left[\tpsi(y,\!y_0)\right]
\simeq - \sqrt{\frac{ \bom(y)}{\bom(y_0)}}\cos[\bar\phi(y,\!y_0)],\label{PolarCoordDdot}\\[8pt]
\dot\bD\mrq(y,\!y_0) =\sqrt{2\I(y,\!y_0) \,\bom(y_0)}\cos\left[\tpsi(y,\!y_0)\right]
\simeq -\sqrt{ \bom(y)\bom(y_0)}\sin[\bar\phi(y,\!y_0)], \label{PolarCoordD''}\\[8pt]
\dot\bD\mrq\left[y, \ty_{2h+1}(y)\right]=0,
\qquad \dot\bD\mrq\left[y, \ty_{2h}(y)\right] \simeq    (-1)^{h}\sqrt{\bom(y)\,\bom[\ty_{2h}(y)]}.
\label{PolarCoordD'''}
\eea
Using (\ref{PolarCoordDdot}), from (\ref{barepsilon}) we  obtain 
\bea
\bar\sigma(y)  \simeq\!\!\int^{y}_0 \!\!\!\!dy_0\!\! \ba{l} \left[1\!-\!\frac{\widetilde{n_0}}{\bar n(y_0)}\right] \ea\!\! \bom^2(y_0) \sqrt{\!\frac{ \bom(y)}{\bom(y_0)}}\cos\!\left[\bar\phi(y,\!y_0)\right]
=\!\!\int^{y}_0 \!\!\!\!dy_0\!\! \ba{l} \left[\bar n(y_0)\!-\! \widetilde{n_0}\right] \ea\!\! \frac{Kl }{\bar\kappa(y_0)}\: \sqrt{\!\frac{ \bom(y)}{\bom(y_0)}}\cos\!\left[\bar\phi(y,\!y_0)\right],
\nonumber
\eea
which changing independent variable $y\mapsto\xi$ and redisplaying $Z$-dependences yields (\ref{hatJsigmaapprox}b).

From (\ref{barepsilon}) we also obtain
\bea
\bar\sigma(y) 
=\left[\dot\bD\mrq(y,\!y_0)\right]_{y_0=0}^{y_0=y}-
\!\!\int^{y}_{0} \!\!\!\!\!  dy_0  \, \frac{\widetilde{n_0}}{\bar n(y_0)}  \dot\bD"(y,\!y_0)
=-\dot\bD\mrq(y,\!0)-\!\!\int^{y}_{0} \!\!\!\!\!  dy_0  \, \frac{\widetilde{n_0}}{\bar n(y_0)} 
 \dot\bD"(y,\!y_0).
  \label{barsigma}
\eea
If  $0\!\in\! [ \ty_{-2}(y),y]$, or equivalently $y\!\in\! [0, \ty_0(0)]$, or $\bar\phi(l,0)\in  [0,\pi/2]$,
 then $\dot\bD"\ge 0$ in all the integration interval, and by (\ref{barsigma})
\bea
\frac{\widetilde{n_0}}{n_d} \dot\bD\mrq(y,\!0)=-\frac{\widetilde{n_0}}{n_d} \int^{y}_{0} \!\!\!\!\!  dy_0  \,  \dot\bD"(y,\!y_0)\le \bar\sigma(y)+\dot\bD\mrq(y,\!0)\le -\frac{\widetilde{n_0}}{n_u} \int^{y}_{0} \!\!\!\!\!  dy_0  \,  \dot\bD"(y,\!y_0)
=\frac{\widetilde{n_0}}{n_u}  \dot\bD\mrq(y,\!0)\nn
\Rightarrow \qquad %\qquad
-\left(\!\frac{\widetilde{n_0}}{n_d}\!-\!1\!\right)  \sqrt{ \bom(y)\bom_0}\sin[\bar\phi(y,\!0)]
\lesssim \bar\sigma(y)
\lesssim\left(\!1\!-\!\frac{\widetilde{n_0}}{n_u}\!\right)  \sqrt{ \bom(y)\bom_0}\sin[\bar\phi(y,\!0)]
\qquad  
\label{boundssigma}
\eea
The inequalities (\ref{barsigma-bounds0}) follow  observing that, by (\ref{defphi}), \ 
$\sqrt{\bom_0\bom[y(\xi)]} =l\sqrt{K}\left[\widetilde{n_0}\check n(\xi)  \hat s^3(\xi)/\mu^2(\xi)\right]^{1/4}$.

\subsubsection*{Two lemmas for proving Proposition \ref{Jsigmabounds}}

\begin{lemma} 
For all \ $y\in [0,y(l)]$, let $k(y)\in\NN_0$ be the natural number such that $y\in[ \ty_{2k-2}(0), \ty_{2k}(0)]$,
or (almost) equivalently \ $\bar\phi(y,0)\in \left[\pi(k\!-\!\frac 12),\pi(k\!+\!\frac 12)\right]$. \ Then
\bea
 -\left(\!\frac{\widetilde{n_0}}{n_d} \!-\!  \frac{\widetilde{n_0}}{n_u}\!\right)\! 
\MN_k(y)-  \left| 1\!-\!\frac{\widetilde{n_0}}{n_{k+1}} \right|  \: \lesssim \: \frac
{\bar\sigma(y)}{\sqrt{\bom_0\,\bom(y)}} \: \lesssim \:   
\left(\!\frac{\widetilde{n_0}}{n_d} \!-\!  \frac{\widetilde{n_0}}{n_u}\!\right)\! 
\MN_k(y)+  \left|1\!-\! \frac{\widetilde{n_0}}{n_k}\right|   
\label{barsigma-long''}
\eea
where \ $\MN_{0}(y)\equiv 0$, and, if $k>0$, \ \ 
$\displaystyle \MN_k(y)\equiv \frac{\sqrt{\bom[\ty_{-2}(y)]}\!+\!...\!+\!\sqrt{\bom[\ty_{-2k}(y)]}}{\sqrt{\bom_0}}$.
\end{lemma} 

\bp{}
If $y\le  \ty_0(0)$, i.e. if $\bar\phi(l,0)\le\pi/2$, the claim immediately follows from (\ref{boundssigma}). 
If $y(l)> \ty_0(0)$, i.e. if $\bar\phi(l,0)>\pi/2$, then for all $y\ge  \ty_0(0)$
we can split  the integral in (\ref{barsigma}) into
\bea
\bar\sigma(y)= -\dot\bD\mrq(y,\!0)-\!\!\int^{y}_{\ty_{-2}} \!\!\!\!\!\!\! dy_0
 \frac{\widetilde{n_0}}{\bar n(y_0)}   \dot\bD"\!(y,\!y_0)-
\!\!\int^{\ty_{-2}}_{\ty_{-4}} \!\!\!\!\!\!\!\!\!\!dy_0
\frac{\widetilde{n_0}}{\bar n(y_0)} \dot\bD"\!(y,\!y_0)-....-
\!\!\int^{\ty_{-2k}}_{0} \!\!\!\!\!\!\!\!\!\! dy_0
\frac{\widetilde{n_0}}{\bar n(y_0)}  \dot\bD"\!(y,\!y_0)\quad
 \label{barsigma-long}
\eea
where we have abbreviated $\ty_{-2h}\equiv \ty_{-2h}(y)$. This implies on one hand
\bea
\bar\sigma(y)\le-\dot\bD\mrq(y,\!0)-\frac{\widetilde{n_0}}{n_u}\! \int^{y}_{\ty_{-2}} \!\!\!\!\!\!\! dy_0
  \dot\bD"(y,\!y_0)-\frac{\widetilde{n_0}}{n_d}
\! \int^{\ty_{-2}}_{\ty_{-4}} \!\!\!\!\!\!\!\!\!\!dy_0
 \dot\bD"(y,\!y_0)-....-\frac{\widetilde{n_0}}{n_k}
\!\int^{\ty_{-2k}}_{0} \!\!\!\!\!\!\!\!\!\! dy_0   \dot\bD"(y,\!y_0)\nn
=-\dot\bD\mrq(y,\!0)-\frac{\widetilde{n_0}}{n_u}\!  \left[\dot\bD\mrq(y,\!y_0)\right]^{y}_{\ty_{-2}}-\frac{\widetilde{n_0}}{n_d}
\!  \left[\dot\bD\mrq(y,\!y_0)\right]^{\ty_{-2}}_{\ty_{-4}}-....-\frac{\widetilde{n_0}}{n_k}
\! \left[\dot\bD\mrq(y,\!y_0)\right]^{\ty_{-2k}}_{0} \nn
=\left(\!\frac{\widetilde{n_0}}{n_u}\!-\!\frac{\widetilde{n_0}}{n_d}\!\right) 
\left[\dot\bD\mrq(y,\!\ty_{-2})-\dot\bD\mrq(y,\!\ty_{-4})+
...-(-1)^k\dot\bD\mrq(y,\!\ty_{-2k}) \right]-
\left(\!1\!-\!\frac{\widetilde{n_0}}{n_k}\!\right) \dot\bD\mrq(y,\!0) \nn 
\simeq \sqrt{\bom(y)}\! \left\{\!\left(\!1\!-\!\frac{\widetilde{n_0}}{n_k}\!\right) 
\sqrt{\bom_0}\sin[\bar\phi(y,\!0)]+\left(\!\frac{\widetilde{n_0}}{n_d}\!-\!\frac{\widetilde{n_0}}{n_u}\!\right) \left[\sqrt{\bom[\ty_{-2}(y)]}\!+\!
...\!+\!\sqrt{\bom[\ty_{-2k}(y)]} \right]\right\}. 
\nonumber
\eea
Dividing this inequality by $\sqrt{\bom_0\,\bom(y)}$ and replacing  the first term in the square bracket  by its maximum we obtain (\ref{barsigma-long''}b).
Eq. (\ref{barsigma-long}) implies also
\bea
\bar\sigma(y)\ge-\dot\bD\mrq(y,\!0)-\frac{\widetilde{n_0}}{n_d}\! \int^{y}_{\ty_{-2}} \!\!\!\!\!\!\! dy_0
  \dot\bD"(y,\!y_0)-\frac{\widetilde{n_0}}{n_u}
\! \int^{\ty_{-2}}_{\ty_{-4}} \!\!\!\!\!\!\!\!\!\!dy_0
 \dot\bD"(y,\!y_0)-....-\frac{\widetilde{n_0}}{n_{k+1}}
\!\int^{\ty_{-2k}}_{0} \!\!\!\!\!\!\!\!\!\! dy_0   \dot\bD"(y,\!y_0)\nn
=-\dot\bD\mrq(y,\!0)-\frac{\widetilde{n_0}}{n_d}\!  \left[\dot\bD\mrq(y,\!y_0)\right]^{y}_{\ty_{-2}}-\frac{\widetilde{n_0}}{n_u}
\!  \left[\dot\bD\mrq(y,\!y_0)\right]^{\ty_{-2}}_{\ty_{-4}}-....-\frac{\widetilde{n_0}}{n_{k+1}}
\! \left[\dot\bD\mrq(y,\!y_0)\right]^{\ty_{-2k}}_{0}\nonumber
\eea
\bea
=\left(\!\frac{\widetilde{n_0}}{n_d}\!-\!\frac{\widetilde{n_0}}{n_u}\!\right) 
\left[\dot\bD\mrq(y,\!\ty_{-2})-\dot\bD\mrq(y,\!\ty_{-4})+
...-(-1)^k\dot\bD\mrq(y,\!\ty_{-2k}) \right]-
\left(\!1\!-\!\frac{\widetilde{n_0}}{n_{k+1}}\!\right) \dot\bD\mrq(y,\!0)\nn
\simeq \sqrt{\bom(y)}\!
\left\{\!\left(\!1\!-\!\frac{\widetilde{n_0}}{n_{k+1}}\!\right) \sqrt{\bom_0}\sin[\bar\phi(y,\!0)]-\left(\!\frac{\widetilde{n_0}}{n_d}\!-\!\frac{\widetilde{n_0}}{n_u}\!\right) \!
\left[\sqrt{\bom[\ty_{-2}(y)]}\!+\!
...\!+\!\sqrt{\bom[\ty_{-2k}(y)]} \right]\!\right\}. \nonumber
\eea
Dividing this inequality by $\sqrt{\bom_0\,\bom(y)}$ and replacing the first term in the square bracket  by its minimum we obtain  (\ref{barsigma-long''}a).
\ep

{\bf Remarks.} 
The last term at the rhs(\ref{barsigma-long''}) vanishes if $n_k=\widetilde{n_0}$, what is the case  
if $k$ is even and  $\check n(\xi,Z)\le \widetilde{n_0}(Z)$  for all $\xi\in[0,l]$
(so that $\bar n(y_0,Z)\le \widetilde{n_0}(Z)$  for all $y_0\in[0,y]$), or if $k$ is odd and  $\check n(\xi,Z)\ge \widetilde{n_0}(Z)$  for all $\xi\in[0,l]$
(so that $\bar n(y_0,Z)\ge \widetilde{n_0}(Z)$ and  for all $y_0\in[0,y]$). 
The last term at the lhs(\ref{barsigma-long''}) vanishes if $n_{k+1}=\widetilde{n_0}$, what is the case  
if $k$ is odd and  $\check n(\xi,Z)\le \widetilde{n_0}(Z)$  for all $\xi\in[0,l]$
(so that $\bar n(y_0,Z)\le \widetilde{n_0}(Z)$  for all $y_0\in[0,y]$), or if $k$ is even and  $\check n(\xi,Z)\ge \widetilde{n_0}(Z)$  for all $\xi\in[0,l]$
(so that $\bar n(y_0,Z)\ge \widetilde{n_0}(Z)$ and  for all $y_0\in[0,y]$). 

To make formulae short, in (\ref{barepsilon-long''}),
(\ref{barsigma-long''}), (\ref{barepsilon-long'''hat}),  (\ref{barsigma-long'''hat}) we have not displayed 
these $Z$-dependences of $\widetilde{n_0},n_u,n_d,n_k,s_u,s_d,\MM_u,\br\!_u,\hat J,\hat\sigma$.

\begin{lemma} 
For all $y\in [0,y(l)]$,  let $k(y)\in\NN_0$ be  such that
$y\in[\cy_{2k}(0),\cy_{2k+2}(0)[$. Then
\bea
\frac{\widetilde{n_0}}{n_u} -
\left(\!\frac{\widetilde{n_0}}{n_d} \!-\!  \frac{\widetilde{n_0}}{n_u}\!\right)\! 
\MM_{k}(y)- \left|1\!-\! \frac{\widetilde{n_0}}{n_k}\right| \rr(y)  \lesssim  \bar J(y) \lesssim  
 \frac{\widetilde{n_0}}{n_d} +\left(\!\frac{\widetilde{n_0}}{n_d} \!-\!  \frac{\widetilde{n_0}}{n_u}\!\right)\! 
\MM_{k}(y)+ \left|1\!-\! \frac{\widetilde{n_0}}{n_{k+1}}\right|\rr(y),
\label{barepsilon-long''}
\eea
where \ $n_k\equiv \left\{\!\!\ba{ll} n_u \: & \mbox{if $k$ is  even}\\
n_d \: & \mbox{if $k$ is  odd} \ea\right.$,  $\MM_{0}(y)\equiv 0$, and, if $k>0$,
\bea
\MM_k(y)\equiv \frac{1}{\sqrt{\bom(y)}}\left[\sqrt{\bom[\cy_{-2}(y)]}
\!+\!\sqrt{\bom[\cy_{-4}(y)]}\!+\!...\!+\!\sqrt{\bom[\cy_{-2k}(y)]}\right],\qquad 
 \rr(y)\equiv\sqrt{\!\frac{\bom_0}{\bom(y)}}. \nonumber
\eea
\end{lemma} 

\bp{}  
The claim immediately 
follows from (\ref{cond0'}) if $y\le \cy_2(0)$, i.e. if $\bar\phi(y,0)\le\pi$. 
If $y(l)\ge\cy_2(0)$, i.e. if $\phi_f=\bar\phi(l,\!0)\ge\pi$, then  $\forall y\ge\cy_2(0)$
we can split the integral in (\ref{barepsilon'}) into
\bea
\bar J(y)= \bD\mrq(y,\!0)+\!\!\int^{y}_{\cy_{-2}} \!\!\!\!\!\!\! dy_0
 \frac{\widetilde{n_0}}{\bar n(y_0)}   \bD"(y,\!y_0)+
\!\!\int^{\cy_{-2}}_{\cy_{-4}} \!\!\!\!\!\!\!\!\!\!dy_0
\frac{\widetilde{n_0}}{\bar n(y_0)} \bD"(y,\!y_0)+....+
\!\!\int^{\cy_{-2k}}_{0} \!\!\!\!\!\!\!\!\!\! dy_0
\frac{\widetilde{n_0}}{\bar n(y_0)}  \bD"(y,\!y_0)
 \label{barepsilon-long'}
\eea
where $\cy_{-2k}\equiv \cy_{-2k}(y)$.
Since $\bD"\ge 0$ in the first, third,... integration interval,
 while $\bD"\le 0$ in the second, fourth,... ones,  it follows on one hand
\bea
\bar J(y) \ge  \bD\mrq(y,\!0)+ 
 \frac{\widetilde{n_0}}{n_u} \int^{y}_{\cy_{-2}} \!\!\!\!\!\!\! dy_0\,  \bD"(y,\!y_0)+
\frac{\widetilde{n_0}}{n_d}  \!\!
\int^{\cy_{-2}}_{\cy_{-4}} \!\!\!\!\!\!\!\!\!\!dy_0\,  \bD"(y,\!y_0)+....+
\frac{\widetilde{n_0}}{n_k} \int^{\cy_{-2k}}_{0} \!\!\!\!\!\!\!\!\!\! dy_0\,
 \bD"(y,\!y_0)\nn
=  \bD\mrq(y,\!0)+  \frac{\widetilde{n_0}}{n_u}   \left[\bD\mrq(y,\!y_0)\right]^{y_0=y}_{y_0=\cy_{-2}} +\frac{\widetilde{n_0}}{n_d}  
\left[\bD\mrq(y,\!y_0)\right]^{y_0=\cy_{-2}}_{y_0=\cy_{-4}}+....+
\frac{\widetilde{n_0}}{n_k} \left[\bD\mrq(y,\!y_0)\right]^{y_0=\cy_{-2k}}_{y_0=0}\nn
= \frac{\widetilde{n_0}}{n_u}+ \left(\!\frac{\widetilde{n_0}}{n_d} \!-\!  \frac{\widetilde{n_0}}{n_u}\!\right)\! \left[ \bD\mrq(y,\!\cy_{-2})- \bD\mrq(y,\!\cy_{-4})
+...-(-1)^k \bD\mrq(y,\!\cy_{-2k})\right]+  \left(\!1- 
\frac{\widetilde{n_0}}{n_k}\! \right)\! \bD\mrq(y,\!0)\nn
\simeq \frac{\widetilde{n_0}}{n_u} -\left(\!\frac{\widetilde{n_0}}{n_d} \!-\!  \frac{\widetilde{n_0}}{n_u}\!\right)\! 
\left[\sqrt{\frac{\bom[\cy_{-2}(y)]}{\bom(y)}}
\!+\!\sqrt{\frac{\bom[\cy_{-4}(y)]}{\bom(y)}}
\!+\!...\!+\!\sqrt{\frac{\bom[\cy_{-2k}(y)]}{\bom(y)}} \right]- \left(\!
\frac{\widetilde{n_0}}{n_k}-1\! \right)\!\sqrt{\!\frac{\bom_0}{\bom(y)}}
\cos\!\left[\bar\phi(y,\!0)\right] \nonumber
\eea
Replacing the last
term at the rhs by its minimum we obtain  (\ref{barepsilon-long''}a).
Eq. (\ref{barepsilon-long'}) implies also
\bea
\bar J(y) \le  \bD\mrq(y,\!0)+ 
 \frac{\widetilde{n_0}}{n_d} \int^{y}_{\cy_{-2}} \!\!\!\!\!\!\! dy_0\,  \bD"(y,\!y_0)+\frac{\widetilde{n_0}}{n_u}  \!\!
\int^{\cy_{-2}}_{\cy_{-4}} \!\!\!\!\!\!\!\!\!\!dy_0\,  \bD"(y,\!y_0)+....+
\frac{\widetilde{n_0}}{n_{k+1}} \int^{\cy_{-2k}}_{0} \!\!\!\!\!\!\!\!\!\! dy_0\,
 \bD"(y,\!y_0)\nn
=  \bD\mrq(y,\!0)+  \frac{\widetilde{n_0}}{n_d}   \left[\bD\mrq(y,\!y_0)\right]^{y_0=y}_{y_0=\cy_{-2}} +\frac{\widetilde{n_0}}{n_u}  
\left[\bD\mrq(y,\!y_0)\right]^{y_0=\cy_{-2}}_{y_0=\cy_{-4}}+....+
\frac{\widetilde{n_0}}{n_{k+1}} \left[\bD\mrq(y,\!y_0)\right]^{y_0=\cy_{-2k}}_{y_0=0}\nn
= \frac{\widetilde{n_0}}{n_d}- \left(\!\frac{\widetilde{n_0}}{n_d} \!-\!  \frac{\widetilde{n_0}}{n_u}\!\right)\! \left[ \bD\mrq(y,\!\cy_{-2})- \bD\mrq(y,\!\cy_{-4})
+...-(-1)^k \bD\mrq(y,\!\cy_{-2k})\right]+  \left(\!1- 
\frac{\widetilde{n_0}}{n_{k+1}}\! \right)\! \bD\mrq(y,\!0)\nn
\simeq \frac{\widetilde{n_0}}{n_d} +\left(\!\frac{\widetilde{n_0}}{n_d} \!-\!  \frac{\widetilde{n_0}}{n_u}\!\right)\! 
\left[\sqrt{\frac{\bom[\cy_{-2}(y)]}{\bom(y)}}
\!+\!\sqrt{\frac{\bom[\cy_{-4}(y)]}{\bom(y)}}
\!+\!...\!+\!\sqrt{\frac{\bom[\cy_{-2k}(y)]}{\bom(y)}} \right]+  \left(\!1- 
\frac{\widetilde{n_0}}{n_{k+1}}\! \right)\! \bD\mrq(y,\!0)
\nonumber
\eea
Replacing the last term at the rhs by its maximum we obtain (\ref{barepsilon-long''}b), as claimed. 
\ep

{\bf Remarks.} 
The last term  at the lhs(\ref{barepsilon-long''})  vanishes if $n_k=\widetilde{n_0}$, 
what is the case if $k$ is even and  $\check n(\xi,Z)\le \widetilde{n_0}(Z)$ 
or if $k$ is odd and  $\check n(\xi,Z)\ge \widetilde{n_0}(Z)$  for all $\xi\in[0,l]$.
The last term at the rhs(\ref{barepsilon-long''}) 
vanishes   if $n_{k+1}=\widetilde{n_0}$,
what is the case if $k$ is even and  $\check n(\xi,Z)\le \widetilde{n_0}(Z)$  for all $\xi\in[0,l]$
or if $k$ is odd and  $\check n(\xi,Z)\ge \widetilde{n_0}(Z)$  for all $\xi\in[0,l]$.

\subsubsection*{Proof of Proposition \ref{Jsigmabounds}}

If $\phi_f\in[0,\pi/2]$ the claim of Proposition \ref{Jsigmabounds}
directly follows from (\ref{condNoWB}), (\ref{barsigma-bounds0}).
If $\phi_f\in[\pi/2,\pi]$ then (\ref{barepsilon-long'''hat})   follows from (\ref{condNoWB'});
(\ref{barsigma-long'''hat}) follows from (\ref{barsigma-long''}).

If $\phi_f>\pi$
the definition of $\tilde\delta$ implies $\left|1\!-\! \frac{\widetilde{n_0}}{n_k}\right|,\left|1\!-\! \frac{\widetilde{n_0}}{n_{k+1}}\right|\le\tilde\delta$. On the other hand, 
 the definitions  of $\rr,\bom^2$ and the inequalities \cite{FioDeAFedGueJov22} $s_u\ge\hs\ge s_d:=1/s_u$ imply for all $y,y'\in[0,y(l)]$
\bea
\rr(y)=\sqrt{\!\frac{\bom_0}{\bom(y)}}=\left[\frac{\widetilde{n_0}\, \bar s^3(y)}{\bar n(y) \bar\mu^2(y)}\right]^{1/4} \le\br\!_u,\qquad
\frac{\bom^2(y')}{\bom^2(y)}=\frac{\bar n(y')\bar s^3(y')\bar  \mu^2(y)}{\bar n(y)\bar s^3(y)\bar \mu^2(y')}
\le \frac{n_u\, s_u^6\,\mu^2_{\scriptscriptstyle M}}{n_d\, }.
\nonumber
\eea
The latter formula and definition of $\MM_k$ imply $\MM_k(y)\le k \left[\frac{n_u\, s_u^6\,(1\!+\!v_{{\scriptscriptstyle M}})}{n_d}\right]^{1/4}\le \check k \left[\frac{n_u\, s_u^6\,(1\!+\!v_{{\scriptscriptstyle M}})}{n_d }\right]^{1/4}$, where  $\check k\!\in\!\NN_0$ is the integer 
defined by the condition $y(l)\!\in\![\cy_{2\check k}(0),\cy_{2\check k+2}(0)]$. 
But by  definition 
$\check k\pi\le -\cpsi[y(l),0] \simeq \bar\phi[y(l),0]=\phi(l,0)=\phi_f$.
Replacing in (\ref{barepsilon-long''}) we obtain 
\bea
\frac{\widetilde{n_0}}{n_u} -
\left(\!\frac{\widetilde{n_0}}{n_d} \!-\!  \frac{\widetilde{n_0}}{n_u}\!\right)\! 
\MM_u-\tilde\delta\,\br\!_u \: \lesssim \: \bar J(y)  \: \lesssim \:  
 \frac{\widetilde{n_0}}{n_d} + \left(\!\frac{\widetilde{n_0}}{n_d} \!-\!  \frac{\widetilde{n_0}}{n_u}\!\right)\! 
\MM_u+ \tilde\delta \, \br\!_u,\qquad 
\label{barepsilon-long'''}
\eea
which amounts to (\ref{barepsilon-long'''hat}) for all $\xi\in[0,l]$.
The definitions of  $\bom^2,\MN_k$ imply for all $y\in[0,y(l)]$
\bea
\sqrt{\!\frac{\bom(y)}{\bom_0}}=\left[\frac{\bar n(y) \bar\mu^2(y)}{\widetilde{n_0}\, \bar s^3(y)}\right]^{1/4} \le\left[\frac{n_u s_u^3\mu^2_{{\scriptscriptstyle M}}}{\widetilde{n_0}}\right]^{1/4},\qquad
\MN_k \le k \left[\frac{n_us_u^3\mu^2_{{\scriptscriptstyle M}}}{\widetilde{n_0}}\right]^{1/4}
\le \tilde k \left[\frac{n_u s_u^3\mu^2_{{\scriptscriptstyle M}}}{\widetilde{n_0}}\right]^{1/4}.
\nonumber
\eea
where  $\tilde k\!\in\!\NN_0$ is the integer defined by the condition $y(l)\in[ \ty_{2\tilde k-2}(0), \ty_{2\tilde k}(0)]$.
But by  definition 
$\tilde k\pi\le -\tpsi[y(l),0] \simeq\bar\phi[y(l),0]+\pi/2=\phi_f+\pi/2$.
Replacing in (\ref{barsigma-long''}) we obtain 
\bea
 -\left(\!\frac{\widetilde{n_0}}{n_d} \!-\!  \frac{\widetilde{n_0}}{n_u}\!\right)\! 
\MN_u-  \tilde\delta  \: \lesssim \: \frac
{\bar\sigma(y)}{\sqrt{\bom_0\,\bom(y)}} \: \lesssim \:   
\left(\!\frac{\widetilde{n_0}}{n_d} \!-\!  \frac{\widetilde{n_0}}{n_u}\!\right)\! 
\MN_u+  \tilde\delta
\label{barsigma-long'''}
\eea
which amounts to (\ref{barsigma-long'''hat}) for all $\xi\in[0,l]$.

\subsubsection*{Acknowledgments}

Work done also in the framework of the activities of G. Fiore within GNFM.

\end{document}